\RequirePackage{fix-cm}
\documentclass[smallextended]{svjour3}       
\smartqed  
\usepackage{graphicx}

\usepackage[utf8x]{inputenc}
\usepackage{todonotes}

\usepackage[numbers]{natbib}
\usepackage{caption}
\usepackage{subcaption}
\usepackage{subfiles} 
\usepackage{amsmath}
\usepackage{csvsimple}
\usepackage{float}
\usepackage{xcolor}
\usepackage{geometry}
\usepackage [autostyle, english = american]{csquotes}
\usepackage[nottoc]{tocbibind}
\usepackage{setspace}
\usepackage{adjustbox}
\usepackage{footmisc}
\usepackage{nameref}
\usepackage[toc,page]{appendix}
\usepackage[noline,ruled]{algorithm2e}
\usepackage[english]{babel}

\usepackage{tocloft}

\newcommand\notaamas[1]{#1}
\newcommand\onlyaamas[1]{}
\newcommand\results{results_analysis/results}
\newcommand\subtopic[1]{\subsection{#1}}
\newcommand\subsubtopic[1]{\paragraph{#1}}
\newcommand\anonymize[2]{#1}

\newfam\hebfam
\font\tmp=rcjhbltx at10pt \textfont\hebfam=\tmp
\font\tmp=rcjhbltx at7pt  \scriptfont\hebfam=\tmp
\font\tmp=rcjhbltx at5pt  \scriptscriptfont\hebfam=\tmp

\edef\declfam{\ifcase\hebfam 
     0\or1\or2\or3\or4\or5\or6\or7\or8\or9\or A\or B\or C\or D\or E\or F\fi}

\mathchardef\shin   = "0\declfam 98 
\mathchardef\aleph  = "0\declfam 27
\mathchardef\beth   = "0\declfam 62
\mathchardef\gimel  = "0\declfam 67
\mathchardef\daleth = "0\declfam 64
\mathchardef\lamed  = "0\declfam 6C
\mathchardef\mim    = "0\declfam 6D
\mathchardef\ayin   = "0\declfam 60
\mathchardef\tsadi  = "0\declfam 76
\mathchardef\qof    = "0\declfam 72

\begin{document}

\title{A Negotiating Strategy for a Hybrid Goal Function in Multilateral Negotiation}


\author{Alon Stern \and
Sarit Kraus \and
David Sarne}

\institute{A. Stern \at
              Bar Ilan University\\
              \email{alon.stern206@gmail.com}           
           \and
           S. Kraus \at
              Bar Ilan University\\
              \email{sarit@cs.biu.ac.il}           
           \and
           D. Sarne \at
              Bar Ilan University\\
              \email{david.sarne@gmail.com}
}

\maketitle

\begin{abstract}

In various multi-agent negotiation settings, a negotiator's utility depends, either partially or fully, on the sum of negotiators' utilities (i.e., social welfare). While the need for effective negotiating-agent designs that take into account social welfare has been acknowledged in recent work, and even established as a category in automated negotiating agent competitions, very few designs have been proposed to date. In this paper, we present the design principles and results of an extensive evaluation of agent HerbT+, a negotiating agent aiming to maximize a linear tradeoff between individual and social welfare. Our evaluation framework relies on the automated negotiating agents competition (ANAC) and includes a thorough comparison of performance with the top 15 agents submitted between 2015-2018 based on negotiations involving 63 agents submitted to these competitions.  We find that, except for a few minor exceptions, when social-welfare plays a substantial role in the agent's goal function, our agent outperforms all other tested designs.

\end{abstract}

\section{Introduction}
\label{introduction}

\notaamas {

A negotiation is a form of interaction between entities who compromise in order to agree on matters of mutual interest. Negotiation is an integral part of our life, carried out on a daily basis, even unknowingly \cite{GettingtoYes}. It can be over the choice of a restaurant for dinner with your spouse, the terms of a contract for a new job, or the conditions set for moving to a new house. Negotiation can take place between people, between companies, or even between states. The impact of negotiation on our lives has led researchers to study the field of automatic negotiation. With today's hardware and advances in computational methods, automated negotiation has the potential to solve complex negotiations that a human cannot solve. A human might find it challenging to participate in and keep track of negotiations whenever the number of negotiated issues is rather large. Furthermore, in environments with continuous changes, such as the stock market, continuous negotiation is required, which is unfeasible for a non-automated agent. In 2006, as part of a large-scale research study aiming to compare the performance of automated agents against human agents, it was shown that, at times, scenario-dependent, automated negotiation outperforms human negotiation \cite{AnAutomatedAgentforBilateralNegotiationwithBoundedRationalAgentswithIncompleteInformation}.

}

\onlyaamas{

Negotiation is an integral part of our life, carried out on a daily basis, even unknowingly \cite{GettingtoYes}. It can be over the choice of a restaurant for dinner with your spouse, the terms of a contract for a new job, or the conditions set for moving to a new house. Negotiation can take place between people, between companies, or even between states. Due to its importance, much research has been devoted over the years to automatic negotiation, carried out by software agents, as a means for overcoming human negotiators' challenges in participating in and keeping track whenever the number of negotiated issues is rather large  \cite{AnAutomatedAgentforBilateralNegotiationwithBoundedRationalAgentswithIncompleteInformation}.

}

A negotiating agent's performance  can be measured by its individual utility and by the social welfare resulting from the negotiations in which it participated. The individual utility of an agent is a representation of how optimal the agreement eventually reached is to the agent\notaamas{. It is calculated based only on the agent's own utility function. S}\onlyaamas{ and s}ocial welfare is a representation of how optimal the agreement is to all parties involved.\notaamas{ It is calculated using the entire population's utility functions.} To date, the majority of the strategies suggested over the years for automatic negotiation have aimed at maximizing the agent's individual utility rather than social welfare, which makes sense as in most negotiation scenarios, the negotiator negotiates in his or her own self-interest.
However, in real life, there are many scenarios where people negotiate, at least to some extent, also for the greater good. For example, consider a charity organization with the novel goal of helping people break out of the cycle of poverty by giving them jobs. While we do expect employees to negotiate the terms of their employment in a way that maximizes their individual utility, the charity will aim to maximize social welfare, either in the form of sum of employee utilities or the number of employees hired within a given budget (or any tradeoff within).   
Or, consider a firm organizing a cooking class as a communal activity to its employees that needs to negotiate the date, menu and other terms with employees representatives from different departments.  Here, each representative aims to maximize the welfare of employees from her department, based on their preferences, whereas the firm aims to maximize overall social welfare.
While the first example above represents bilateral negotiation and the second represents multilateral negotiation, in both examples, as well as in many other representative scenarios, the environment of agents is mixed. Some agents' goal might be to maximize their individual utility and some agents' goal might be to maximize the overall social welfare, or a function of the two. Furthermore, even when negotiators do not have any explicit interest in caring for other negotiators' welfare, their behavior is often influenced by the latter factor. For example, in a survey from 2006 conducted on pro-social behaviors, primarily using field tests, it was shown that people's behavior is generally pro-social rather than  fully self-interested \cite{SurveyofEconomic}.  The appeal of social-welfare maximizing agreements is also explained by people's preference of agreements that will long last. An agreement that is excellent for one side and terrible for the other will not last. For example, a worker who agreed to poor working conditions will very soon search for a better job.
In this paper, we present and provide an extensive evaluation of an automatic negotiation agent strategy aiming to maximize a tradeoff of social and individual welfare. We focus in linear tradeoffs between the two (i.e., a weighted sum), ranging from caring only for own individual utility to a fully social goal function.  \notaamas{, i.e., a weighted sum of social welfare and individual utility where the weight of either one is set by the user according to his will.}  One major challenge in the development of such agent is the modeling of negotiators' utility from a given agreement. This is because typically there is no way to tell which agent is trying to maximize individual utility and which agent is interested also in social welfare (and to what extent), as revealing one's true preferences provides a leverage to the other negotiators.  In that sense, the proposed strategy contributes several innovative concepts that we believe can be useful for other automated negotiation agents' designs. First, in our strategy, we show that an accurate estimate for an opponent's bid acceptance probability can be used as a good measure for its utility from the bid, based on the correlation between the two. Second, we show that separately maximizing individual utility and social welfare and taking the weighted sum of the two, results in an effective hybrid strategy for any trade-off between individual utility and social welfare (see \onlyaamas{Section}\notaamas{Chapter}~\ref{strategy_design} for more information). 
Through an extensive evaluation, based on the Genius infrastructure \cite{Genius}, we show that when fully maximizing social welfare, our strategy was able to outperform all of the top agents in the Automated Negotiating Agents Competition (ANAC) \cite{ANAC} between the years 2015 - 2018, in most tested domains.  Furthermore, we show that our agent's achievements are cross-domain stable and persistent and in most cases only influenced by the discount factor of the domain. A preliminary version of our strategy won first place in ANAC \anonymize{2018 repeated multilateral negotiation league \cite{ANAC2018}}{XXX (removed for double-blind rule)} in the social welfare category.


\section{Related Work}

\notaamas{

In recent years there is a growing interest in strategies for agents engaged in multilateral negotiation, either in the form of extending existing bilateral negotiation protocols   \cite{Monotonicconcessionprotocolformultilateralnegotiation,AlternatingOffersProtocolsforMultilateralNegotiation} or developing new ones \cite{Multiagentnegotiationonmultipleissueswithincompleteinformation}.  The transition from bilateral to multilateral negotiation suggests many challenges, among which managing a concession process that involves several parties and generally longer negotiation rounds. 

}


The problem of maximizing the social welfare of multiple entities was often solved by using a centralized approach. In this approach, there is a centralized entity (aka mediator approach) that is aware of all the agents and their preferences.  This entity uses this information to determine the next course of action that will maximize social welfare. However, this approach is not always feasible and has a few drawbacks. It has no notion of privacy as each agent is obligated to reveal its preferences to the centralized entity \cite{MultiagentResourceNegotiationforSocialWelfare}. In addition, sometimes, a centralized approach is computationally too expensive as the number of agents might be substantial \cite{ApplicationofAutomatedNegotiationtoDistributedTaskAllocation}.
To overcome these drawbacks there have been attempts to come up with decentralized approaches using automated negotiation \cite{MultiagentResourceNegotiationforSocialWelfare,MultiAgentResourceNegotiationfortheUtilitarianWelfare,ApplicationofAutomatedNegotiationtoDistributedTaskAllocation}.
However, the assumptions used in these works are quite strict. In particular, it is assumed that\notaamas{:
\begin{itemize}
    \item The aim of all of the agents is to maximize social welfare. 
    \item Each agent has some knowledge about the preferences of at least some of the other agents.
\end{itemize}}\onlyaamas{ the aim of all of the agents is to maximize social welfare and each agent has some knowledge about the preferences of at least some of the other agents.}
These two assumptions usually come together. Indeed, in an environment where everyone tries to maximize social welfare, there is no interest for an agent to hide information from other agents as they all have a common goal. However, in a mixed environment of agents who try to maximize social welfare and agents who try to maximize their individual utility, the situation is different. Even if an agent seemingly reveals its preferences, there is no way to know whether its goal is to maximize social welfare and the disclosed preferences are its real preferences, or if its goal is to maximize its individual utility and the disclosed preferences are not its real preferences but rather meant to deceive the other agents, thereby causing them to agree to an offer which is good for it.
\onlyaamas{
Most designs for agents in such settings aiming to maximize social welfare were suggested within the context of ANAC \cite{ANAC}.
}

\notaamas{
Social-welfare maximization \cite{AgreeingOnSocialOutcomesUsingIndividualCPnets, ConditionalPreferenceNetworksSupportMultiissueNegotiationswithMediator} was also researched in the context of negotiations with a mediator in mediator-based protocols such as Mediated Single Text Negotiation \cite{ProtocolsforNegotiatingComplexContracts}. The mediator is an unbiased entity without individual preferences whose role is to mediate between the other agents by proposing an offer that each agent either accepts or rejects. The mediator tries to find an offer that will be agreed on by all and will maximize the social welfare of the population. In this case, the agents' aim is to maximize their individual utility. These strategies also assume that the mediator has some knowledge about the agents' preferences.  Most of the work that was done on negotiation with preference uncertainty was done in an attempt to maximize the negotiator's individual utility.
}
ANAC was founded in 2010 \cite{baarslag2012first}. This competition allows researchers to submit an agent that will compete against other agents in different negotiation domains. ANAC aims to bring together researchers from the negotiation community and provide a benchmark and a research agenda for automated negotiation.
\notaamas{ 
Until 2014 the agents in ANAC were evaluated based solely on their achieved individual utility. Since 2014 the agents have also been evaluated based on the average social welfare in negotiations in which they participated.
In 2015 the competition settings were changed from bilateral negotiation to multilateral negotiation \cite{ANAC2015}.

Most of the work on social welfare for multilateral negotiation with preference uncertainty was done for agents who participated in this competition. 
}
We hereby review several strategies that were designed for agents who participated in ANAC and got to the finals in the social welfare category.

\notaamas{
\citeauthor{Agent33} (\citeyear{Agent33}) proposed a negotiation strategy that aims to find bids around the Nash bargaining solution for maximizing social welfare \cite{Agent33}. 
They evaluate what issues and what values are more valuable to their opponents by applying statistical analysis to calculate the standard deviation of the values of the issues and the standard deviation of each issue that was offered by their opponents. Then, they assume that the higher the standard deviation is, the more valuable the issue is to their opponent.
}
Agent Atlas3, proposed by Mori and Ito (2017) \cite{Atlas3}, won in both the social welfare and the individual utility category in ANAC 2015. The agent's strategy consists of a bids searching method and a compromising strategy. The searching method is based on the relative utility compared to the maximum possible individual utility of the agent and the offered values' frequencies.  The compromising strategy divides the negotiation flow to alternative offer phases and a final offer phase, and bases the current concession value on the final offer phase's expected utility.
Gu and Ito (2017) proposed an agent  which tries to maximize social welfare in closed multilateral negotiation \cite{AgentX}. Their agent strategy is strictly built for 3-agent scenarios and is composed of an agreement behavior and a conceder behavior. The agreement behavior tries to make a partial agreement with one agent before it tries to come to an agreement with both agents. It tries to estimate the number of bids with which the opponent may agree by calculating the average and standard deviation of the opponent's last ten bids. The conceder behavior is based on a formula by Faratin and Sierra (1998) \cite{NegotiationDecisionFunctionsforAutonomousAgents} for determining the agreement threshold, which is modified to concede more as time goes by.
Hayashi and Ito (2017) proposed a strategy \cite{AgentH} for maximizing social welfare.
Their agent accepts an offer if the offer's utility is higher than a threshold value that decreases as the negotiation proceeds. The threshold function is tuned using the offered utilities, the time passed, and a constant value that was found by experimental negotiations. Their bidding strategy consists of choosing the highest utility bid that was offered or accepted by the opponents and randomly changing the value of one of its issues in order to increase its utility for the agent.
\notaamas{
\citeauthor{JonnyBlack} (\citeyear{JonnyBlack}) proposed a strategy \cite{JonnyBlack} that acts as a mediator to maximize social welfare in close multilateral negotiation.\notaamas{ Not to be confused with the above discussion of mediators, their agent only sees itself as a mediator and is not really a mediator, as the protocol is not a mediator-based one---the agent has the same position as any other agent in the negotiation.} Their strategy uses the frequencies of the bids offered by the opponents to estimate the utility function of their opponents. Then it tries to find a bid that will be accepted by all the opponents and will maximize the agent's utility. 
}
There are some common properties in the strategies mentioned above, such as the use of a heuristic method on the frequencies of the values of the suggested bids for opponent modeling, a concession strategy based on a threshold that decreases as the negotiation goes on, and the use of constant values chosen empirically based on experimental negotiations. In our work, we tried different approaches. First, our social welfare strategy does not rely on a concession threshold as it has disadvantages for maximizing social-welfare (more information is provided in \onlyaamas{Section}\notaamas{Section}~\ref{strategy_design}).
In addition, for modeling the opponent, we use a machine learning model with some adaptation to its initialization for improving its predictions in automated negotiation settings (see \onlyaamas{Section}\notaamas{Section}~\ref{strategy_design} for more information).




\notaamas{
A key component of our proposed agent design is the usage of machine learning for opponent modeling. The idea is not new, and several prior works use machine learning for opponent modeling as part of automated negotiation.

\citeauthor{WeightingEstimationMethodsforOpponentsUtilityFunctionsUsingBoostinginMultiTimeNegotiations} (\citeyear{WeightingEstimationMethodsforOpponentsUtilityFunctionsUsingBoostinginMultiTimeNegotiations}) proposed a method \cite{WeightingEstimationMethodsforOpponentsUtilityFunctionsUsingBoostinginMultiTimeNegotiations} for estimating the utility function of the opponents in a closed multi-issue negotiation. Their method uses the Boosting algorithm, which tries to combine several "weak learners" into one "strong learner" by weighing each one of the learners using the data collected from the agents' offers during the negotiation. Their paper is focused more on creating a good model with high accuracy and less about providing a negotiation strategy. Their method is only effective when the negotiations are under the same opponents and domains are repeated multiple times, unlike our design that becomes effective from the first negotiation.
\notaamas{
Gaussian process regression \cite{GaussianProcessesforMachineLearning} is also a quite popular model for opponent modeling \cite{Phoenix,OptimizingcomplexautomatednegotiationusingsparsepseudoinputGaussianprocesses,UsingGaussianProcessestoOptimiseConcessioninComplexNegotiationsagainstUnknownOpponents,IAMhaggler,IAMhaggler2011,AgentBuyog}. The usage of this model is a good example for a machine learning model that tries to learn a different thing - here the model is not used to learn the utility function of each opponent or whether the opponent would accept or reject a bid, but instead tries to learn the opponent's concession strategy.
}
\citeauthor{OpponentModellingBayesianLearning} (\citeyear{OpponentModellingBayesianLearning}) proposed an opponent modeling technique using Bayesian Learning\cite{OpponentModellingBayesianLearning} to estimate their opponent’s utility function. However, \notaamas{their model assumes the values of each issue of a domain are ordered and that there is a value with a maximum utility, such that every other value's utility is linearly decreased according to the distance from the maximum value. In our environment, there are two types of issues: integer issue and discrete issue. While this assumption applies to integer issues, it does not apply to discrete issues as the values of discrete issues are not ordered.
}\onlyaamas{their model only supports integer type of issues, and unlike our agent, it does not support discrete issues (see \ref{negotiation_environment} for information about the issue types)}
In addition, their paper focuses only on bilateral negotiation and their strategy's goal is to maximize the agent's individual utility.

Overall it seems that the vast majority of agents' strategies described in prior work do not use machine learning models and that the agents prefer to use heuristic methods and experimental negotiations to build their strategy. It is understandable as machine learning models require a certain amount of data that can only start being collected during the negotiation itself, and when there is a discount factor, the more data one allows himself to collect, the more utility he loses. However, the strategy proposed in this paper, despite this difficulty, was proven efficient for maximizing a trade-off function between social welfare and individual utility.
}

\section{Negotiation Environment}

As a basis for our negotiation environment, we adopt the negotiation model used in ANAC \cite{ANAC}, which is based on the multilateral alternating offers protocol \cite{AlternatingOffersProtocolsforMultilateralNegotiation,baarslag2013evaluating}. In this protocol, which is fully sequential, the agents follow some pre-specified cyclic order which determines their turn to take an action. The actions available to each agent on its turn are: (a) accept the last offer (denoted \emph{bid} onward) made; (b) make a new offer and thus reject the last offer made; or (c) end the negotiation with no agreement.  If an offer made is accepted by all other agents in their subsequent turn, then the negotiation terminates and the utility of each agent is determined according to the accepted bid.  
The negotiation is divided into $T$ rounds, where each round contains one turn for each agent. 
At the end of each round all utilities are discounted as explained below.
If no agreement is reached within $T$ rounds, or if one of the agents decides to terminate the negotiation at round $t\leq T$ with no agreement, then each agent receives some pre-set default utility (termed \emph{reservation value}), respectively discounted.

Each negotiation is associated with a negotiation domain. A negotiation domain is composed of a set of issues which values are the focus of the negotiation, meaning that each bid placed specifies a suggested set of values for the different negotiation issues.  In our environment, there are two types of issues: an integer issue and a discrete issue. An integer issue describes a value that can be any integer between a range of two integers [x, y] where x and y are predetermined in the issue. A discrete issue describes a value that can be any value from a set of values \{x1, x2, x3...\} where the size and the values of the set are predetermined in the issue.  
Each agent $A_i$ is assigned with a utility function $u_{A_i}$, which maps each possible bid to the utility the agent gains if the negotiation terminates with the acceptance of that bid. The utility functions are linear, i.e., can be computed as a weighted sum of the utilities associated with each value of each issue. Thus the utility encapsulated in a bid $b$ to agent $A_i$, denoted $u_{A_i}(b)$ is given by $\sum_{j=1}^{n}w_ju_{A_i}(b_j)$, where $n$ is the number of negotiated issues, $w_j$ is the weight assigned to the $j$th issue and $u_{A_i}(b_j)$ is the utility of Agent $A_i$ from the value $b_j$ assigned to issue $j$ in bid $b$. All weights and utilities from specific issues are picked such that the resulting utility from a bid is necessarily within the interval $(0,1)$. 
Each domain also specifies a discount factor, $df$, which dictates the rate of the decline in the utility of a bid as the negotiation progresses. A utility $U\leq 1$ obtained at round $t\leq T$ is discounted to a value $U\cdot df^{t/T}$.  
\notaamas{\footnote{For example, a bid with a utility of 0.7 in a domain with a discount factor of 0.5 will have a utility of 0.35 in the last round.}}

While all above information about the domain is public and available to all agents in the negotiation, the individual utility functions are considered private and each agent knows only its own utility function.\footnote{In some years, ANAC allowed agents to use some information from previous negotiations (specifically, the identity of the bid upon which the agents agreed). While we allow such knowledge for the other agents in our evaluation, we intentionally preclude such information in our own agent design in order to increase its applicability.}
The goal of the agent is to maximize some goal function which is a discounted weighted sum of the individual utility and social welfare resulting from the bid eventually agreed upon (or the reservation value if the negotiation terminates without an agreement). We use $0\leq \beta \leq 1$ as the weight assigned to the social welfare (and consequently the weight $1-\beta$ to the individual utility component). This enables a wide spectrum of agent designs with respect to the agent's tendency towards social welfare, ranging from a completely self-interested agent ($\beta=0$) to a fully social agent ($\beta=1$). The social welfare is taken to be the sum of all of the agents' utilities (utilitarian social welfare) divided by the number of agents. 
\notaamas{, i.e.: 
\[
\begin{adjustbox}{max width=\columnwidth}{
$utilitarian\_welfare(bid)=\sum_{i}^{number\_of\_agents}u_i(bid)$
}
\end{adjustbox}
\]
Where $u_i$ is the utility function of the i-th agent} 
We divide the utilitarian welfare by the number of agents so the scale of the social welfare values and the individual utility values will be the same, both between 0 and 1.

\notaamas{
Similar to the settings of ANAC 2015 - 2018, we assume that it is possible to iterate over (and evaluate) all of the solution space (i.e., all plausible  bids) in a feasible time. We roughly estimate a feasible time to be one minute, although we are aware that the real run time is affected by additional parameters such as the machine's available resources. The same time limit is enforced by default in the framework we use.
}

\label{negotiation_environment}

\section{Strategy Design}
\label{strategy_design}

We base our agent's architecture on BOA\notaamas{ (bidding strategy, opponent model, and acceptance condition)} architecture, introduced by Baarslag et al. (2014) \cite{DecouplingNegotiatingAgentstoExploretheSpaceofNegotiationStrategies}\onlyaamas{ (see Figure~\ref{architecture} in the appendix)}, which has become a common framework for developing negotiating agents \cite{BraveCat,TNR,IQSon,ParsAgent}. 

\notaamas{
\begin{figure}[H]
    \resizebox{\columnwidth}{!}{
    \includegraphics{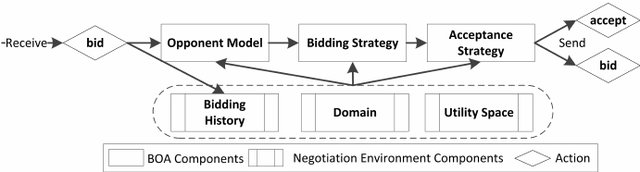}
    }
    
    \caption{BOA architecture (taken from Baarslag et al (\citeyear{DecouplingNegotiatingAgentstoExploretheSpaceofNegotiationStrategies})).}
    \label{architecture}
    
\end{figure}
}

The BOA architecture defines the different components in an agent's strategy used for its decision-making process, which are: opponent model, bidding strategy and acceptance condition. The opponent model component aims to model the opponents' preferences based on their actions thus far in the negotiation. These models are then used for evaluating the social welfare components of bids received and for generating the bid the agent can offer as an alternative.  According to this framework, the agent's decision whether to accept or reject a newly received bid is made based on comparing the utility encapsulated in that bid and the expected utility from proposing an alternative bid of its own.  In the following paragraphs, we provide a detailed description of HerbT+'s specific implementation of the above components.\onlyaamas{ A figure describing the BOA architecture and some more technical details about the strategy are provided in the Appendix.}

\subtopic{Opponent Model}
\label{opponent_model}

In the absence of any information about the other agents' design, strategies and goals, iHerbT+ generates and maintains a different opponent model for each negotiating opponent. Each one of those models is made to predict the probability that the opponent will accept a given offer. 

\notaamas{\paragraph{Learning Algorithm.}}
For constructing the opponent models we use Logistic regression, which is a standard supervised learning regression algorithm used to predict the probability of a target variable \cite{GeneralizedLinearModels}. The nature of a target variable is dichotomous, which fits the nature of the bid acceptance variable in our case, enabling mapping any new bid to a value between 0 and 1, representing the probability of its acceptance by the opponent.
The choice of using logistic regression in our setting heavily relies on the strong correlation between an opponent's utility from a bid and the probability that it will accept that bid\notaamas{ (Figure~\ref{pearson_correlation} demonstrates this correlation)}. Since individual utility functions in our application domain are linear, there is much advantage in using a linear prediction model such as logistic regression.  Furthermore, regardless of the extent to which an opponent's utility depends on social welfare, its goal function maps to a linear combination of the different opponents' individual direct utilities from the bid.  Using a non-linear model in this case might result in overfitting and slower learning \cite{DeterminingOptimumStructureforArtificialNeuralNetworks}.

\notaamas{
\paragraph{Assembling the Data.}
\label{assembling_the_data}
}
\onlyaamas{For assembling the data, a}\notaamas{A}ny newly received bid from an opponent or an opponent's decision to accept or reject a bid received is added to the bidding history and augments the input used for constructing the model of that opponent.  A new bid is a positive example for an offer that the bidding opponent will accept, and so is the acceptance of a bid by an opponent. A decision to reject a bid is a negative example for a bid that will not be accepted by that opponent. Hence a rejection of an offer and consequently the proposal of a counter offer result in two additional inputs (one negative and one positive) for the modeling of that opponent. 

\notaamas{
The processing of the input involves converting each bid to a vector, in a way that values of discrete nature are represented as a sub-vector, using one-hot encoding \cite{ComparativeStudyofCategoricalVariableEncodingTechniquesforNeuralNetworkClassifiers}, where only the value picked is true and all others are false, and values of parameters defined over a range are normalized to values in the range 0-1. The above usage of sub-vectors for discrete values enables correlating the weights assigned with specific values of the issue of concern rather than using a fixed weight.  As for issues which values are taken from a range, the transition to values within the range 0-1 is a common practice in machine learning, aiming to provide a common scale, hence boost convergence of assigned weights by starting with similar initial weights. Each such bid input also includes the indication of whether to be accepted or rejected as the classification result. 
}

\notaamas{
\paragraph{Model Construction.}
}

For model construction, we provide an innovative mechanism, specifically suited for automated negotiation scenarios, to improve the model's prediction based on randomly initializing and training the model from scratch in each round.  In our model construction, we use the Stochastic Gradient Descent Algorithm (SGD) \cite{SGD}\notaamas{, which minimizes the loss function by computing its gradient after each training sample, slightly pushing the weights in the right direction (the opposite direction of the gradient)}. \onlyaamas{When iterating over the samples, rather}\notaamas{Rather} than choosing a single random sample at a time,\notaamas{ moving the weights so as to improve performance on that single sample,} we provide the samples (bids) in the order in which they have been collected by the agent. This choice is made primarily because it is possible that opponents' acceptance strategies evolve over time.\notaamas{ This can be either because they acquire and make use of the information unfolded along the negotiation, due to modeling of their opponents, which keeps evolving throughout, or simply because of using complex strategies with different acceptance and concession strategies for different phases of the negotiation.}  Providing the examples in the order in which they were received thus assigns greater weight to most recent observations.  

\notaamas{
Formally, we consider the hypothesis space $h_\theta(x) = sigmoid(Wx + b)$, where $\theta = (W, b)$ and $W \in R^{|x|}, b \in R$. $x$ is the input bid, converted to a vector as mentioned earlier in the current section, and $h_\theta(x)$ is the predicted acceptance probability of the input bid. We search for $\theta$ which minimizes the cost function: \[
Cost(h_\theta(x), y) =
\begin{cases}
    -log(h_\theta(x))& y = 1\\
    -log(1 - h_\theta(x))& y = 0
\end{cases}
\]
Where $y$ is the actual outcome for the input bid, i.e., 1 if the bid was accepted (or offered) and 0 if it was rejected.

To search for $\theta$ that minimizes the cost function, as mentioned above, we use SGD with the modifications we mentioned earlier:

\DontPrintSemicolon
\begin{algorithm}
\caption{Stochastic gradient descent (SGD)}

\KwIn{Training data $S$, learning rate $\eta$}

\KwOut{Model parameters $\Theta$}

\;

$\theta \gets |x| + 1 \text{ random negative values}$

\For{$(x,y) \in S$}
{
    $\theta \gets \theta -\eta \nabla Cost(h_\theta(x), y)$
}

\end{algorithm}
\vspace{-6mm}

Where $\nabla Cost(h_\theta(x), y)$ is the gradient of the cost function.
In our implementation, we use a learning rate of 0.5.
}

On each turn, we train the model from scratch (as opposed to updating the model based on the last observation received), using random initial weights.
This design approach relates to a specific characterizing property of our setting--while typically the learned model is used for classifying all upcoming examples, in our case the goal is to maximize the benefit from the one bid we place during any of the coming turns that will be accepted by the opponents. Meaning that instead of being relatively accurate with all predictions made, it is far better to be able to accurately predict acceptance probability in at least one turn such that a highly affected bid will be produced and accepted by the opponents. Indeed, any model constructed from scratch using SGD when following all observations in the order received is just as good as the one constructed based on the current model with the necessary changes to weights resulting from the new observation according to the gradient descent, as they only differ in the initial weights assigned at the beginning of the process. However, with a new model during each turn, we substantially increase the probability that one of these new models will be far more accurate than the one that is continuously being updated with the new observations, hence the advantage of our approach. A comparison of our approach against continuously updating the same model is provided in the Appendix.
At the beginning of the negotiation, when the amount of available data about bids and acceptances is very small, an accurate prediction is precluded. To handle this problem, we initialize the logistic regression model weights with random negative values. Doing so guarantees that the predicted acceptance probability for most offers evaluated is substantially small, hence the dominating part in the agent's goal function will be its own direct utility (see below for a more detailed explanation)\onlyaamas{, thus the repercussions of incorrect predictions will be reduced.}\notaamas{. This design choice is made to enable further learning, as much insight will be gained from the acceptance or rejection of such a-typical bids. This changes quite quickly---with the very first bids placed by the opponent or her decision to accept or reject a bid placed by one of the other agents the negative values are changed by the training procedure of the logistic regression. The weight of a value that was accepted by an opponent will increase and the weight of a value that was rejected will decrease further.}
\notaamas{Figure~\ref{mean_absolute_error}, shows the mean absolute error of the model in each round of the negotiation.}

\subtopic{Bid Valuation}
\label{bid_valuation}

As defined in Section~\ref{negotiation_environment}, the agent aims to maximize a discounted trade-off between individual and social welfare which is being captured by the parameter $\beta$.
Knowing the direct utility $u_i(b)$ that each agent gains from a bid $b$ enables calculating the direct individual utility and social welfare offered by this bid, if accepted. Still, the agent only knows its own direct utility function. Indeed, the opponent modeling applied enables estimating the acceptance probability of the bid by the other agents, and since this probability is highly correlated with individual utility it can be used, with some manipulation, to estimate the other agent's utility as part of the social welfare calculation.  Unfortunately, even if one has a good estimation for the weighted utility encapsulated in each bid, it is not necessarily the bid associated with the maximum predicted utility that needs to be picked. Instead, the selection process needs to take into consideration the probability that the bid will be accepted and the (discounted) utility that will eventually be obtained in case it is rejected. 

Considering the bid's chance of acceptance as part of the selection process is highly challenging, as predictions are highly noisy at early stages and the influence of an inaccurate estimate can be devastating. Estimating the expected utility from the continued negotiation in case the bid is not accepted is even more challenging, as the opponents' negotiation strategies are unknown to the agent and therefore difficult to model. We therefore turn to an alternative approach, one that treats each of the two components of the agent's goal function (individual utility and social welfare) differently. The idea is to assign a score to the bid and choose the bid associated with the maximum score. While the score we assign to bids is expressed in terms of utility, its calculation does not adhere to the traditional expected utility calculation principles, as we explain in the following paragraphs.\notaamas{\footnote{We attempted to create a score based on expected utility calculation, but the results were lower than the results of the current approach. Still, for completeness, this approach is reported in the Appendices chapter.}}
For the direct individual utility part of a bid, which is known to the agent at the time it makes its decision, we enforce a utility threshold, i.e., consider a value for this part only if the direct individual utility is above the threshold. The threshold starts at a value of 1 and linearly decreases over time until reaching the discounted reservation value, meaning that there will always be a bid above the threshold since there is always a bid with a utility value of 1. For bids offering a direct individual utility exceeding the threshold, we take their contribution to the bid's overall score to be the average acceptance probability of the bid by the opponents (which, as discussed above, is a good measure for their utility from that bid). This is done in order to increase the score of bids that are likely to be accepted, among those that offer an individual utility greater than the threshold set. \notaamas{Otherwise, i.e., if the bid offers the agent a direct individual utility smaller than the threshold set}\onlyaamas{For other bids}, the individual utility component adds no contribution to the score.\notaamas{ This means that in such case, the bid must get an exceptionally high score in the social welfare component in order to compensate for the relatively low individual utility, in order to be picked as the winning bid.}  For the social-welfare part of the bid, which is highly noisy (and hence has a high potential for substantially damaging the score), we use a heuristics score function. The function estimates the average of the utilities that each agent gains from the bid. For our agent, we use our utility function, $u_{ours}(bid)$, as we know it. For the opponents, using the correlation assumption mentioned above, we use our opponent modeling output for a bid as an estimate of the utility for the opponent. To reduce the effect of estimation inaccuracies we take the square of the opponent modeling output for the score calculation. Thus, the calculation of $social\_score(bid)$ is:
\[
    \frac{u_{ours}(bid) + \sum_{A_i\in \{ opponents \}} (u'_i(bid))^2}{number of opponents + 1}
\]
where $u'_i(bid)$ is the estimated acceptance probability according to the opponent modeling.
\notaamas{
This approach for bid valuation is very different from most strategies for automated negotiation with preference uncertainty as it is not based on a concession threshold. Most strategies, including social welfare-based ones \cite{Agent33,Atlas3,AgentX,AgentH,JonnyBlack}, maintain a threshold that decreases along the negotiation, such that the agent will not accept or propose offers with a utility lower than the threshold. Our strategy does not use such a threshold.\footnote{indeed it uses a threshold for the determination of the individual utility score, yet for different considerations as explained above.} Instead, it always turn to the bid that maximizes expected score, where the only thing that changes along the negotiation rounds is the agent's modeling of the opponents' willingness to accept the different bids and consequently the utilities they encapsulate. Thus, as the negotiation progresses, our agent will offer bids that are more likely to be accepted by its opponents.
}
\onlyaamas{
Unlike most other negotiation strategies \cite{Agent33,Atlas3,AgentX,AgentH,JonnyBlack}, our approach for social welfare maximization does not include a concession threshold.
Instead, it only relies on changes in the opponents' models for its concessions, which makes our agent offer bids that are more likely to be accepted as the negotiation progresses.
}
This innovative approach is effective for reaching early agreements since the use of a concession threshold strictly forces the agent to reject a bid (until a certain point), regardless of the domain's preference homogeneity\notaamas{ (preference homogeneity is described in Section~\ref{experimental_domains})} or the willingness of the opponents to compromise. As discussed in Section~\ref{discount_factor_implications}, reaching early agreements is extremely effective for maximizing social welfare.
The two score components above are then weighed according to $\beta$. Thus the calculation of $score(bid)$ is:
\begin{adjustbox}{max width=\columnwidth} {
$\beta*social\_score(bid) + (1-\beta)*individual\_score(bid)$ 
}
\end{adjustbox}
and the bid associated with the highest score (in a tie, one of the top bids is chosen randomly) is the one to be used as the alternative when evaluating the currently received bid\notaamas{ (i.e., the bid that the agent needs to decide whether to accept or reject)}. Searching for the bid with the highest score presents a drawback: we assume it is feasible to evaluate the entire bid space in one turn. While this is true for the settings in ANAC between 2015 and 2018, it is problematic in other settings where the space is too big to iterate it fully. Still, in most real-life negotiation environments, such as the scenarios described in the introduction, the negotiation space is reasonable, and the negotiation cycles are long enough to reason about the entire space \cite{automatednegotiationwithdecommitmentfordynamicresourceallocationincloudcomputing,AnApplicationofAutomatedNegotiationtoDistributedTaskAllocation}.

\subtopic{Acceptance Condition}
Upon receiving a bid, our agent compares the discounted score of the bid against the discounted score of the next bid our agent will offer according to the bidding strategy. The discounting is calculated according to the time each one of these bids might get accepted by all agents. The agent accepts the proposed bid only if its score is higher than or equal to the next bid's discounted score.\notaamas{ If the current bid and the new bid both apply to the same round, then no discounting takes place.}

\section{Experimental Design}

A preliminary version of our negotiation agent\anonymize{ (AgentHerb)}{} was submitted to \notaamas{the ninth Automated Negotiating Agents Competition (ANAC)}\onlyaamas{ANAC\anonymize{ 2018}{}} and won first place in the social-welfare category. The core difference between the submitted agent and HerbT+, which evaluation is described in the following paragraphs, is that while the latter fully implements the design proposed in the former section, the first is a specific instance that does not make any tradeoff between individual utility and social welfare (which is equivalent to using $\beta=1$).
We hereby detail the evaluation process of HerbT+, as well as the choices made during the evaluation and their reasoning.

\notaamas{

\subtopic{Framework}
The evaluation was run using the GENIUS framework \cite{Genius}. GENIUS \notaamas{-- \textbf{G}eneral \textbf{E}nvironment for \textbf{N}egotiation with \textbf{I}ntelligent multi-purpose \textbf{U}sage \textbf{S}imulation }is a framework for designing and evaluating agents for automatic negotiations. It is used as the primary framework for the ANAC competitions and is actively used in academic research. Using this framework, one can create an agent that can negotiate in any domain. The GENIUS framework also allows the user to define a domain for the negotiation and to define utility functions which represent different preferences in the domain.
Once a domain and agents are chosen for the negotiation, each utility function of the domain is assigned to an agent. The framework allows the user to run the negotiation with the agents he chose over the domains he chose, and analyze the results.\notaamas{ In the framework, a domain is composed of several issues. As mentioned in Section \ref{negotiation_environment}, each issue can be either an integer issue or a discrete issue.} GENIUS framework provides a comprehensive set of agents and domain scenarios that the user can use to evaluate his strategy. Most of these domains and agents are based on submissions from previous years of ANAC. Using this framework and the agents it provides, we ran an extensive set of negotiations to evaluate our strategy.

\begin{figure}[H]
    \resizebox{\columnwidth}{!} {
    \includegraphics{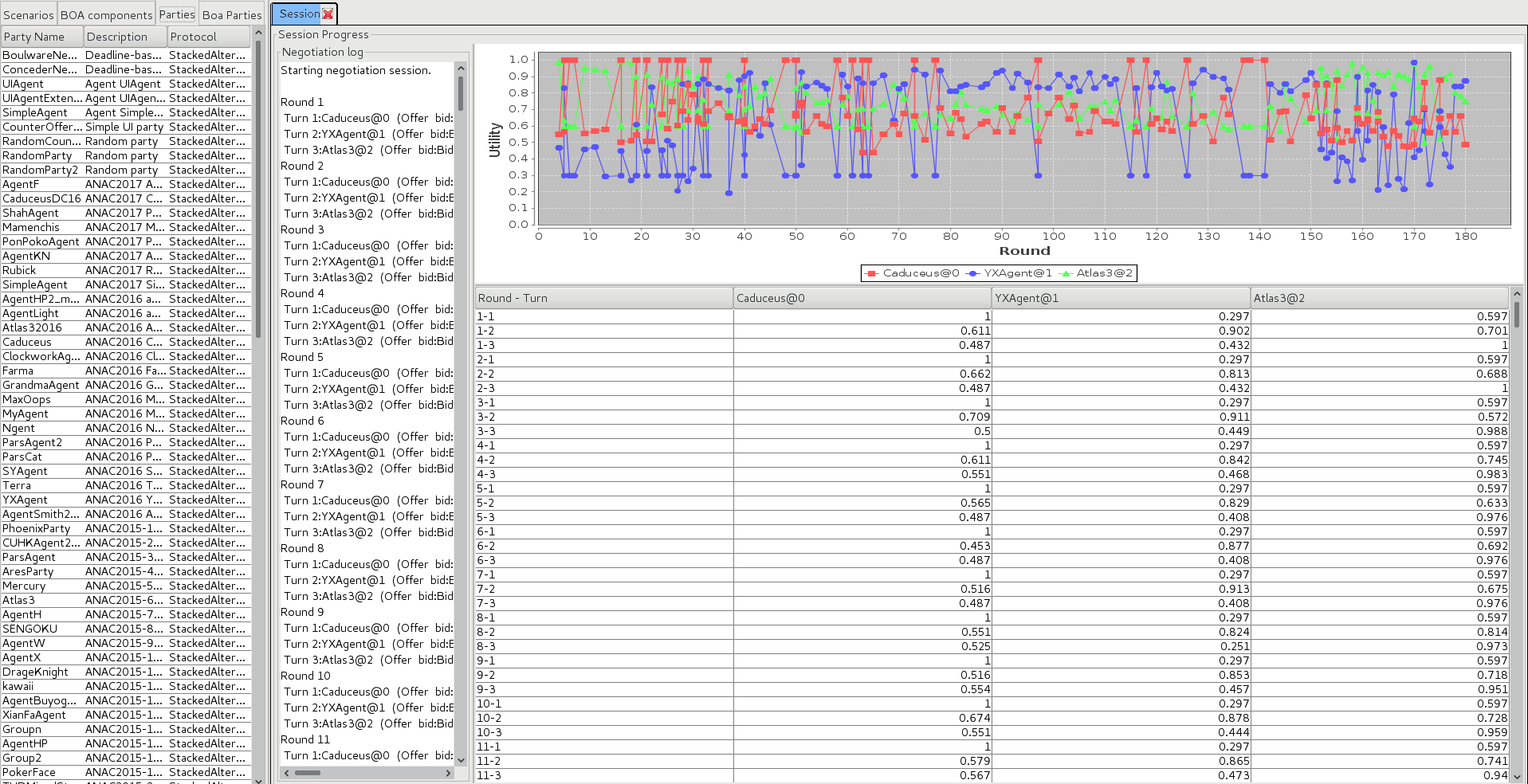}
    }
    \caption{GENIUS framework \cite{Genius}}
\end{figure}

}

\notaamas{
To enable the execution of a large volume of negotiations based on which the evaluation takes place, we modified the source code of GENIUS to enable us to run the pools procedure, which is described in Figure~\ref{pools_procedure}. In addition, for analyzing the results, we modified the GENIUS's source code to receive additional information from the negotiation process about the actions of the agents during the negotiation.
}

The evaluation used seven different domains, each characterizing different classes of negotiation settings, and 63 negotiation agents (other than HerbT+).
\onlyaamas{
To run the evaluation we used the GENIUS framework \cite{Genius}. GENIUS is a framework for designing and evaluating agents for automatic negotiations. It is used as the primary framework for the ANAC competitions and is actively used in academic research.
}
In ANAC, the time limit of each negotiation was 180 seconds. In our case, a limit of 180 seconds per negotiation would be impossible to run as we ran millions of negotiations in our evaluation. Therefore, we ran our experiment with a limit of 180 rounds, which is the default limit in the GENIUS framework. To ensure the difference between the time limit and the round limit is not critical to our agent performance, we did run some experiments with a time limit of 180 seconds.

\notaamas{
\subsubtopic{Domains}
\label{experimental_domains}
}
\onlyaamas{In our experiment, a domain is characterized based on six parameters:}\notaamas{A domain characterizes a class of negotiation settings based on six parameters:} \onlyaamas{the number of issues in the domain, the size of the domain (i.e., the number of possible outcomes in the domain calculated by the multiplication of the number of the possible values of each issue in the domain), the discount factor of the domain, the reservation value of the domain, the types of issues in the domain (i.e., integer or discrete), and the preference homogeneity in the domain.} 
\notaamas{
\begin{itemize}
    \item The number of issues in the domain.
    \item The size of the domain (i.e., the number of possible outcomes in the domain).
    \item The discount factor of the domain.
    \item The reservation value of the domain.
    \item The types of issues in the domain, i.e., integer or discrete.
    \item The preference homogeneity in the domain.
\end{itemize}
}

Most of these characteristics were also used in ANAC \cite{ANAC2015}.
The last parameter, preference homogeneity, reflects the extent to which the different agents' utility functions align with each other. It is measured as the maximum social welfare the agents can reach in the domain\onlyaamas{.}\notaamas{, i.e.,   

\[
\begin{adjustbox}{max width=\columnwidth}{
$preference\_homogeneity(domain)=\max_{bid \in domain}\sum_{i}^{number\_of\_agents}u_i(bid)$
}
\end{adjustbox}
\]

}

\notaamas{
The greater the maximum possible social welfare, the greater the homogeneity between the utility functions, whereas low preference homogeneity means there are more conflicts between the utility functions used, hence it will be harder for the agents to reach an agreement.}
For the evaluation, we used the following domains:
\begin{table}[H]
\onlyaamas{
\vspace{-6mm}
}

\centering
\resizebox{\columnwidth}{!}{
\begin{filecontents*}{experimental_design/domains.csv}
Name,Number of Issues,Domain Size,Discount Factor,Reservation Value,Issues Type,Preference Homogeneity
Domain A,2,25,0.2,0,discrete,2.21
Domain B,4,320,0.5,0.2,discrete,1.59
Domain C,4,320,0.8,0,discrete,2.66
Domain D,2,25,1,0.2,discrete,2.6
Domain E,16,65536,0.5,0.2,discrete,2.68
Domain F,2,49,0.5,0,integer,2.57
Domain G,2,25,0.8,0.7,discrete,2.21
\end{filecontents*}
\csvautotabular{experimental_design/domains.csv}

}
\caption{The domains used for the evaluation.}
\label{domains}

\onlyaamas{
\vspace{-10mm}
}
\end{table}
The above domains are based on a subset of the prebuilt domains and preferences from GENIUS 8.0.4: "Domain2", "Domain4", "Domain16", and "testIntDomain". In order to reach a highly varied set of settings, including those typically used in ANAC, we changed the discount factor and reservation value in a few of the prebuilt domains. This choice of domains aimed to provide a balanced mixture of small, medium and large solution spaces (yet still iterable in a feasible time), different kinds of discount factors, discrete and integer issues, low, medium and high reservation values and low and high preference homogeneity values.
\notaamas{
\subsubtopic{Agents}
\label{experimental_agents}
}
For the evaluation, we used agents from ANAC 2015 - 2018, which were available to us via the GENIUS framework \cite{Genius}\notaamas{ (see Table~\ref{Tabl:agents})}. Agents from ANAC competitions which took place before 2015 were designed for bilateral negotiation and therefore are not suitable for multilateral negotiation. As for more recent ANAC competitions, after ANAC 2018 more constraints were added to the contest framework, such as having incomplete knowledge about the agent's own utility function. Therefore agents designed for those competitions could not be used either.
\notaamas{In all these years, except for 2016, ANAC had two categories for the multilateral negotiation league: individual utility and social welfare. The winners of the individual utility category and the winners of the social welfare category are the ones who achieved the highest individual utility and social welfare, respectively.

\begin{table}[H]
\centering
\begin{filecontents*}{experimental_design/agents.csv}
Competition,Number of agents,Competition Categories
ANAC 2015,22,social welfare and individual utility
ANAC 2016,15,individual utility
ANAC 2017,8,social welfare and individual utility
ANAC 2018,18,social welfare and individual utility
\end{filecontents*}
\csvautotabular{experimental_design/agents.csv}
\caption{Agents chosen from evaluation pool} \label{Tabl:agents}
\end{table}

}
In the evaluation, we wanted to assess our agent's performance compared to all types of agents, those who achieve higher utilities relative to others and those who do not. Therefore we want to let our agent negotiate against all of the agents available to us. Most multilateral agents support negotiating with two agents and are evaluated in three-agent scenarios, as those are the scenarios that are used in ANAC. Therefore, in our evaluation, we used three-agent scenarios as well.
In order to compare the performance of all of these agents, a run of every scenario of three agents from the set of agents is required. With an upper limit of 180 rounds per negotiation, which is the default round limit in our framework, the time required to run all of these negotiations becomes impractical. In order to reduce the number of negotiations, we chose to compare our agent's performance against only the top agents. However, each of those agents will negotiate with all of the available agents.
\onlyaamas{
Thus we will only need to run the negotiations containing at least one of the top agents (i.e., the agents in the top places in both social welfare category and individual utility category in ANAC) instead of running all of the negotiations. We chose the top 15 agents with which to compare our agents from a total of 63 agents used for the evaluation, not including our agent, from ANAC 2015 - 2018. From the top agents, 7 of them excelled in the individual utility category, 7 of them excelled in the social welfare category and 1 in both. Using these agents, we ran over 10,000,000 negotiations in our experiment.
}
\notaamas{
Therefore we create two pools. The first pool, i.e., evaluation pool, is used for evaluating a single agent's performance and includes all of the agents available to us. An agent evaluated using this pool will negotiate against every agent from this pool. The second pool, i.e., comparison pool, is one that contains the agents for whom we are interested in comparing their evaluations. The agents from the comparison pool are the agents who will be evaluated using the evaluation pool and compared to each other. In addition, since we are also interested in evaluating an agent against the top agents, the comparison pool is contained inside the evaluation pool. Using these two pools, we run every scenario of agents from the evaluation pool only if it contains at least one agent from the comparison pool. 

\begin{figure}[H]
    \resizebox{\columnwidth}{!} {
    \includegraphics{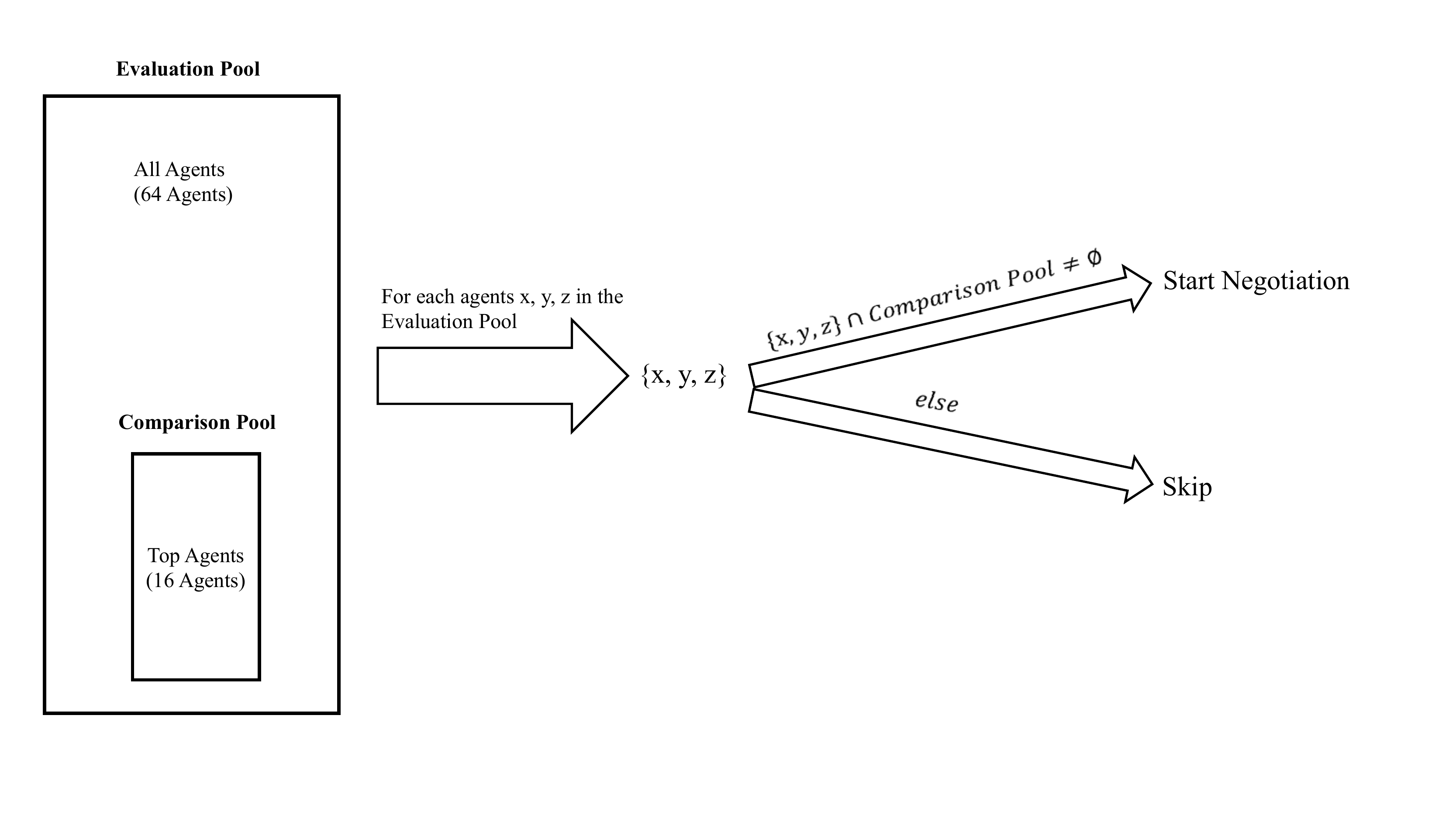}
    }
    \caption{The evaluation process of the top agents.}
    \label{pools_procedure}
\end{figure}

For the evaluation pool, we chose all of the agents available to us, a total of 63 agents, not including our agent, from ANAC 2015 - 2018. For the comparison pool, we chose our agent and 15 more agents who reached the top places in these years, either in the social welfare category or in the individual utility category. 
The following agents were used for the comparison pool:

\begin{table}[H]
\centering
\begin{filecontents*}{experimental_design/top_agents.csv}
Agent,Competeion,Acheivemnt
Agent33,ANAC 2018,social welfare
Agent36,ANAC 2018,individual utility
AgentH,ANAC 2015,social welfare
AgentKN,ANAC 2017,social welfare
AgentX,ANAC 2015,social welfare
AgreeableAgent2018,ANAC 2018,individual utility
Atlas3,ANAC 2015,individual utility and social welfare
Caduceus,ANAC 2016,individual utility
JonnyBlack,ANAC 2015,social welfare
ParsAgent,ANAC 2015,individual utility
PonPokoAgent,ANAC 2017,individual utility
Rubick,ANAC 2017,individual utility
ShahAgent,ANAC 2017,social welfare
Sontag,ANAC 2018,social welfare
AgentYX,ANAC 2016,individual utility
\end{filecontents*}
\csvautotabular{experimental_design/top_agents.csv}
\caption{Top agent chosen from the comparison pool}
\label{top_agents}
\end{table}

Using this procedure, we significantly reduce the number of negotiations without compromising on the quality of the evaluation of each agent, as each agent from the top agents is still evaluated against all the types of agents available to us.
}

\section{Results Analysis}

In the following \onlyaamas{subsections}\notaamas{sections}, we show the performance of our agent, HerbT+, compared to the performance of the chosen top agents. Furthermore, we provide an extensive analysis that sheds light on the contribution of several important design choices made and their contribution for the agent's success. \notaamas{ We will start by providing the overall performance of our agent and then proceeds to discuss the observations and insights we have made about our agent from the results of the evaluation.}

\subtopic{Agent Performance}

In the following paragraphs, we compare how well each agent manages to make the tradeoff between social welfare and individual utility. To do so, we define $beta\ score$ to be the score of an agreement given the tradeoff coefficient $\beta$ between social welfare and individual utility: 
\notaamas{$beta\_score(bid)$ is calculated using the following equation: 
\[
\begin{adjustbox}{max width=\columnwidth}{
$\beta*social\_welfare(bid) + (1-\beta)*individual\_utility(bid)$
}
\end{adjustbox}
\]
}
\onlyaamas{$beta\_score(bid)=\beta*social\_welfare(bid) + (1-\beta)*individual\_utility(bid)$.  
}
\notaamas{$\beta$ of 1 means we look only at the social welfare of a bid, and $\beta$ of 0 means we look only at the individual utility of a bid. }
The graphs in Figure~\ref{beta_score} show the beta score of \onlyaamas{the top 11 agents}\notaamas{each agent} for each beta with jumps of 0.1\onlyaamas{ in domain A and domain E\footnote{For space considerations, we provide the results of additional domains in the Appendix.}, i.e., a small sized domain and a high sized domain.}

\notaamas{

\begin{figure}[H]
    \centering
    \begin{subfigure}[b]{0.45\columnwidth}
        \centering
        \includegraphics[width=\columnwidth]{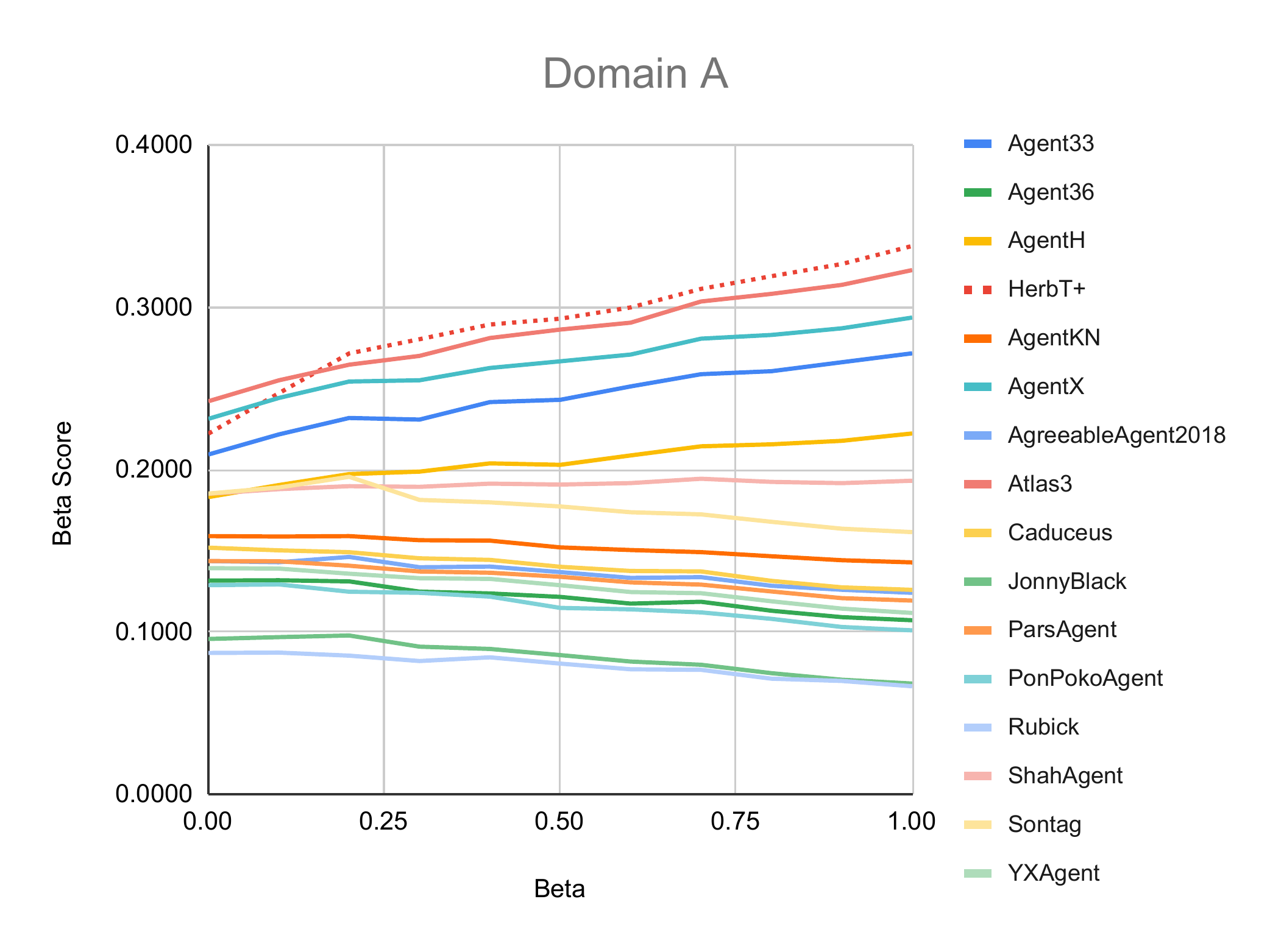}
    \end{subfigure}
    \hfill
    \begin{subfigure}[b]{0.45\columnwidth}
        \centering
        \includegraphics[width=\columnwidth]{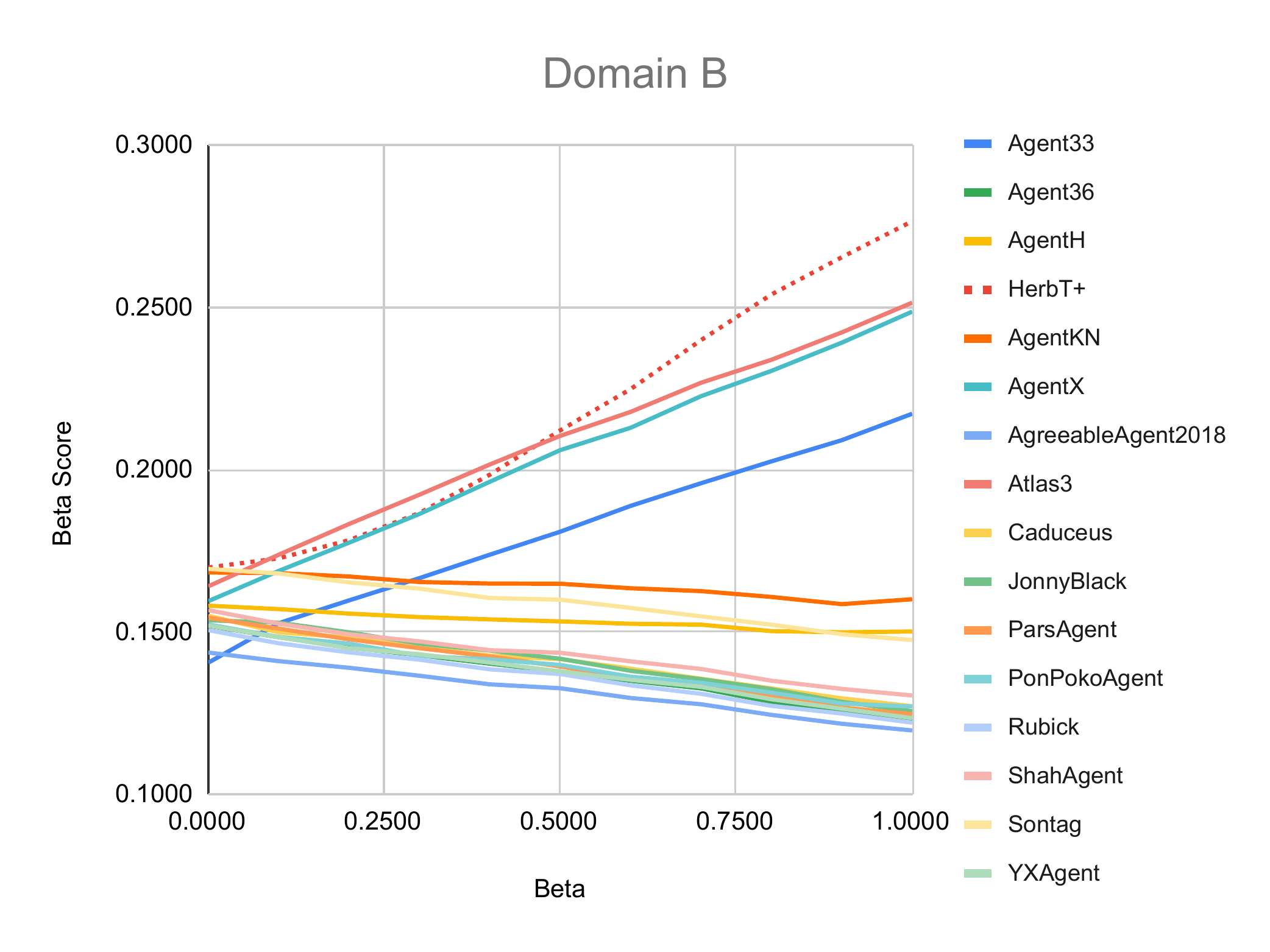}
    \end{subfigure}
    \hfill
    \begin{subfigure}[b]{0.45\columnwidth}
        \centering
        \includegraphics[width=\columnwidth]{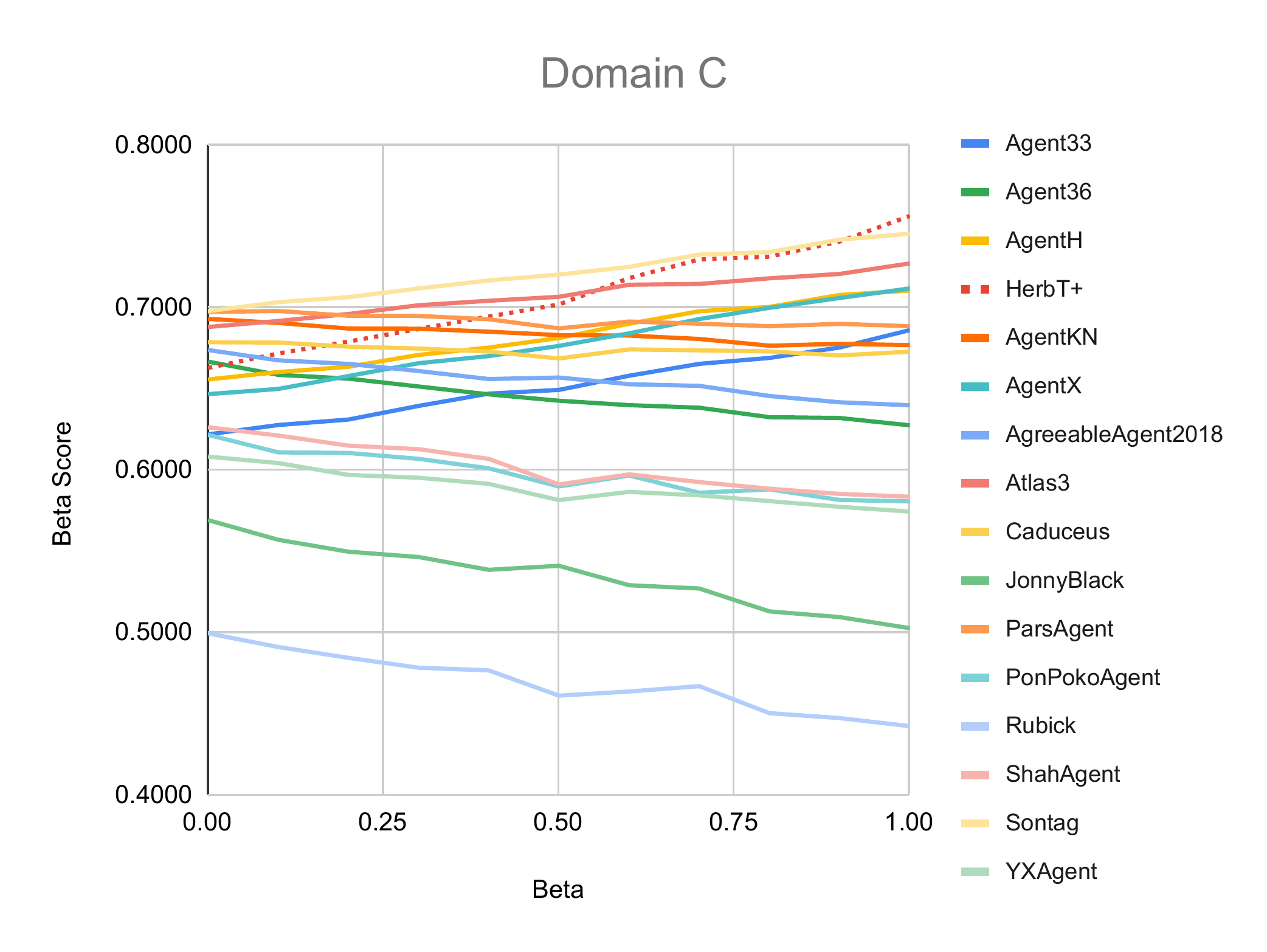}
    \end{subfigure}
    \hfill
    \begin{subfigure}[b]{0.45\columnwidth}
        \centering
        \includegraphics[width=\columnwidth]{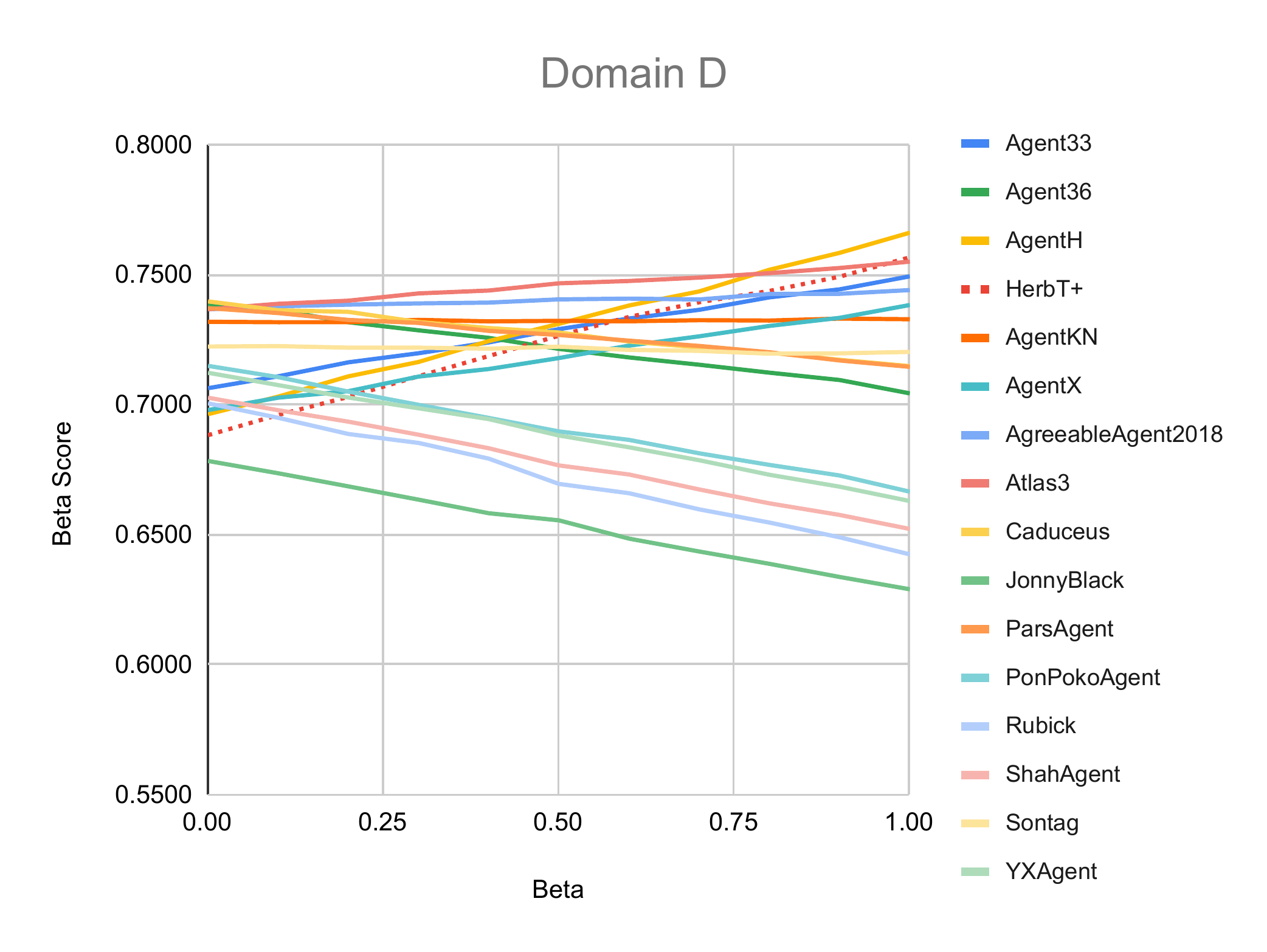}
    \end{subfigure}
    \hfill
    \begin{subfigure}[b]{0.45\columnwidth}
        \centering
        \includegraphics[width=\columnwidth]{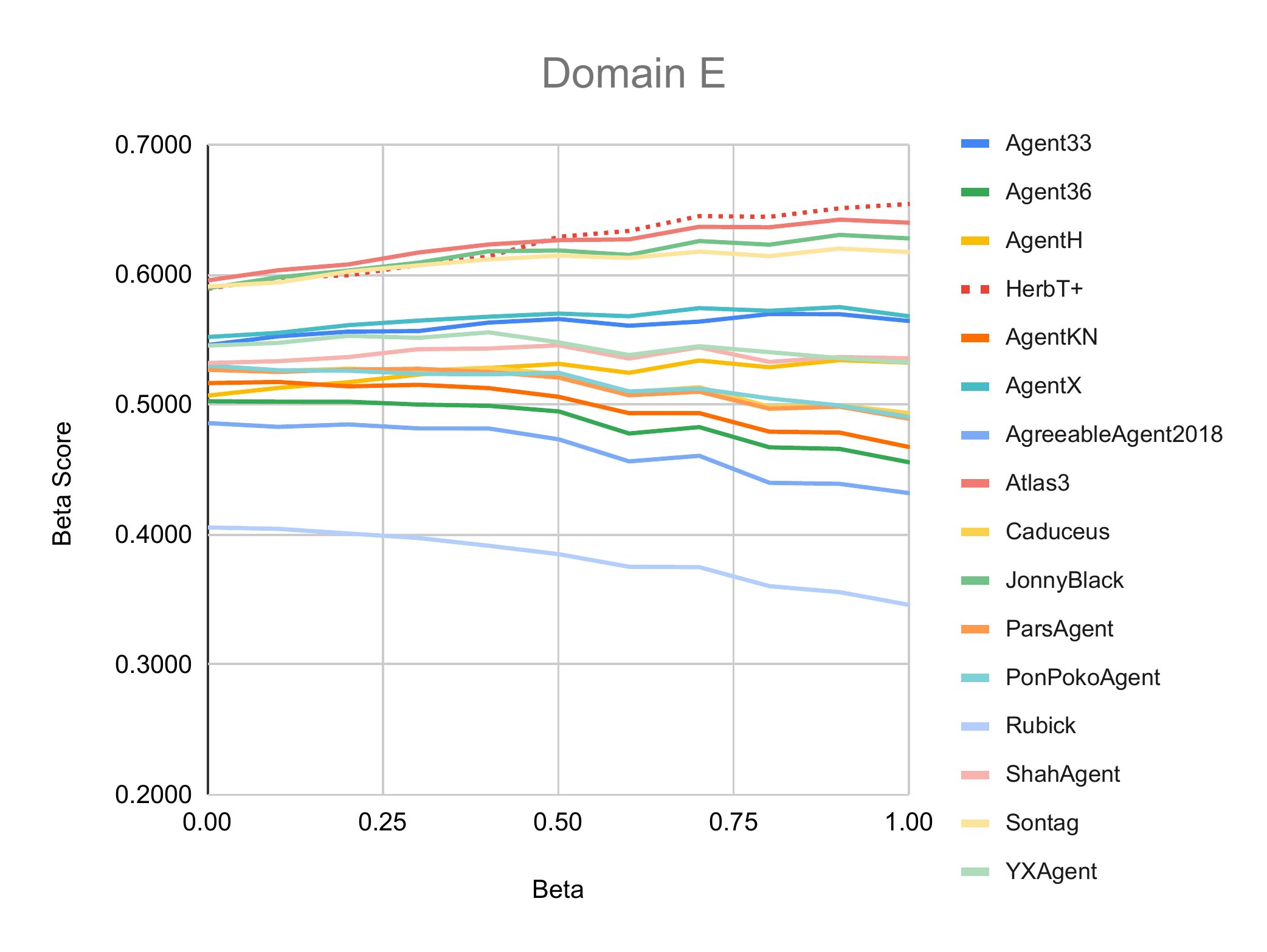}
    \end{subfigure}
    \hfill
    \begin{subfigure}[b]{0.45\columnwidth}
        \centering
        \includegraphics[width=\columnwidth]{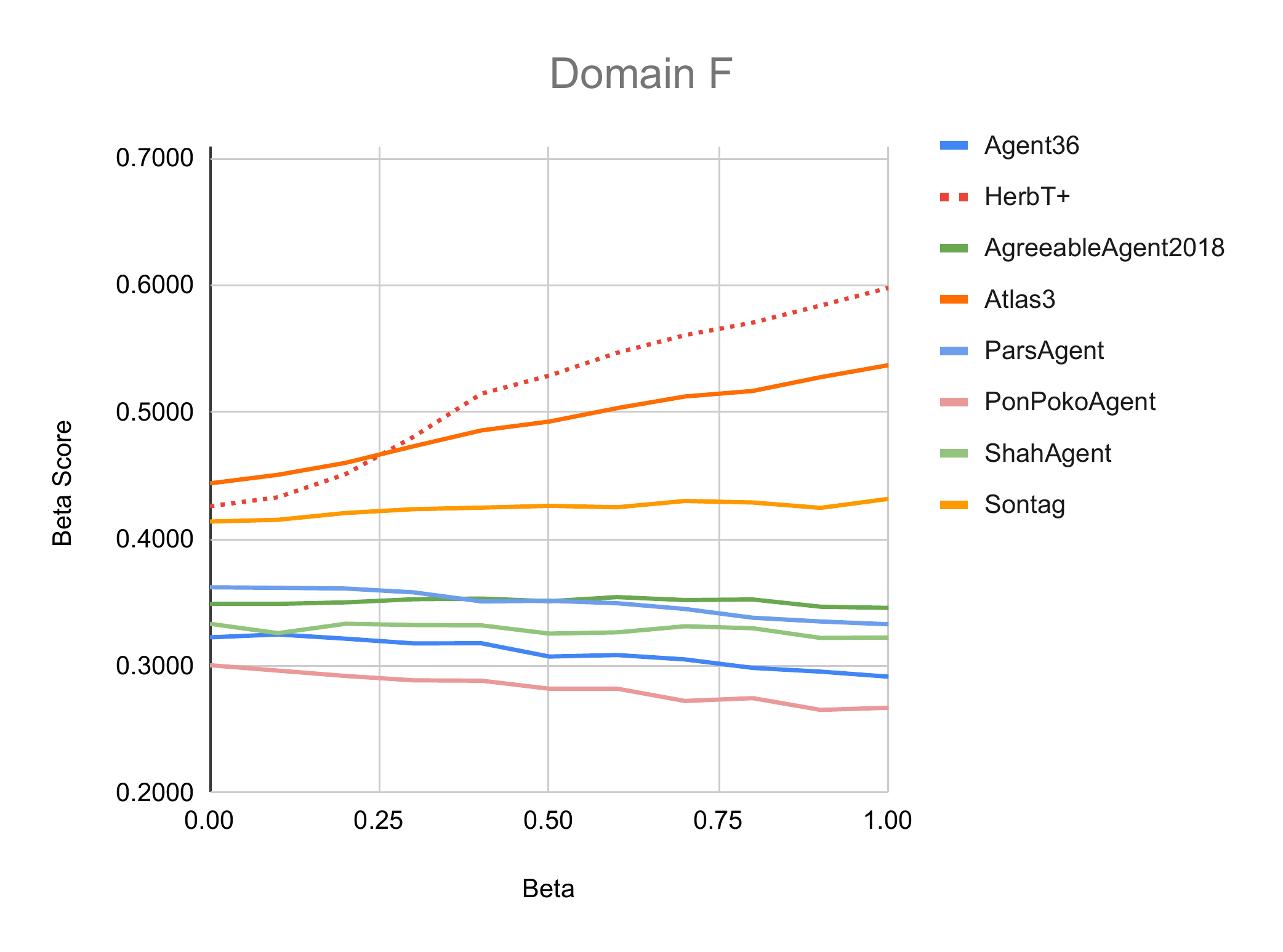}
    \end{subfigure}
    \hfill
    \caption{Beta score of the top agent for each beta}
\end{figure}
\begin{figure}[H]\ContinuedFloat
    \centering
    \begin{subfigure}[b]{0.45\columnwidth}
        \centering
        \includegraphics[width=\columnwidth]{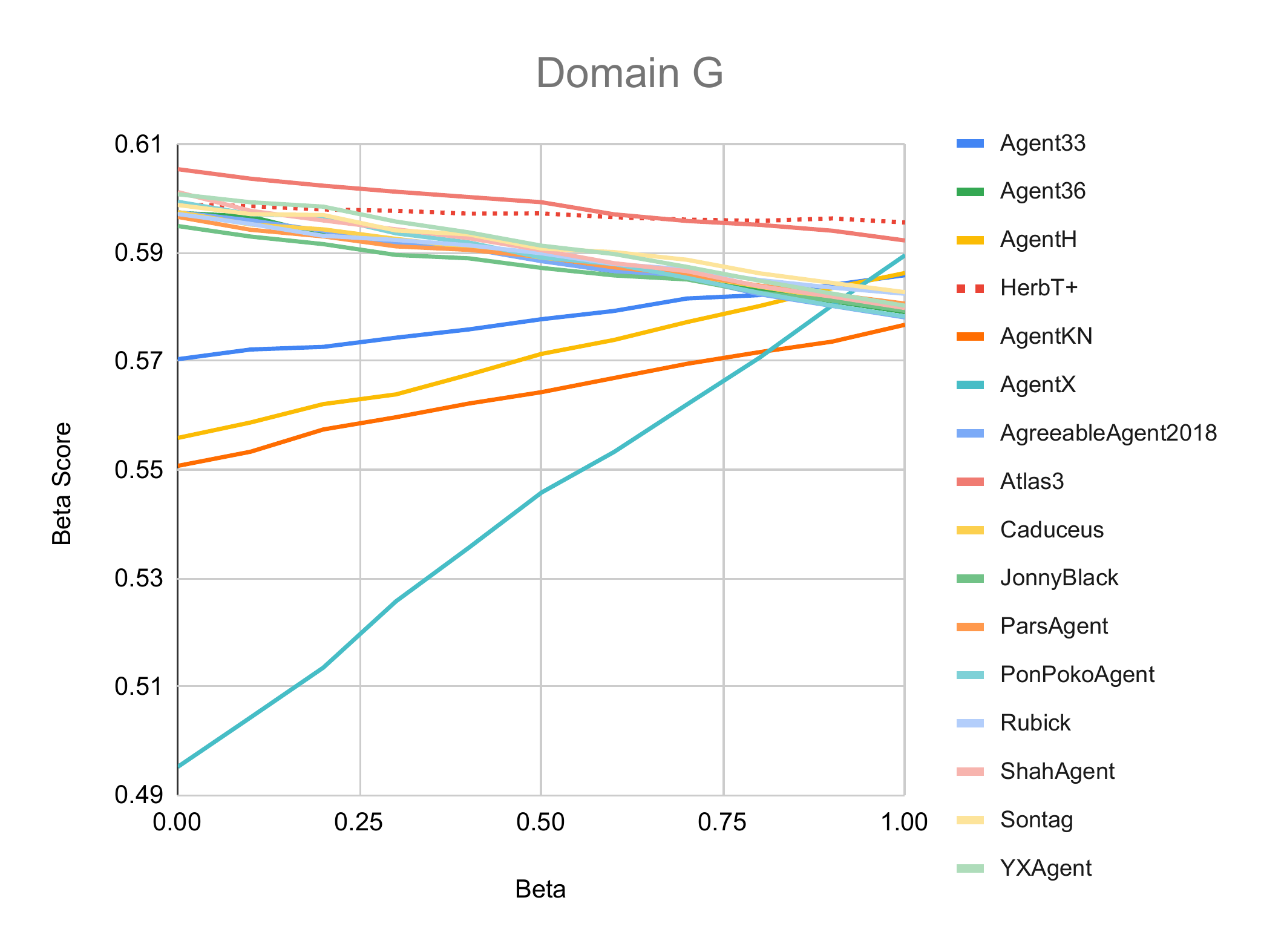}
    \end{subfigure}
    \caption{Beta score of the top agent for each beta}
    \label{beta_score}
\end{figure}

}

\onlyaamas{
\begin{figure}[ht]
    \vspace{-4mm}
    \centering
    \begin{subfigure}[b]{0.44\columnwidth}
        \centering
        \includegraphics[width=\columnwidth]{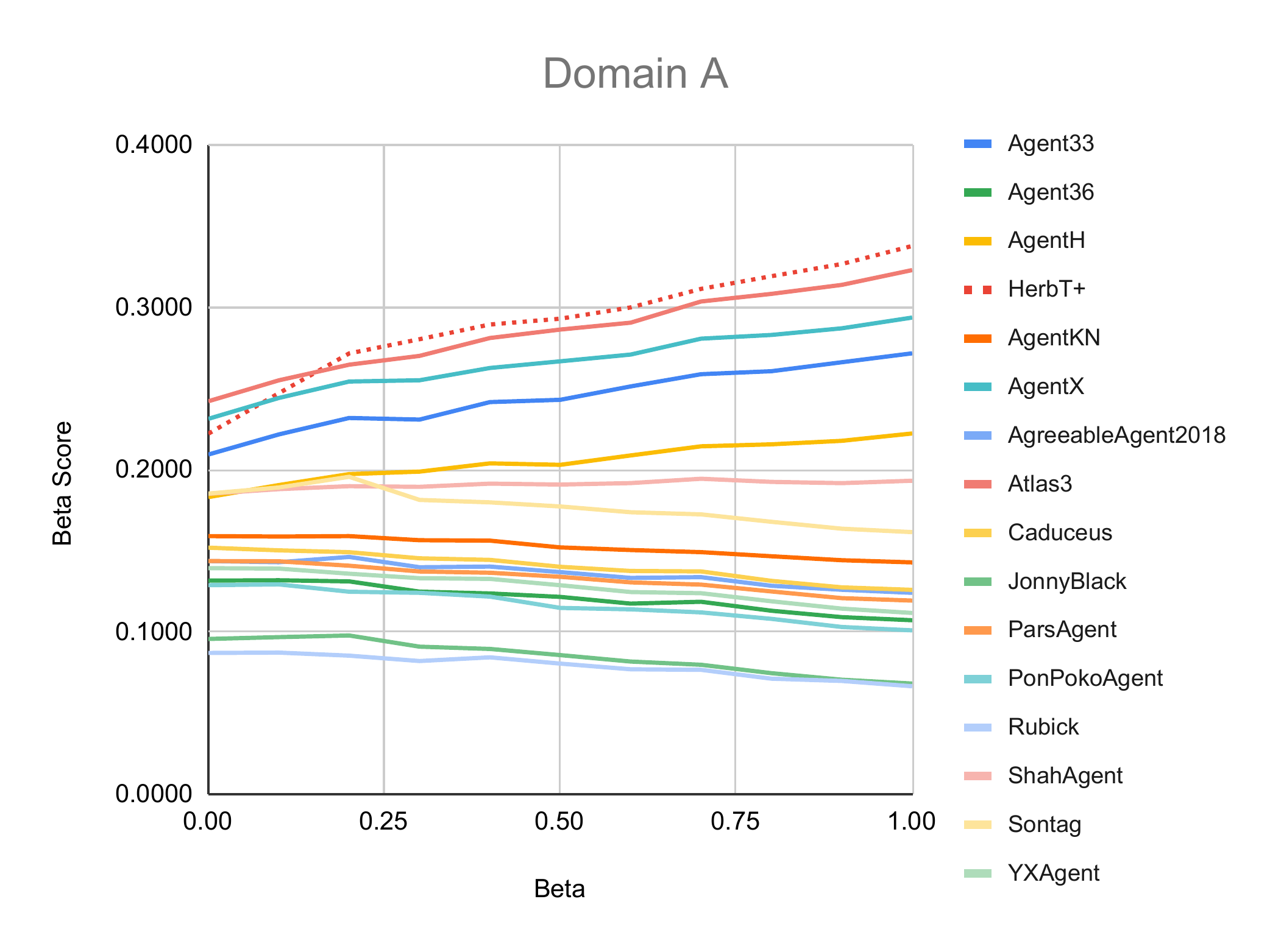}
    \end{subfigure}
    \hfill
    \begin{subfigure}[b]{0.44\columnwidth}
        \centering
        \includegraphics[width=\columnwidth]{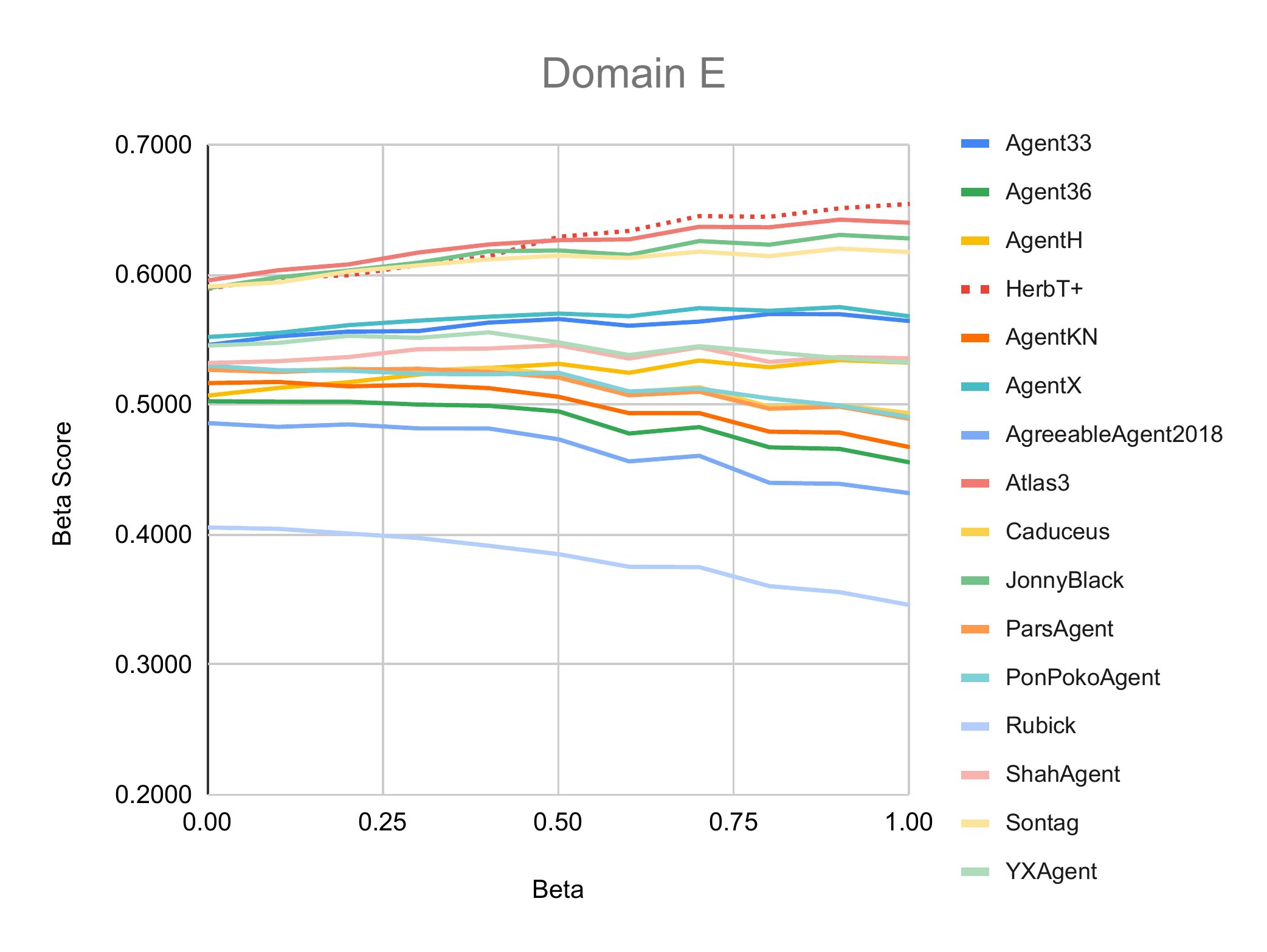}
    \end{subfigure}
    \caption{Beta score of the top agents for each beta}
    \label{beta_score}
    \vspace{-8mm}
\end{figure}
}

According to our results, except for domain D, where the discount factor is 1, our agent managed to outperform most of the agents. The lower the discount factor in the domain, the higher the beta score obtained by our agent.
\notaamas{ In fact, in domain A in which the discount factor is 0.2, our agent was able to achieve the highest beta score for almost every $\beta$ (every $\beta$ above 0.1).} In addition, as can be seen, the closer the beta is to 1, i.e., the higher the weight of social welfare, the better the performance of our agent relative to the other agents. Yet, even when $\beta$ is equal to 0, i.e., when our agent's goal is to maximize its individual utility, in most domains, our agent manages to outperform most of the agents\notaamas{ and in domain B our agent managed to achieve the highest individual welfare among the agents}.\notaamas{ In most values of $\beta$ in the graphs, Atlas3 \cite{Atlas3} had the closest beta score to our agent, and in most domains, he was able to achieve a higher beta score than our agent did when $\beta$ was close to 0. Atlas3's beta score can be explained by the fact that Atlas3 was proven effective for both social welfare and individual utility maximization (as he won both categories in ANAC 2015 \cite{ANAC2015}), unlike the other agents who were found effective for at most one of them.}
In order to further verify the performance of our agent, we ran a dependent t-test \cite{ttest}
on the results of the two agents whose sum of the beta scores over all the negotiations is the highest. The first agent is our agent, HerbT+, and the second is Atlas3. Using t-test, we found the difference between the two agents' performance to be statistically significant ($p<0.0001$).
Table~\ref{social_welfare_results} summarizes the social welfare score of \notaamas{all the}\onlyaamas{the top 4} agents in all of the domains when $\beta$ equals 1, i.e., when the goal is to maximize the social welfare.\footref{empty_values}

\begin{table}[ht]
\centering
\resizebox{0.9\columnwidth}{!}{
\csvautotabular{\results/social_welfare.csv}
}
\caption{Social welfare results}
\label{social_welfare_results}

\onlyaamas{
\vspace{-10mm}
}

\end{table}

As can be seen from the table, the strategy described in this research was able to achieve the highest social welfare among the agents in all domains except for domain D (where the discount factor is 1) in which AgentH \cite{AgentH} achieved the highest social welfare.
Even though our evaluation is based on limiting the number of rounds rather than the number of seconds allotted, we provide a comparison based on the latter measure (180 seconds) for domain B. Domain B was chosen because most of its characteristics are average compared to the rest of the domains. Here, again, the performance of our agent is best among all the agents tested, resulting in social welfare of 0.3221, compared to 0.3142 (JonnyBlack), 0.2926 (AgentX) and 0.2921 (Atlas3), which are the best agents in this category (see complete table in the Appendix). 
\notaamas{
From those results, we made several conclusions about the scenarios we mentioned in section \ref{experimental_domains}:

\begin{itemize}
    \item From the results in domain A and domain E, we conclude that a low number of possible outcomes (in our evaluation, 25 possible outcomes) and a high number of possible outcomes (in our evaluation, 65,536 possible outcomes) does not have a negative effect on our agent's performance.
    
    \item From the results of domain F, i.e., the domain with the integer issues type, and the results of domains A, B, C, E, and G, i.e., the domains with the discrete issues type, we can see that our agent manages to handle both domains with discrete issues and domains with integer issues.
    
    \item From the results of domains A, C, and F, i.e., the domains with a low reservation value (reservation value of 0), and the results of domain G, i.e., the domain with high reservation value (reservation value of 0.7), we can see that both low reservation values and high reservation values do not have a negative effect on the agent's performance, as in both scenarios, our agent was able to achieve the best results.
    
    \item From the results in domain B, we can see that although naturally, our agent's performance is lower on domains with low preference homogeneity than the other domains, it is still high relative to the other agents.

    \item From the results of domain B with time limit, we can see that the difference between time limit and rounds limit does not have a negative effect on our agent's performance.

    \item From the results of domain D, we conclude that our agent is sensitive to high discount values, and specifically, our agent's performance is lower on domains with a discount factor of 1, i.e., domains without discounting. More about that will be covered in the following sections.
\end{itemize}
}
\onlyaamas{
Overall, we observe that the performance of HerbT+ are robust cross domains and setting parameters.  The only domain parameter which seems to have a significant influence on the agent's performance is the discount factor of the domain. The effect of the discount factor will be covered in the following paragraphs. 
}

\subtopic{Discount Factor Implications}
\label{discount_factor_implications}

Since in domain D, which was the only one where HerbT+ was not ranked first, the discount factor is 1, we extended the evaluation to verify that indeed the discount factor fully accounts to the difference.  Specifically, we  used domains A - D with $\beta=1$ and $\beta=0.5$ for all discount factors with jumps of 0.2. \notaamas{The results are presented in Table~\ref{social_welfare_by_discount_factor} and Table~\ref{social_welfare_by_discount_factor_0.5}.}
\onlyaamas{The average beta score in these settings are given in Table~\ref{social_welfare_by_discount_factor}.\footnote{The agents presented in this table, as well as the agents in the rest of the tables in this section, are the agents who achieved top results according to $\beta$.}}

{
\notaamas{
\setstretch{1.0}
}

\notaamas{
\begin{table}[H]
    \centering
    
    \begin{subtable}[b]{\columnwidth}
        \centering
        \begin{adjustbox}{max width=\columnwidth}
            \csvautotabular{\results/discounts_social_welfare/domain_a.csv}
        \end{adjustbox}
        \caption{Domain A}
    \end{subtable}
    \begin{subtable}[b]{\columnwidth}
        \centering
        \begin{adjustbox}{max width=\columnwidth}
            \csvautotabular{\results/discounts_social_welfare/domain_b.csv}
        \end{adjustbox}
        \caption{Domain B}
    \end{subtable}
\caption{Beta score by discount factor for $\beta=1$}
\end{table}
\begin{table}[H]\ContinuedFloat
    \begin{subtable}[b]{\columnwidth}
        \centering
        \begin{adjustbox}{max width=\columnwidth}
            \csvautotabular{\results/discounts_social_welfare/domain_c.csv}
        \end{adjustbox}
        \caption{Domain C}
    \end{subtable}
    \begin{subtable}[b]{\columnwidth}
        \centering
        \begin{adjustbox}{max width=\columnwidth}
            \csvautotabular{\results/discounts_social_welfare/domain_d.csv}
        \end{adjustbox}
        \caption{Domain D}
    \end{subtable}
    \caption{Beta score by discount factor for $\beta=1$}
    \label{social_welfare_by_discount_factor}
\end{table}

\begin{table}[H]
    \centering
    
    \begin{subtable}[b]{\columnwidth}
        \centering
        \begin{adjustbox}{max width=\columnwidth}
            \begin{filecontents*}{\results/beta_0.5/score_a.csv}
Agent,df=0.2,df=0.4,df=0.6,df=0.8,df=1.0
HerbT+,\textbf{0.2963},\textbf{0.3432},\textbf{0.3830},\textbf{0.4334},0.4174
Agent33,0.2447,0.2995,0.3514,0.4034,0.4433
Agent36,0.1215,0.1790,0.2293,0.2706,0.2934
AgentH,0.2058,0.2669,0.3190,0.3769,0.4148
AgentKN,0.1519,0.2486,0.3287,0.4117,0.4839
AgentX,0.2677,0.3265,0.3774,0.4215,0.4726
AgreeableAgent2018,0.1357,0.2179,0.2900,0.3672,0.4168
Atlas3,0.2878,0.3301,0.3700,0.4265,\textbf{0.4825}
Caduceus,0.1398,0.2032,0.2537,0.3052,0.3244
JonnyBlack,0.0848,0.1199,0.1400,0.1816,0.1759
ParsAgent,0.1327,0.1951,0.2440,0.2899,0.3104
PonPokoAgent,0.1178,0.1602,0.1975,0.2251,0.2447
Rubick,0.0789,0.1132,0.1328,0.1642,0.1886
ShahAgent,0.1910,0.2466,0.2114,0.2193,0.2143
Sontag,0.1768,0.2147,0.2633,0.3094,0.3241
YXAgent,0.1292,0.1605,0.1937,0.2142,0.2349
\end{filecontents*}
\csvautotabular{\results/beta_0.5/score_a.csv}
        \end{adjustbox}
        \caption{Domain A}
    \end{subtable}
    \begin{subtable}[b]{\columnwidth}
        \centering
        \begin{adjustbox}{max width=\columnwidth}
            \begin{filecontents*}{\results/beta_0.5/score_b.csv}
Agent,df=0.2,df=0.4,df=0.6,df=0.8,df=1.0
HerbT+,\textbf{0.1791},\textbf{0.1972},0.2606,0.2834,0.2997
Agent33,0.1401,0.1607,0.2378,0.2662,0.3031
Agent36,0.0881,0.1140,0.1992,0.2255,0.2512
AgentH,0.1073,0.1311,0.2135,0.2429,0.2748
AgentKN,0.1062,0.1374,0.2344,0.2702,0.3069
AgentX,0.1536,0.1829,0.2640,0.2939,0.3283
AgreeableAgent2018,0.0843,0.1097,0.1919,0.2176,0.2433
Atlas3,0.1686,0.1899,\textbf{0.2684},\textbf{0.3024},\textbf{0.3429}
Caduceus,0.0908,0.1178,0.2016,0.2293,0.2507
JonnyBlack,0.0904,0.1173,0.2064,0.2341,0.2579
ParsAgent,0.0896,0.1158,0.2003,0.2267,0.2506
PonPokoAgent,0.0922,0.1164,0.1979,0.2271,0.2534
Rubick,0.0854,0.1125,0.1974,0.2265,0.2543
ShahAgent,0.1077,0.1346,0.1979,0.2202,0.2415
Sontag,0.1123,0.1387,0.2195,0.2504,0.2832
YXAgent,0.0920,0.1154,0.1951,0.2228,0.2496
\end{filecontents*}
\csvautotabular{\results/beta_0.5/score_b.csv}
        \end{adjustbox}
        \caption{Domain B}
    \end{subtable}
    \caption{Beta score by discount factor for $\beta=0.5$}
\end{table}
\begin{table}[H]\ContinuedFloat

    \begin{subtable}[b]{\columnwidth}
        \centering
        \begin{adjustbox}{max width=\columnwidth}
            \begin{filecontents*}{\results/beta_0.5/score_c.csv}
Agent,df=0.2,df=0.4,df=0.6,df=0.8,df=1.0
HerbT+,\textbf{0.5010},\textbf{0.5917},0.6838,0.7041,0.7125
Agent33,0.4495,0.5248,0.6297,0.6532,0.6871
Agent36,0.2686,0.4050,0.6098,0.6452,0.7520
AgentH,0.3671,0.4868,0.6545,0.6824,0.7673
AgentKN,0.3021,0.4446,0.6468,0.6856,0.7647
AgentX,0.4946,0.5634,0.6611,0.6766,0.7300
AgreeableAgent2018,0.2820,0.4177,0.6160,0.6556,0.7576
Atlas3,0.5265,0.5889,\textbf{0.6965},0.7093,0.7739
Caduceus,0.3277,0.4597,0.6406,0.6736,0.7535
JonnyBlack,0.2567,0.3645,0.5067,0.5348,0.6015
ParsAgent,0.3294,0.4688,0.6588,0.6943,0.7774
PonPokoAgent,0.3374,0.4409,0.5734,0.6002,0.6640
Rubick,0.2172,0.3192,0.4485,0.4764,0.5620
ShahAgent,0.3700,0.4989,0.5952,0.6097,0.5687
Sontag,0.4676,0.5648,0.6880,\textbf{0.7210},\textbf{0.7843}
YXAgent,0.3671,0.4489,0.5712,0.5898,0.6533
\end{filecontents*}
\csvautotabular{\results/beta_0.5/score_c.csv}
        \end{adjustbox}
        \caption{Domain C}
    \end{subtable}
    \begin{subtable}[b]{\columnwidth}
        \centering
        \begin{adjustbox}{max width=\columnwidth}
            \begin{filecontents*}{\results/beta_0.5/score_d.csv}
Agent,df=0.2,df=0.4,df=0.6,df=0.8,df=1.0
HerbT+,\textbf{0.5026},\textbf{0.5553},\textbf{0.6272},\textbf{0.6741},0.7273
Agent33,0.4468,0.5168,0.6009,0.6455,0.7276
Agent36,0.2712,0.3789,0.5245,0.6055,0.7229
AgentH,0.3897,0.4820,0.5933,0.6512,0.7307
AgentKN,0.2988,0.4037,0.5409,0.6182,0.7323
AgentX,0.4080,0.4873,0.5819,0.6420,0.7180
AgreeableAgent2018,0.3021,0.4029,0.5384,0.6264,0.7404
Atlas3,0.4954,0.5269,0.6000,0.6346,\textbf{0.7455}
Caduceus,0.3056,0.4067,0.5425,0.6168,0.7267
JonnyBlack,0.2821,0.3725,0.4817,0.5562,0.6538
ParsAgent,0.3120,0.4186,0.5498,0.6254,0.7253
PonPokoAgent,0.3041,0.3980,0.5206,0.5899,0.6913
Rubick,0.2714,0.3619,0.4737,0.5543,0.6719
ShahAgent,0.3251,0.4358,0.5125,0.5765,0.6782
Sontag,0.3562,0.4549,0.5731,0.6331,0.7215
YXAgent,0.4955,0.4991,0.5221,0.5875,0.6890
\end{filecontents*}
\csvautotabular{\results/beta_0.5/score_d.csv}
        \end{adjustbox}
        \caption{Domain D}
    \end{subtable}
    \caption{Beta score by discount factor for $\beta=0.5$}
    \label{social_welfare_by_discount_factor_0.5}
\end{table}
}
}

\onlyaamas{

\begin{table}[ht]
    \vspace{-6mm}

    \centering
    \begin{subtable}[b]{0.49\columnwidth}
        \centering
        \begin{adjustbox}{max width=\columnwidth}
        \begin{filecontents*}{\results/discounts_social_welfare/average.csv}
Agent/Beta,0,0.1,0.2,0.3,0.4,0.5,0.6,0.7,0.8,0.9,1
HerbT+,0.3575,0.3745,0.3842,0.4086,\textbf{0.4239},\textbf{0.4355},\textbf{0.4513},\textbf{0.4687},\textbf{0.4841},\textbf{0.5053},\textbf{0.5059}
AgentX,0.3969,0.3869,0.3920,0.4026,0.4006,0.4042,0.3984,0.3899,0.4016,0.4041,0.3990
Atlas3,\textbf{0.4102},\textbf{0.4097},\textbf{0.4096},\textbf{0.4099},0.4107,0.4107,0.4100,0.4097,0.4097,0.4090,0.4096
\end{filecontents*}
\csvautotabular{\results/discounts_social_welfare/average.csv}
        \end{adjustbox}
        \caption{$\beta=1$}
    \end{subtable}
    \hfill
    \begin{subtable}[b]{0.49\columnwidth}
        \centering
        \begin{adjustbox}{max width=\columnwidth}
        \csvautotabular{\results/discounts_social_welfare/average.csv}
        \end{adjustbox}
        \caption{$\beta=0.5$}
    \end{subtable}
    \vspace{-2mm}
    \caption{Beta score by discount factor}
    \label{social_welfare_by_discount_factor}
    \vspace{-10mm}
\end{table}

}

Indeed, according to our results, a discount factor of 1 decreases the performance of our agent relative to other agents.

\notaamas{ In addition, the results emphasize the stability of our agent. Our agent's achievements seem to be only influenced by the discount factor, unlike other agents whose results are harder to predict and seem to be influenced by many parameters since in each domain we see that another agent got the highest beta score.

}
The difference between the performance in discounted and undiscounted domains indicates that our agent manages to keep its utility from being substantially reduced by the discount factor and therefore increases the social welfare it achieves at the end of the negotiation.
Table~\ref{social_welfare_undiscounted} shows the undiscounted social welfare of \notaamas{all the agents}\onlyaamas{the top 5 agents} in all the domains when our agent only tries to maximize social welfare.\footnote{\label{empty_values}Some agents do not support integer issues, hence the N/A values in domain F.}

\begin{table}[ht]

\centering
\resizebox{0.9\columnwidth}{!}{
\begin{filecontents*}{\results/social_welfare_undiscounted.csv}
Agent,Domain A,Domain B,Domain C,Domain D,Domain E,Domain F,Domain G
HerbT+,0.5858,\textbf{0.4166},0.8172,0.7444,0.8210,\textbf{0.7355},0.6958
Agent33,0.5720,0.3630,0.7649,0.7581,0.7959,N/A,0.6969
Agent36,0.3674,0.2379,0.7496,0.6930,0.8167,0.5204,0.6961
AgentH,0.5155,0.2711,0.8060,\textbf{0.7627},0.8216,N/A,0.6920
AgentKN,0.5352,0.3087,0.8031,0.7311,0.8223,N/A,0.6911
AgentX,\textbf{0.5907},0.4089,0.7770,0.7423,0.7931,N/A,0.6795
AgreeableAgent2018,0.4750,0.2329,0.7652,0.7458,0.8000,0.6077,0.6954
Atlas3,0.5892,0.4067,0.8123,0.7501,0.8338,0.7306,\textbf{0.6989}
Caduceus,0.4118,0.2442,0.7984,0.7130,0.8339,N/A,0.6964
JonnyBlack,0.2162,0.2436,0.5978,0.6104,0.8304,N/A,0.6976
ParsAgent,0.3968,0.2407,0.8039,0.7130,0.8320,0.5586,0.6967
PonPokoAgent,0.3147,0.2439,0.6713,0.6752,0.7902,0.4327,0.6961
Rubick,0.2314,0.2382,0.5303,0.6491,0.6273,N/A,0.6981
ShahAgent,0.5375,0.2472,0.6775,0.6581,0.8320,0.5437,0.6970
Sontag,0.4140,0.2741,\textbf{0.8275},0.7216,\textbf{0.8393},0.6710,0.6930
YXAgent,0.3074,0.2374,0.6562,0.6767,0.7873,N/A,0.6962
\end{filecontents*}
\csvautotabular{\results/social_welfare_undiscounted.csv}
}
\caption{Undiscounted social welfare results}
\label{social_welfare_undiscounted}

\onlyaamas{
\vspace{-10mm}
}
\end{table}

From the table, we can see that when removing the effect of the discount factor on the results, our agent's score is still among the top agents\onlyaamas{, but not the top one in some domains, as with the discounted results.}\notaamas{. However, for example, in domain A, domain C, domain E, and domain G, we can see that there are agents with higher undiscounted social welfare than our agent.}
This difference strengthens our agent's capability of avoiding a significant discount factor reduction, relative to other agents, and explains a significant part in its success in maximizing the beta score.
To understand the difference between our agent's discounted social welfare and undiscounted social welfare, first, we look at the average negotiation length of each agent, as provided in Table~\ref{average_agreement_round_by_discount_factor}\onlyaamas{.}\notaamas{ and Table~\ref{average_agreement_round_by_discount_factor_0.5}.\footref{empty_values} The lower the average length is, the smaller the effect of the discount factor.}

\notaamas{

\begin{table}[H]

    \centering
    
    \begin{subtable}[b]{\columnwidth}
        \centering
        \begin{adjustbox}{max width=\columnwidth}
            \begin{filecontents*}{\results/rounds/rounds_all.csv}
Agent,Domain A,Domain B,Domain C,Domain D,Domain E,Domain F,Domain G
HerbT+,\textbf{85.125},\textbf{127.434},\textbf{72.923},\textbf{114.533},\textbf{62.043},\textbf{75.794},\textbf{111.389}
Agent33,126.063,152.606,100.945,147.354,94.251,N/A,143.353
Agent36,166.299,178.350,149.840,168.837,140.371,162.480,151.934
AgentH,136.669,169.397,105.920,123.210,105.619,N/A,136.991
AgentKN,159.450,176.452,143.952,159.520,136.731,N/A,149.481
AgentX,119.059,143.997,81.912,115.286,91.793,N/A,118.288
AgreeableAgent2018,165.897,178.691,150.918,163.835,149.567,157.982,152.468
Atlas3,108.318,141.514,97.888,158.638,70.251,96.420,115.843
Caduceus,157.968,177.035,142.681,164.611,126.131,N/A,150.863
JonnyBlack,169.793,178.218,153.947,167.152,73.204,N/A,153.327
ParsAgent,161.997,177.697,131.548,158.421,126.852,152.117,150.368
PonPokoAgent,159.488,177.079,134.590,158.620,122.080,154.664,152.994
Rubick,167.821,178.862,162.165,163.009,153.995,N/A,149.639
ShahAgent,138.571,175.482,136.006,165.761,107.597,145.000,151.873
Sontag,149.042,172.452,91.464,144.624,78.087,129.471,143.246
YXAgent,162.292,177.431,126.917,156.416,104.814,N/A,150.393
\end{filecontents*}
\csvautotabular{\results/rounds/rounds_all.csv}
        \end{adjustbox}
        \caption{All domains}
    \end{subtable}
    \begin{subtable}[b]{\columnwidth}
        \centering
        \begin{adjustbox}{max width=\columnwidth}
            \begin{filecontents*}{\results/rounds/rounds_a.csv}
Agent,df=0.2,df=0.4,df=0.6,df=0.8,df=1
HerbT+,\textbf{85.125},\textbf{94.051},\textbf{106.077},\textbf{121.524},\textbf{131.978}
Agent33,126.063,124.424,132.770,149.806,160.754
Agent36,166.299,166.651,170.720,173.379,175.677
AgentH,136.669,132.999,143.053,150.276,157.140
AgentKN,159.450,164.040,167.986,170.487,172.729
AgentX,119.059,108.964,118.291,127.841,138.754
AgreeableAgent2018,165.897,168.001,172.403,174.880,176.946
Atlas3,108.318,105.785,133.035,156.070,170.467
Caduceus,157.968,162.281,167.031,170.542,174.759
JonnyBlack,169.793,167.645,174.730,176.513,178.108
ParsAgent,161.997,162.243,167.729,170.421,174.284
PonPokoAgent,159.488,166.458,170.612,173.237,174.546
Rubick,167.821,173.212,176.475,178.052,178.317
ShahAgent,138.571,146.672,167.731,174.007,175.357
Sontag,149.042,152.486,157.898,162.213,167.589
YXAgent,162.292,163.501,169.138,171.555,174.058
\end{filecontents*}
\csvautotabular{\results/rounds/rounds_a.csv}
        \end{adjustbox}
        \caption{Domain A}
    \end{subtable}
\caption{Average agreement round by discount factor for $\beta=1$}
\end{table}
\begin{table}[H]\ContinuedFloat
    \begin{subtable}[b]{\columnwidth}
        \centering
        \begin{adjustbox}{max width=\columnwidth}
            \begin{filecontents*}{\results/rounds/rounds_b.csv}
Agent,df=0.2,df=0.4,df=0.6,df=0.8,df=1
HerbT+,\textbf{111.462},\textbf{121.710},\textbf{131.579},\textbf{144.253},\textbf{160.708}
Agent33,137.850,149.285,155.471,163.158,169.301
Agent36,173.376,176.722,178.925,179.542,180.193
AgentH,159.572,166.663,170.450,172.805,175.290
AgentKN,171.190,175.027,177.311,178.029,178.969
AgentX,133.503,140.152,146.440,151.551,162.422
AgreeableAgent2018,173.428,176.871,179.063,179.675,180.330
Atlas3,119.337,133.609,148.639,167.625,176.594
Caduceus,171.223,175.174,177.825,178.959,180.281
JonnyBlack,173.334,176.610,178.754,179.470,180.069
ParsAgent,172.490,176.174,178.384,179.292,180.119
PonPokoAgent,171.178,175.447,178.037,178.763,179.589
Rubick,173.765,177.182,179.221,179.820,180.275
ShahAgent,163.429,169.353,177.642,178.872,179.920
Sontag,164.485,169.741,173.620,175.131,176.928
YXAgent,171.171,175.719,178.097,178.962,179.621
\end{filecontents*}
\csvautotabular{\results/rounds/rounds_b.csv}
        \end{adjustbox}
        \caption{Domain B}
    \end{subtable}

    \begin{subtable}[b]{\columnwidth}
        \centering
        \begin{adjustbox}{max width=\columnwidth}
            \begin{filecontents*}{\results/rounds/rounds_c.csv}
Agent,df=0.2,df=0.4,df=0.6,df=0.8,df=1
HerbT+,\textbf{59.946},\textbf{61.975},\textbf{63.567},\textbf{72.923},\textbf{98.055}
Agent33,78.774,86.484,93.312,100.945,132.342
Agent36,139.620,143.315,147.446,149.840,159.473
AgentH,95.896,102.834,107.360,105.920,120.963
AgentKN,131.101,135.895,140.778,143.952,150.778
AgentX,68.607,74.541,80.887,81.912,100.888
AgreeableAgent2018,138.616,144.421,148.401,150.918,159.328
Atlas3,64.483,73.687,83.838,97.888,151.529
Caduceus,118.518,126.531,133.612,142.681,160.886
JonnyBlack,140.326,144.918,150.906,153.947,161.592
ParsAgent,118.981,123.672,128.287,131.548,146.624
PonPokoAgent,118.136,126.113,132.496,134.590,146.151
Rubick,148.989,153.979,158.319,162.165,166.065
ShahAgent,101.037,110.401,120.881,136.006,154.354
Sontag,77.308,81.854,86.945,91.464,106.075
YXAgent,109.195,114.821,121.517,126.917,140.051%
\end{filecontents*}
\csvautotabular{\results/rounds/rounds_c.csv}
        \end{adjustbox}
        \caption{Domain C}
    \end{subtable}
\caption{Average agreement round by discount factor for $\beta=1$}
\end{table}
\begin{table}[H]\ContinuedFloat
    \begin{subtable}[b]{\columnwidth}
        \centering
        \begin{adjustbox}{max width=\columnwidth}
            \begin{filecontents*}{\results/rounds/rounds_d.csv}
Agent,df=0.2,df=0.4,df=0.6,df=0.8,df=1
HerbT+,\textbf{49.726},\textbf{59.652},\textbf{75.106},\textbf{85.102},\textbf{114.533}
Agent33,74.491,80.755,95.066,113.155,147.354
Agent36,131.371,136.426,156.344,158.962,168.837
AgentH,84.421,90.601,103.120,108.119,123.210
AgentKN,120.495,130.536,149.736,151.678,159.520
AgentX,89.313,94.111,107.953,111.078,115.286
AgreeableAgent2018,121.639,132.314,151.763,153.576,163.835
Atlas3,57.394,61.563,79.159,92.321,158.638
Caduceus,115.349,126.214,146.345,151.728,164.611
JonnyBlack,123.633,134.156,154.294,156.722,167.152
ParsAgent,115.341,124.523,142.637,146.123,158.421
PonPokoAgent,117.246,126.885,146.482,150.884,158.620
Rubick,124.493,135.240,159.112,161.865,163.009
ShahAgent,100.457,108.682,148.911,157.108,165.761
Sontag,101.362,107.083,122.043,129.270,144.624
YXAgent,120.933,132.433,143.601,148.966,156.416
\end{filecontents*}
\csvautotabular{\results/rounds/rounds_d.csv}
        \end{adjustbox}
        \caption{Domain D}
    \end{subtable}
    \caption{Average agreement round by discount factor for $\beta=1$}
    \label{average_agreement_round_by_discount_factor}
\end{table}

\begin{table}[H]
    \centering
    
    \begin{subtable}[b]{\columnwidth}
        \centering
        \begin{adjustbox}{max width=\columnwidth}
            \begin{filecontents*}{\results/beta_0.5/round_a.csv}
Agent,df=0.2,df=0.4,df=0.6,df=0.8,df=1.0
HerbT+,\textbf{87.944},\textbf{100.090},\textbf{115.543},\textbf{130.515},149.148
Agent33,112.125,122.747,132.883,150.198,162.879
Agent36,163.332,167.645,170.578,173.305,176.277
AgentH,126.806,136.259,144.416,150.954,159.007
AgentKN,161.264,164.583,168.256,170.748,173.049
AgentX,103.273,111.474,119.401,128.599,\textbf{141.566}
AgreeableAgent2018,166.460,169.588,172.878,174.756,177.030
Atlas3,92.734,109.854,133.154,156.528,171.056
Caduceus,157.324,162.568,167.154,170.627,175.157
JonnyBlack,168.840,172.051,175.094,176.364,178.394
ParsAgent,159.260,163.702,167.625,170.779,174.593
PonPokoAgent,161.743,166.804,170.666,173.923,176.339
Rubick,171.404,174.013,176.550,178.027,178.801
ShahAgent,134.872,148.036,168.140,174.054,176.954
Sontag,142.644,153.309,158.227,163.164,168.556
YXAgent,156.880,164.184,168.694,172.274,175.123
\end{filecontents*}
\csvautotabular{\results/beta_0.5/round_a.csv}
        \end{adjustbox}
        \caption{Domain A}
    \end{subtable}
    \caption{Average agreement round by discount factor for $\beta=0.5$}
\end{table}
\begin{table}[H]\ContinuedFloat
    \begin{subtable}[b]{\columnwidth}
        \centering
        \begin{adjustbox}{max width=\columnwidth}
            \begin{filecontents*}{\results/beta_0.5/round_b.csv}
Agent,df=0.2,df=0.4,df=0.6,df=0.8,df=1.0
HerbT+,\textbf{117.044},\textbf{125.804},155.657,163.722,170.727
Agent33,142.618,149.279,161.604,165.880,169.735
Agent36,174.769,176.678,179.535,179.891,180.329
AgentH,162.592,166.980,172.719,173.925,175.318
AgentKN,172.722,175.014,177.997,178.554,179.067
AgentX,136.035,140.054,\textbf{151.147},\textbf{154.950},\textbf{159.921}
AgreeableAgent2018,175.239,176.926,179.645,180.079,180.437
Atlas3,125.124,133.782,162.747,170.853,176.900
Caduceus,173.163,174.993,178.996,179.562,180.362
JonnyBlack,174.573,176.602,179.491,179.826,180.223
ParsAgent,174.212,176.172,179.119,179.770,180.208
PonPokoAgent,172.944,175.529,178.837,179.144,179.727
Rubick,175.418,177.208,179.866,180.054,180.358
ShahAgent,166.224,169.709,178.866,179.528,180.108
Sontag,166.749,169.678,175.222,176.258,177.219
YXAgent,173.449,175.654,178.972,179.232,179.774
\end{filecontents*}
\csvautotabular{\results/beta_0.5/round_b.csv}
        \end{adjustbox}
        \caption{Domain B}
    \end{subtable}
    \begin{subtable}[b]{\columnwidth}
        \centering
        \begin{adjustbox}{max width=\columnwidth}
            \begin{filecontents*}{\results/beta_0.5/round_c.csv}
Agent,df=0.2,df=0.4,df=0.6,df=0.8,df=1.0
HerbT+,\textbf{59.611},\textbf{61.122},\textbf{80.220},\textbf{84.126},114.524
Agent33,79.367,86.247,98.975,102.458,132.658
Agent36,140.010,143.662,149.573,151.413,159.464
AgentH,96.013,101.113,107.055,108.803,120.647
AgentKN,131.051,134.275,142.429,143.927,150.952
AgentX,69.129,75.103,82.495,85.209,\textbf{101.437}
AgreeableAgent2018,138.480,142.504,150.021,151.590,159.313
Atlas3,59.713,68.911,91.972,99.582,151.722
Caduceus,118.817,124.911,140.013,143.173,160.947
JonnyBlack,140.616,144.834,153.543,154.970,161.978
ParsAgent,119.425,122.889,130.041,131.688,146.879
PonPokoAgent,118.176,122.943,132.937,134.406,146.281
Rubick,148.950,152.863,160.844,162.384,166.281
ShahAgent,101.551,105.981,131.511,136.276,154.363
Sontag,77.576,82.292,90.757,92.179,106.606
YXAgent,109.630,116.239,126.317,129.456,140.422
\end{filecontents*}
\csvautotabular{\results/beta_0.5/round_c.csv}
        \end{adjustbox}
        \caption{Domain C}
    \end{subtable}
\caption{Average agreement round by discount factor  for $\beta=0.5$}
\end{table}
\begin{table}[H]\ContinuedFloat
    \begin{subtable}[b]{\columnwidth}
        \centering
        \begin{adjustbox}{max width=\columnwidth}
            \begin{filecontents*}{\results/beta_0.5/round_d.csv}
Agent,df=0.2,df=0.4,df=0.6,df=0.8,df=1.0
HerbT+,55.237,65.248,\textbf{84.605},\textbf{88.527},\textbf{106.107}
Agent33,73.316,80.052,102.644,113.045,151.985
Agent36,128.207,136.331,158.246,159.092,167.569
AgentH,84.536,90.945,107.357,108.678,124.251
AgentKN,122.482,130.472,151.286,151.742,161.236
AgentX,88.099,94.491,109.651,110.251,119.094
AgreeableAgent2018,123.328,132.550,153.307,153.718,162.732
Atlas3,\textbf{47.450},\textbf{59.497},89.418,92.709,160.806
Caduceus,116.685,126.220,149.477,151.872,164.689
JonnyBlack,125.733,134.270,156.527,156.801,166.460
ParsAgent,116.761,124.403,145.824,146.220,159.665
PonPokoAgent,119.534,127.429,148.886,151.104,160.840
Rubick,126.116,135.341,159.310,161.991,164.238
ShahAgent,100.893,108.843,152.000,157.093,166.748
Sontag,100.622,106.805,126.341,129.392,145.867
YXAgent,110.437,127.113,146.729,149.201,159.383
\end{filecontents*}
\csvautotabular{\results/beta_0.5/round_d.csv}
        \end{adjustbox}
        \caption{Domain D}
    \end{subtable}
    \caption{Average agreement round by discount factor  for $\beta=0.5$}
    \label{average_agreement_round_by_discount_factor_0.5}
\end{table}

}

\onlyaamas{
\begin{table}[ht]
    \vspace{-6mm}

    \centering
    \begin{subtable}[b]{0.49\columnwidth}
        \centering
        \begin{adjustbox}{max width=\columnwidth}
        \begin{filecontents*}{\results/rounds/rounds_avg.csv}
Agent,df=0.2,df=0.4,df=0.6,df=0.8,df=1
HerbT+,76.565,84.347,94.082,105.950,126.319
AgentX,102.620,104.442,113.393,118.096,129.337
Atlas3,87.383,93.661,111.168,128.476,164.307
Sontag,123.049,127.791,135.127,139.520,148.804
\end{filecontents*}
\csvautotabular{\results/rounds/rounds_avg.csv}
        \end{adjustbox}
        \caption{$\beta=1$}
    \end{subtable}
    \hfill
    \begin{subtable}[b]{0.49\columnwidth}
        \centering
        \begin{adjustbox}{max width=\columnwidth}
            \csvautotabular{\results/beta_0.5/round_avg.csv}
        \end{adjustbox}
        \caption{$\beta=0.5$}
    \end{subtable}
    \vspace{-2mm}
    \caption{Average agreement round by discount factor}
    \label{average_agreement_round_by_discount_factor}
    \vspace{-10mm}
\end{table}
}

According to our results, all of the agents who achieved good results in terms of social welfare also have a relatively low agreement round, e.g., AgentX, AgentH, Atlas3.
These results genuinely emphasize the importance of finding agreement fast and finishing the negotiation early in order to maximize social welfare. Furthermore, in most domains, our agent has the lowest average agreement round, and when $\beta=1$, it has the lowest one in all of the domains. Our agent's ability to quickly find agreements with a high beta score, with a low number of rounds, is one of the key aspects for its success in maximizing the beta score in domains with discount factors that are different than one. To understand the reasons for our agent's low agreement round, we collected statistics from the negotiation process of the agents about the agents' behaviors.
The decline rate of an agent is the rate at which the agent rejects a received bid and proposes a counteroffer\onlyaamas{, i.e., the number of declines an agent made out of the total number of its actions}. 
\notaamas{
It is calculated using the following equation:

\[decline\_rate(agent)=\frac{number\_of\_declines(agent)}{number\_of\_actions(agent)}\]

$number\_of\_declines(agent)$ and $number\_of\_actions(agent)$ are the number of time $agent$ rejected an offer and all the number of times $agent$ had to make a decision on accepting or rejecting  of an agreement, respectively, in all the negotiations we ran on a domain.

}
Table~\ref{decline_rate_by_discount_factor} \notaamas{and Table~\ref{decline_rate_by_discount_factor_0.5} describe each agent's decline rate by discount factor}\onlyaamas{describes the average decline rate of each agent for different discount factor values}.
\notaamas{
\begin{table}[H]
    \centering
    
    \begin{subtable}[b]{\columnwidth}
        \centering
        \begin{adjustbox}{max width=\columnwidth}
            \csvautotabular{\results/decline_rate/decline_rate_a.csv}
        \end{adjustbox}
        \caption{Domain A}
    \end{subtable}
\caption{Decline rate by discount factor for $\beta=1$}
\end{table}
\begin{table}[H]\ContinuedFloat
    \begin{subtable}[b]{\columnwidth}
        \centering
        \begin{adjustbox}{max width=\columnwidth}
            \csvautotabular{\results/decline_rate/decline_rate_b.csv}
        \end{adjustbox}
        \caption{Domain B}
    \end{subtable}
    \begin{subtable}[b]{\columnwidth}
        \centering
        \begin{adjustbox}{max width=\columnwidth}
            \csvautotabular{\results/decline_rate/decline_rate_c.csv}
        \end{adjustbox}
        \caption{Domain C}
    \end{subtable}
\caption{Decline rate by discount factor for $\beta=1$}
\end{table}
\begin{table}[H]\ContinuedFloat
    \centering
    \begin{subtable}[b]{\columnwidth}
        \centering
        \begin{adjustbox}{max width=\columnwidth}
            \csvautotabular{\results/decline_rate/decline_rate_d.csv}
        \end{adjustbox}
        \caption{Domain D}
    \end{subtable}
    \caption{Decline rate by discount factor for $\beta=1$.}
    \label{decline_rate_by_discount_factor}
\end{table}

\begin{table}[H]
    \centering
    
    \begin{subtable}[b]{\columnwidth}
        \centering
        \begin{adjustbox}{max width=\columnwidth}
            \csvautotabular{\results/beta_0.5/decline_rate_a.csv}
        \end{adjustbox}
        \caption{Domain A}
    \end{subtable}
\caption{Decline rate by discount factor for $\beta=0.5$}
\end{table}
\begin{table}[H]\ContinuedFloat
    \begin{subtable}[b]{\columnwidth}
        \centering
        \begin{adjustbox}{max width=\columnwidth}
            \csvautotabular{\results/beta_0.5/decline_rate_b.csv}
        \end{adjustbox}
        \caption{Domain B}
    \end{subtable}
    \begin{subtable}[b]{\columnwidth}
        \centering
        \begin{adjustbox}{max width=\columnwidth}
            \csvautotabular{\results/beta_0.5/decline_rate_c.csv}
        \end{adjustbox}
        \caption{Domain C}
    \end{subtable}
\caption{Decline rate by discount factor for $\beta=0.5$.}
\end{table}
\begin{table}[H]\ContinuedFloat
    \begin{subtable}[b]{\columnwidth}
        \centering
        \begin{adjustbox}{max width=\columnwidth}
            \csvautotabular{\results/beta_0.5/decline_rate_d.csv}
        \end{adjustbox}
        \caption{Domain D}
    \end{subtable}
    \caption{Decline rate by discount factor for $\beta=0.5$}
    \label{decline_rate_by_discount_factor_0.5}
\end{table}
}

\onlyaamas{
\begin{table}[ht]
    \vspace{-6mm}

    \centering
    
    \begin{adjustbox}{max width=0.9\columnwidth}
    \csvautotabular{\results/decline_rate/decline_rate_average.csv}
    \end{adjustbox}

    \caption{Decline rate by discount factor for $\beta=1$}
    \label{decline_rate_by_discount_factor}
    \vspace{-10mm}
\end{table}
}

Further analysis of the results reveals that HerbT+ has the lowest decline rate among the agents in most of the domains and discount factors. In domains with a discount factor of 0.2, it even seems that our agent almost always accepts. Our agent's low decline rate relative to that of other agents explains our agent's low agreement round. As with average agreement rounds, the agents who achieved good social welfare among the top agents have a relatively low decline rate, which indicates that a low decline rate plays a substantial role for agents who want to maximize social welfare.
In Table~\ref{average_agreement_round_by_discount_factor} and Table~\ref{decline_rate_by_discount_factor}, we can see that even when the discount factor equals 1, our agent's average agreement round is low relative to other agents. However, since there is no discount factor reduction, an early agreement does not improve the negotiation's outcome, which causes our agent to lose that advantage over the other agents. In addition, when there is no discount factor reduction at all, agents have no incentive to make concessions and change their original offer until the late rounds in the negotiation, which makes our attempt to model the agent's utility function more difficult. These two reasons explain the lower results of our agent with a discount factor of 1. With that being said, while in ANAC all agents have the same discount factor, the situation might be different in other settings. In settings where each agent has its own discount factor, one that is not known to the other agents, there is an advantage for maximizing social welfare in early agreements, even if our agent's discount factor is 1, as we do not know how much the social welfare is reduced due to other agents' discount factors.

\subtopic{Always Accept Policy}

Our agent's low decline rate, which is a key factor in its success, raises an important question regarding the need in efficient opponent modeling and bidding strategy---if the bids our agent rejects and the offers our agent makes are negligible for the performance of the agent, any agent who agrees on everything could achieve the same results. To verify the contribution encapsulated in our agent's opponent modeling and bidding strategy, \notaamas{we created a new agent to compare our agent against. Our new agent will have the same opponent model and bidding strategy, but its acceptance strategy will be to always accept, and therefore the new agent opponent model and bidding strategy will be used only when the new agent has to make an offer, i.e., when the new agent is the first to make an action in the negotiation. By comparing the new agent performance against the other top agents, we will be able to see the contribution of a low decline rate to the social welfare, as our new agent will have the lowest possible rate.
To evaluate the agent we created, we added it to \notaamas{both the evaluation pool and comparison pool, as mentioned in section \ref{experimental_agents}}\onlyaamas{to the evaluation} under the name $AlwaysAcceptAgent$.}\onlyaamas{we created a new agent that always accepts offered bids and added it to the evaluation under the name $AlwaysAcceptAgent$.} We ran the evaluation with all of the top agents, including our agent, HerbT+, and the new agent on domains A-D, with $beta=1$ (where the decline rate of our agent is the lowest), and with all of the discount factors with jumps of 0.2. \notaamas{The results are described in Table~\ref{social_welfare_always_accept}.}
\onlyaamas{The average results of each agent are given in Table~\ref{social_welfare_always_accept}.}
\notaamas{
\begin{table}[H]
    \centering
    
    \begin{subtable}[b]{\columnwidth}
        \centering
        \csvautotabular{\results/always_accept/social_welfare/domain_a.csv}
        \caption{Domain A}
    \end{subtable}
    \hfill
\caption{Social welfare with AlwaysAcceptAgent by discount factor}
\end{table}
\begin{table}[H]\ContinuedFloat
    \begin{subtable}[b]{\columnwidth}
        \centering
        \csvautotabular{\results/always_accept/social_welfare/domain_b.csv}
        \caption{Domain B}
    \end{subtable}
    \hfill
    \centering
    \begin{subtable}[b]{\columnwidth}
        \centering
        \csvautotabular{\results/always_accept/social_welfare/domain_c.csv}
        \caption{Domain C}
    \end{subtable}
    \hfill
\caption{Social welfare with AlwaysAcceptAgent by discount factor}
\end{table}
\begin{table}[H]\ContinuedFloat
    \begin{subtable}[b]{\columnwidth}
        \centering
        \csvautotabular{\results/always_accept/social_welfare/domain_d.csv}
        \caption{Domain D}
    \end{subtable}
    \hfill
    \caption{Social welfare with AlwaysAcceptAgent by discount factor}
    \label{social_welfare_always_accept}
\end{table}
}
\onlyaamas{
\begin{table}[ht]
    \vspace{-6mm}

    \centering
    \begin{adjustbox}{max width=0.9\columnwidth}
        \csvautotabular{\results/always_accept/social_welfare/average.csv}
    \end{adjustbox}
    \caption{Social welfare with AlwaysAcceptAgent by discount factor}
    \label{social_welfare_always_accept}
    \vspace{-10mm}
\end{table}
}
According to our results, AlwaysAcceptAgent managed to outperform most of the agents in terms of social welfare, especially in domains with a low discount factor. \notaamas{From these results, we learn that in some scenarios, in order to maximize social welfare in multilateral negotiation, a strategy to consider, which is simple and achieves reasonably good results, is to agree on everything, i.e., taking yourself out of the picture and thus making it easier for the other agents to find an agreement.}\onlyaamas{These results verify the effectiveness of a low decline rate for social welfare maximization in multilateral negotiation settings (such as the settings in ANAC).}
In addition, the results above demonstrate the effectiveness of the opponent model and the bidding strategy of our agent, HerbT+. Our agent was able to surpass AlwaysAcceptAgent's performance in all domains and discount factors\onlyaamas{.}\notaamas{ except for a discount factor of 0.2, where they achieved similar results.}
To further explore the differences between the two agents, we provide the average social welfare and the average individual utility of the offers that "AlwaysAcceptAgent" and "HerbT+" accepted, over all the negotiations we ran on the domains (denoted \emph{acceptance social welfare} and \emph{accepted individual utility}, respectively). \notaamas{The average acceptance social welfare and average acceptance individual utility are calculated using the following equations:

\[
\begin{adjustbox}{max width=\columnwidth}
    $acceptance\_social\_welfare(agent) = \frac{\sum_{bid \in accepted\_bids(agent)}social\_welfare(bid)}{|accepted\_bids(agent)|}$
\end{adjustbox}
\]

\[
\begin{adjustbox}{max width=\columnwidth}
    $acceptance\_individual\_utility(agent) = \frac{\sum_{bid \in accepted\_bids(agent)}individual\_utility(bid)}{|accepted\_bids(agent)|}$
\end{adjustbox}
\]

$accepted\_bids(agent)$ is a group of all the bids $agent$ has accepted throughout the negotiations. The values of $individual\_utility(bid)$ and $social\_welfare(bid)$ are according to the negotiations the bid was offered in.

The results are described in Table~\ref{acceptance_social_welfare_always_accept}.
}
\onlyaamas{
The average cross-domain results  are given in Table~\ref{acceptance_social_welfare_always_accept}.
}

\notaamas{

\begin{table}[H]
    \centering

    \begin{subtable}[b]{\columnwidth}
        \centering
        \begin{adjustbox}{max width=\columnwidth}
        \csvautotabular{\results/always_accept/acceptance_social_welfare/domain_a.csv}
        \end{adjustbox}
        \caption{Domain A - social welfare}
    \end{subtable}
    \hfill
    \begin{subtable}[b]{\columnwidth}
        \centering       
        \begin{adjustbox}{max width=\columnwidth}
        \csvautotabular{\results/always_accept/acceptance_individual_utility/domain_a.csv}
        \end{adjustbox}
        \caption{Domain A - individual utility}
    \end{subtable}
    \hfill
    \begin{subtable}[b]{\columnwidth}
        \centering
        \csvautotabular{\results/always_accept/acceptance_social_welfare/domain_b.csv}
        \caption{Domain B - social welfare}
    \end{subtable}
    \hfill
    \begin{subtable}[b]{\columnwidth}
        \centering
        \csvautotabular{\results/always_accept/acceptance_individual_utility/domain_b.csv}
        \caption{Domain B - individual utility}
    \end{subtable}
    \hfill
    \caption{Acceptance social welfare and individual utility by discount factor}
\end{table}
\begin{table}[H]\ContinuedFloat
    \begin{subtable}[b]{\columnwidth}
        \centering
        \csvautotabular{\results/always_accept/acceptance_social_welfare/domain_c.csv}
        \caption{Domain C - social welfare}
    \end{subtable}
    \hfill
    \begin{subtable}[b]{\columnwidth}
        \centering
        \csvautotabular{\results/always_accept/acceptance_individual_utility/domain_c.csv}
        \caption{Domain C - individual utility}
    \end{subtable}
    \hfill
        \begin{subtable}[b]{\columnwidth}
        \centering
        \csvautotabular{\results/always_accept/acceptance_social_welfare/domain_d.csv}
        \caption{Domain D - social welfare}
    \end{subtable}
    \hfill
    \begin{subtable}[b]{\columnwidth}
        \centering
        \csvautotabular{\results/always_accept/acceptance_individual_utility/domain_d.csv}
        \caption{Domain D - individual utility}
    \end{subtable}
    \hfill
    \caption{Acceptance social welfare and individual utility by discount factor}
    \label{acceptance_social_welfare_always_accept}
\end{table}

}

\onlyaamas{

\begin{table}[ht]
    \vspace{-6mm}

    \centering
    \begin{subtable}[b]{0.49\columnwidth}
        \centering
        \begin{adjustbox}{max width=\columnwidth}
        \csvautotabular{\results/always_accept/acceptance_social_welfare/average.csv}
        \end{adjustbox}
        \caption{social welfare}
    \end{subtable}
    \hfill
    \begin{subtable}[b]{0.49\columnwidth}
        \centering
        \begin{adjustbox}{max width=\columnwidth}
        \csvautotabular{\results/always_accept/acceptance_individual_utility/average.csv}
        \end{adjustbox}
        \caption{individual utility}
    \end{subtable}
    \vspace{-2mm}

    \caption{Acceptance social welfare and individual utility by discount factor}
    \label{acceptance_social_welfare_always_accept}
    \vspace{-10mm}
\end{table}

}

According to our results, with most discount factors, our agent, HerbT+, does manage to accept bids with a higher social welfare than AlwaysAcceptAgent. Interestingly, the individual utility of the bids that HerbT+ accepts, in most discount factors, is much higher than the individual utility of the bids AlwaysAcceptAgent accepts even though HerbT+ is tuned to maximizing only social welfare ($\beta=1$). This actually makes sense as our agent's individual utility is part of the overall social welfare, so, in order to maximize social welfare, one of our agent's interests is also to have a high individual utility for itself. From those results we learn that one reason for the difference between our agent's performance and that of AlwaysAcceptAgent is that, unlike AlwaysAcceptAgent, our agent also tries to achieve a high individual utility for itself, as doing so helps it to increase the overall social welfare.

\notaamas{
\subtopic{Offers/Accepts Comparison}

To understand the contribution of the offers our agent makes compared to the offers our agent accepts, we first provide the $negotiation\ agreement\ rate$ of the agent. The negotiation agreement rate of an agent is defined as the likelihood of a negotiation to end with an agreement of the agent given the agent participated in the negotiation. We calculate it as the number of negotiations that ended with an agreement offered by the agent divided by the number of negotiation sessions in which the agent participated. \notaamas{The results are described in Table~\ref{negotiation_agreement_rate}.}
\onlyaamas{The average negotiation agreement rate of the domains is described in Table~\ref{negotiation_agreement_rate}.}

\notaamas{

\begin{table}[H]
    \centering
    
    \begin{subtable}[b]{\columnwidth}
        \centering
        \resizebox{\columnwidth}{!}{
        \csvautotabular{\results/negotiation_agreement_rate/domain_a.csv}
        }
        \caption{Domain A}
    \end{subtable}
    \hfill    
    \caption{Negotiation agreement rate per $\beta$}
\end{table}
\begin{table}[H]\ContinuedFloat
    \begin{subtable}[b]{\columnwidth}
        \centering
        \resizebox{\columnwidth}{!}{
        \csvautotabular{\results/negotiation_agreement_rate/domain_b.csv}
        }
        \caption{Domain B}
    \end{subtable}
    \hfill
    \begin{subtable}[b]{\columnwidth}
        \centering
        \resizebox{\columnwidth}{!}{
        \csvautotabular{\results/negotiation_agreement_rate/domain_c.csv}
        }
        \caption{Domain C}
    \end{subtable}
    \hfill
    \centering
    \begin{subtable}[b]{\columnwidth}
        \centering
        \resizebox{\columnwidth}{!}{
        \csvautotabular{\results/negotiation_agreement_rate/domain_d.csv}
        }
        \caption{Domain D}
    \end{subtable}
    \hfill
    \caption{Negotiation agreement rate per $\beta$}
    \label{negotiation_agreement_rate}
\end{table}

}

\onlyaamas{

\begin{table}[ht]
    \vspace{-6mm}

    \centering
    \begin{adjustbox}{max width=\columnwidth}
        \csvautotabular{\results/negotiation_agreement_rate/average.csv}
    \end{adjustbox}
    \caption{Negotiation agreement rate per $\beta$}
    \label{negotiation_agreement_rate}
    \vspace{-10mm}
\end{table}

}

From the results, we can see that, especially when $\beta$ is closer to 1, the chances of a negotiation ending with an agreement that was offered by our agent are relatively low. Therefore, our agent's performance is more influenced by the offers it accepts rather than the offers it makes. In addition, \notaamas{from the results of other agents, we learn that an agent's negotiation agreement rate is not strongly related to its performance.
Both an aggressive agent, i.e., an agent that tends to end a negotiation with an offer it proposed, and a non-aggressive one that tends to accept other agents' offers can achieve good results.
For example, while $Sontag$ has a relatively low negotiation agreement rate and $Agent33$ has a relatively high negotiation agreement rate, both achieve relatively good results. 
To further investigate the contribution of the bids our agent accepts,} we measured the acceptance beta score (i.e., the average beta score of the accepted bids of an agent) of all of the agents. \notaamas{The results are shown in Table~\ref{acceptance_social_welfare} and Table~\ref{acceptance_social_welfare_0.5}}
\onlyaamas{The average results are shown in Table~\ref{acceptance_social_welfare}.}

\notaamas{

\begin{table}[H]
    \centering
    
    \begin{subtable}[b]{\columnwidth}
        \centering
        \begin{adjustbox}{max width=\columnwidth}
        \csvautotabular{\results/acceptance_social_welfare/domain_a.csv}
        \end{adjustbox}
        \caption{Domain A}
    \end{subtable}
    \caption{Acceptance beta score by discount factor for $\beta=1$}
\end{table}
\begin{table}[H]\ContinuedFloat
    \begin{subtable}[b]{\columnwidth}
        \centering
        \begin{adjustbox}{max width=\columnwidth}
        \csvautotabular{\results/acceptance_social_welfare/domain_b.csv}
        \end{adjustbox}
        \caption{Domain B}
    \end{subtable}

    \centering
    \begin{subtable}[b]{\columnwidth}
        \centering
        \begin{adjustbox}{max width=\columnwidth}
        \csvautotabular{\results/acceptance_social_welfare/domain_c.csv}
        \end{adjustbox}
        \caption{Domain C}
    \end{subtable}
    
    \caption{Acceptance beta score by discount factor for $\beta=1$}
\end{table}
\begin{table}[H]\ContinuedFloat
    \begin{subtable}[b]{\columnwidth}
        \centering
        \begin{adjustbox}{max width=\columnwidth}
        \csvautotabular{\results/acceptance_social_welfare/domain_d.csv}
        \end{adjustbox}
        \caption{Domain D}
    \end{subtable}
    \caption{Acceptance beta score by discount factor for $\beta=1$}
    \label{acceptance_social_welfare}
\end{table}

\begin{table}[H]
    \centering
    
    \begin{subtable}[b]{\columnwidth}
        \centering
        \begin{adjustbox}{max width=\columnwidth}
        \begin{filecontents*}{\results/beta_0.5/acceptance_a.csv}
Agent,df=0.2,df=0.4,df=0.6,df=0.8,df=1.0
HerbT+,\textbf{0.3738},0.4280,0.4706,0.5445,0.6399
Agent33,0.3369,0.4060,0.4423,0.5243,0.6235
Agent36,0.1372,0.2727,0.3952,0.5431,0.6781
AgentH,0.1608,0.2768,0.3704,0.4851,0.5763
AgentKN,0.2292,0.3243,0.4119,0.5247,0.6253
AgentX,0.3181,0.4138,0.4854,0.5512,0.6065
AgreeableAgent2018,0.3117,0.3792,0.4502,0.5579,0.6520
Atlas3,0.3181,0.4044,0.4853,0.5703,0.6561
Caduceus,0.2464,0.3428,0.4310,0.5516,0.6572
JonnyBlack,0.3664,\textbf{0.4690},\textbf{0.5667},\textbf{0.6747},0.7577
ParsAgent,0.2648,0.3669,0.4597,0.5737,0.6831
PonPokoAgent,0.3086,0.4230,0.5057,0.6290,0.7173
Rubick,0.0992,0.1355,0.3417,0.5667,0.7550
ShahAgent,0.1879,0.3157,0.4677,0.6325,\textbf{0.7616}
Sontag,0.3199,0.4383,0.5382,0.6304,0.7267
YXAgent,0.2691,0.3579,0.3668,0.4328,0.5533
\end{filecontents*}
\csvautotabular{\results/beta_0.5/acceptance_a.csv}
        \end{adjustbox}
        \caption{Domain A}
    \end{subtable}
    \caption{Acceptance beta score by discount factor for $\beta=0.5$}
\end{table}
\begin{table}[H]\ContinuedFloat
    \begin{subtable}[b]{\columnwidth}
        \centering
        \begin{adjustbox}{max width=\columnwidth}
        \begin{filecontents*}{\results/beta_0.5/acceptance_b.csv}
Agent,df=0.2,df=0.4,df=0.6,df=0.8,df=1.0
HerbT+,0.2952,0.3159,0.3682,0.3954,0.4811
Agent33,0.2110,0.2254,0.2796,0.3550,0.3659
Agent36,0.1822,0.2542,0.4647,0.5580,0.5331
AgentH,0.1682,0.2085,0.3562,0.4222,0.4672
AgentKN,0.1681,0.2042,0.3429,0.4098,0.4540
AgentX,0.1899,0.2289,0.3315,0.3655,0.4030
AgreeableAgent2018,\textbf{0.4018},\textbf{0.4674},\textbf{0.5649},0.6157,0.5440
Atlas3,0.2097,0.2383,0.3311,0.3720,0.4901
Caduceus,0.2331,0.3011,0.4865,0.5507,0.5294
JonnyBlack,0.2860,0.3158,0.4716,0.5351,0.4657
ParsAgent,0.2536,0.3307,0.5041,0.5680,0.5354
PonPokoAgent,0.3308,0.3797,0.5266,0.5826,0.5275
Rubick,0.1103,0.1823,0.4293,\textbf{0.6605},0.5518
ShahAgent,0.1891,0.2489,0.5052,0.5587,0.5548
Sontag,0.2929,0.3458,0.5050,0.5599,\textbf{0.6149}
YXAgent,0.1723,0.2183,0.2623,0.2825,0.5230
\end{filecontents*}
\csvautotabular{\results/beta_0.5/acceptance_b.csv}
        \end{adjustbox}
        \caption{Domain B}
    \end{subtable}
    \centering

    \begin{subtable}[b]{\columnwidth}
        \centering
        \begin{adjustbox}{max width=\columnwidth}
        \begin{filecontents*}{\results/beta_0.5/acceptance_c.csv}
Agent,df=0.2,df=0.4,df=0.6,df=0.8,df=1.0
HerbT+,\textbf{0.4655},0.5151,0.6651,0.6922,0.8425
Agent33,0.4222,0.5055,0.6416,0.6951,0.7695
Agent36,0.1610,0.3160,0.5726,0.6237,0.7665
AgentH,0.2316,0.3759,0.5745,0.6131,0.7151
AgentKN,0.3287,0.4660,0.6526,0.6919,0.7864
AgentX,0.4207,0.5237,0.6424,0.6582,0.7145
AgreeableAgent2018,0.3254,0.4725,0.6511,0.6889,0.7850
Atlas3,0.4041,0.5123,0.6744,0.7047,0.8015
Caduceus,0.2693,0.4057,0.5971,0.6429,0.7442
JonnyBlack,0.3927,0.5232,0.6902,0.7275,0.8109
ParsAgent,0.3019,0.4468,0.6416,0.6795,0.7806
PonPokoAgent,0.4053,0.5310,0.6867,0.7196,0.7946
Rubick,0.2754,0.4374,0.6426,0.6345,0.8255
ShahAgent,0.2395,0.4141,0.6454,0.6938,0.8211
Sontag,0.4549,\textbf{0.5879},\textbf{0.7464},\textbf{0.7725},\textbf{0.8480}
YXAgent,0.3795,0.4413,0.6011,0.6766,0.7138
\end{filecontents*}
\csvautotabular{\results/beta_0.5/acceptance_c.csv}
        \end{adjustbox}
        \caption{Domain C}
    \end{subtable}
    \caption{Acceptance beta score by discount factor for $\beta=0.5$}
\end{table}
\begin{table}[H]\ContinuedFloat
    \begin{subtable}[b]{\columnwidth}
        \centering
        \begin{adjustbox}{max width=\columnwidth}
        \begin{filecontents*}{\results/beta_0.5/acceptance_d.csv}
Agent,df=0.2,df=0.4,df=0.6,df=0.8,df=1.0
HerbT+,0.5672,0.6401,0.7185,0.7971,0.9146
Agent33,0.4775,0.5770,0.6667,0.7305,0.9250
Agent36,0.1661,0.3277,0.5434,0.6473,0.8980
AgentH,0.2496,0.4011,0.5695,0.6460,0.8497
AgentKN,0.5184,0.6189,0.7094,0.7669,0.9282
AgentX,0.5624,0.6478,0.7191,0.7536,0.7908
AgreeableAgent2018,\textbf{0.6156},\textbf{0.6886},0.7467,0.7899,0.9268
Atlas3,0.5023,0.6175,0.7348,0.7915,0.9000
Caduceus,0.3968,0.4759,0.5695,0.6699,0.8894
JonnyBlack,0.4895,0.6021,0.7043,0.7669,0.9330
ParsAgent,0.5076,0.6094,0.7020,0.7597,0.9244
PonPokoAgent,0.5241,0.6221,0.7117,0.7666,0.9267
Rubick,0.1600,0.3164,\textbf{0.8531},0.6479,\textbf{0.9332}
ShahAgent,0.2218,0.3843,0.5829,0.7089,0.9327
Sontag,0.5086,0.6288,0.7487,\textbf{0.8045},0.8858
YXAgent,0.3901,0.5240,0.6454,0.7190,0.8891
\end{filecontents*}
\csvautotabular{\results/beta_0.5/acceptance_d.csv}
        \end{adjustbox}
        \caption{Domain D}
    \end{subtable}
    \caption{Acceptance beta score by discount factor for $\beta=0.5$}
    \label{acceptance_social_welfare_0.5}
\end{table}

}

\onlyaamas{
\begin{table}[ht]
    \vspace{-6mm}

    \centering
    
    \begin{adjustbox}{max width=0.9\columnwidth}
        \csvautotabular{\results/acceptance_social_welfare/average.csv}
    \end{adjustbox}
    
    \caption{Acceptance beta score by discount factor for $\beta=1$}
    \label{acceptance_social_welfare}
    \vspace{-10mm}
\end{table}
}

According to our results, in most scenarios, our agent, HerbT+, accepts bids with a relatively high beta score but usually not the highest. However, in Table~\ref{negotiation_agreement_rate}, we saw that our agent's performance is more affected by the bids it accepts, i.e., its acceptance beta score, than the bids it offers. Therefore, a good acceptance beta score will increase its overall performance more than it would increase other agents' performance. From this result, we learn that a significant reason for our agent's success is a combination of a low decline rate and a low negotiation agreement rate while still maintaining a high acceptance beta score, especially when $\beta=1$. In other words, our agent manages to achieve good results by accepting many bids relative to other agents and still manages to keep a high average of the beta score of the bids it accepts.
With that being said, it would also be interesting to see the results of negotiations that do end with an agreement that was proposed by our agent. We measured for each agent the average beta score of the negotiations that ended with an offer of the agent. \notaamas{The results are shown in Table~\ref{chosen_social_welfare} and Table~\ref{chosen_social_welfare_0.5}}
\onlyaamas{The average results are shown in Table~\ref{chosen_social_welfare}.}

\notaamas{

\begin{table}[H]
    \centering
    
    \begin{subtable}[b]{\columnwidth}
        \centering
        \begin{adjustbox}{max width=\columnwidth}
        \csvautotabular{\results/chosen_offers_social_welfare/domain_a.csv}
        \end{adjustbox}
        \caption{Domain A}
    \end{subtable}
    \begin{subtable}[b]{\columnwidth}
        \centering
        \begin{adjustbox}{max width=\columnwidth}
        \csvautotabular{\results/chosen_offers_social_welfare/domain_b.csv}
        \end{adjustbox}
        \caption{Domain B}
    \end{subtable}
    \centering
    \caption{Beta score of the negotiations that ended with an offer of a specific agent for $\beta=1$}
\end{table}
\begin{table}[H]\ContinuedFloat
    \begin{subtable}[b]{\columnwidth}
        \centering
        \begin{adjustbox}{max width=\columnwidth}
        \csvautotabular{\results/chosen_offers_social_welfare/domain_c.csv}
        \end{adjustbox}
        \caption{Domain C}
    \end{subtable}

    \begin{subtable}[b]{\columnwidth}
        \centering
        \begin{adjustbox}{max width=\columnwidth}
        \csvautotabular{\results/chosen_offers_social_welfare/domain_d.csv}
        \end{adjustbox}
        \caption{Domain D}
    \end{subtable}
    \caption{Beta score of the negotiations that ended with an offer of a specific agent for $\beta=1$}
    \label{chosen_social_welfare}
\end{table}

\begin{table}[H]
    \centering
    
    \begin{subtable}[b]{\columnwidth}
        \centering
        \begin{adjustbox}{max width=\columnwidth}
        \begin{filecontents*}{\results/beta_0.5/chosen_offer_a.csv}
Agent,df=0.2,df=0.4,df=0.6,df=0.8,df=1.0
HerbT+,\textbf{0.5857},0.3517,0.4318,0.5138,0.6086
Agent33,0.2979,0.3286,0.3575,0.4452,0.4988
Agent36,0.2464,0.3444,0.4256,0.5519,0.6520
AgentH,0.3805,0.4124,0.4409,0.5053,0.5447
AgentKN,0.2427,0.3397,0.4051,0.5138,0.5911
AgentX,0.3686,\textbf{0.4504},\textbf{0.5385},0.6148,0.7138
AgreeableAgent2018,0.2230,0.3300,0.4151,0.5448,0.6454
Atlas3,0.2496,0.3305,0.4172,0.5390,0.6723
Caduceus,0.2441,0.3391,0.4183,0.5482,0.6519
JonnyBlack,0.2475,0.3583,0.4593,\textbf{0.6092},\textbf{0.7323}
ParsAgent,0.2377,0.3361,0.4154,0.5409,0.6418
PonPokoAgent,0.2415,0.3358,0.4222,0.5546,0.6646
Rubick,0.2293,0.3442,0.4408,0.5836,0.7090
ShahAgent,0.4101,0.4100,0.4370,0.5655,0.6806
Sontag,0.2572,0.3542,0.4682,0.5861,0.7081
YXAgent,0.2561,0.3404,0.4238,0.5496,0.6576
\end{filecontents*}
\csvautotabular{\results/beta_0.5/chosen_offer_a.csv}
        \end{adjustbox}
        \caption{Domain A}
    \end{subtable}
    \begin{subtable}[b]{\columnwidth}
        \centering
        \begin{adjustbox}{max width=\columnwidth}
        \begin{filecontents*}{\results/beta_0.5/chosen_offer_b.csv}
Agent,df=0.2,df=0.4,df=0.6,df=0.8,df=1.0
HerbT+,\textbf{0.4507},\textbf{0.3515},0.3747,0.3529,0.4233
Agent33,0.2007,0.2296,0.3182,0.3528,0.3998
Agent36,0.2641,0.3437,0.4077,0.4821,0.5385
AgentH,0.2829,0.3348,0.3235,0.3772,0.3976
AgentKN,0.2239,0.2932,0.3090,0.3593,0.3770
AgentX,0.3253,0.3480,0.4471,0.4733,0.5365
AgreeableAgent2018,0.2579,0.3459,\textbf{0.4644},\textbf{0.5204},\textbf{0.5980}
Atlas3,0.2432,0.2613,0.3651,0.4130,0.4790
Caduceus,0.2509,0.3268,0.3887,0.4593,0.5162
JonnyBlack,0.2538,0.3306,0.4096,0.4868,0.5287
ParsAgent,0.2569,0.3407,0.4101,0.4845,0.5379
PonPokoAgent,0.2345,0.3073,0.3499,0.4109,0.4436
Rubick,0.2482,0.3307,0.4055,0.4794,0.5412
ShahAgent,0.2969,0.3308,0.3836,0.4575,0.4985
Sontag,0.2278,0.2881,0.4546,0.5130,0.5883
YXAgent,0.2578,0.3356,0.3960,0.4657,0.5220
\end{filecontents*}
\csvautotabular{\results/beta_0.5/chosen_offer_b.csv}
        \end{adjustbox}
        \caption{Domain B}
    \end{subtable}
    \caption{Beta score of the negotiations that ended with an offer of a specific agent for $\beta=0.5$}
\end{table}
\begin{table}[H]\ContinuedFloat
    
    \centering
    \begin{subtable}[b]{\columnwidth}
        \centering
        \begin{adjustbox}{max width=\columnwidth}
        \begin{filecontents*}{\results/beta_0.5/chosen_offer_c.csv}
Agent,df=0.2,df=0.4,df=0.6,df=0.8,df=1.0
HerbT+,\textbf{0.8103},\textbf{0.6372},\textbf{0.7557},\textbf{0.7655},0.8186
Agent33,0.4869,0.5716,0.6942,0.7148,0.7964
Agent36,0.3432,0.4884,0.7109,0.7532,0.8784
AgentH,0.4757,0.5663,0.6950,0.7180,0.7817
AgentKN,0.3941,0.5306,0.7302,0.7276,0.8847
AgentX,0.4925,0.5740,0.7329,0.7632,0.8669
AgreeableAgent2018,0.3045,0.4581,0.6834,0.7299,0.8592
Atlas3,0.6337,0.6309,0.7268,0.7430,0.8689
Caduceus,0.3919,0.5125,0.6920,0.7309,0.8503
JonnyBlack,0.3451,0.4869,0.7213,0.7570,\textbf{0.9152}
ParsAgent,0.3834,0.5191,0.7155,0.7518,0.8616
PonPokoAgent,0.4170,0.5288,0.7214,0.7598,0.8793
Rubick,0.3245,0.4736,0.7152,0.7495,0.8761
ShahAgent,0.5875,0.6155,0.7523,0.7754,0.8841
Sontag,0.5190,0.5855,0.7173,0.7382,0.8161
YXAgent,0.3622,0.4833,0.6951,0.7364,0.8680
\end{filecontents*}
\csvautotabular{\results/beta_0.5/chosen_offer_c.csv}
        \end{adjustbox}
        \caption{Domain C}
    \end{subtable}
    \begin{subtable}[b]{\columnwidth}
        \centering
        \begin{adjustbox}{max width=\columnwidth}
        \begin{filecontents*}{\results/beta_0.5/chosen_offer_d.csv}
Agent,df=0.2,df=0.4,df=0.6,df=0.8,df=1.0
HerbT+,\textbf{0.7090},0.5319,\textbf{0.7263},0.7493,0.8315
Agent33,0.5543,0.6030,0.6612,0.6697,0.7530
Agent36,0.3533,0.4769,0.6280,0.7207,0.8452
AgentH,0.5628,0.6198,0.6822,0.7200,0.7636
AgentKN,0.3717,0.5009,0.6628,0.7634,0.9030
AgentX,0.3982,0.5226,0.6848,\textbf{0.7807},\textbf{0.9301}
AgreeableAgent2018,0.3267,0.4589,0.6084,0.7169,0.8469
Atlas3,0.5710,\textbf{0.6125},0.7023,0.7124,0.7949
Caduceus,0.3844,0.5021,0.6505,0.7340,0.8627
JonnyBlack,0.4064,0.5330,0.6958,0.7868,0.9299
ParsAgent,0.3414,0.4684,0.6216,0.7123,0.8362
PonPokoAgent,0.3235,0.4480,0.6100,0.6960,0.8320
Rubick,0.3787,0.5058,0.6484,0.7663,0.9044
ShahAgent,0.5700,0.6103,0.6818,0.7282,0.8545
Sontag,0.3856,0.4714,0.6063,0.6724,0.7693
YXAgent,0.3928,0.5127,0.6270,0.7066,0.8485
\end{filecontents*}
\csvautotabular{\results/beta_0.5/chosen_offer_d.csv}
        \end{adjustbox}
        \caption{Domain D}
    \end{subtable}
    \caption{Beta score of the negotiations that ended with an offer of a specific agent for $\beta=0.5$}
    \label{chosen_social_welfare_0.5}
\end{table}

}

\onlyaamas{
\begin{table}[ht]
    \vspace{-6mm}

    \centering
    
    \begin{adjustbox}{max width=0.9\columnwidth}
        \csvautotabular{\results/chosen_offers_social_welfare/average.csv}
    \end{adjustbox}
        
    \caption{Beta score of the negotiations that ended with an offer of a specific agent for $\beta=1$}
    \label{chosen_social_welfare}
    \vspace{-10mm}
\end{table}
}

According to our results, although not many negotiations end with an agreement that was offered by our agent, the ones that do end with a relatively high beta score, especially when $\beta=1$.
}

\notaamas{

\subtopic{Unique Populations}

In the following section, we provide an analysis of our agent's evaluation in different and unique agent populations.

\paragraph{Only Social}
First, we used only the agents who try to maximize social welfare (i.e., the agents who had an achievement in the social welfare category according to Table~\ref{top_agents}) as our population. Using only negotiations with the agents in this population, we compared our agent's social welfare with $\beta=1$ to the social welfare achieved by each of those agents. The results are provided in Table~\ref{only_social}.

\begin{table}[ht]
    \centering
    
    \begin{adjustbox}{max width=0.9\columnwidth}
        \begin{filecontents*}{\results/only_social/all.csv}
Agent,Domain A,Domain B,Domain C,Domain D,Domain E,Domain F,Domain G
HerbT+,\textbf{0.4827},\textbf{0.3726},\textbf{0.7895},\textbf{0.7755},\textbf{0.7319},\textbf{0.7200},\textbf{0.6293}
Agent33,0.3666,0.3050,0.7246,0.7380,0.6790,N/A,0.6065
AgentH,0.3161,0.2081,0.7328,0.7638,0.5941,N/A,0.6051
AgentKN,0.2192,0.2080,0.7050,0.7476,0.5627,N/A,0.5870
AgentX,0.3984,0.3250,0.7452,0.7457,0.7162,N/A,0.6144
Atlas3,0.4509,0.3409,0.7514,0.7686,0.7347,0.6711,0.6242
JonnyBlack,0.1654,0.1772,0.5483,0.6543,0.7160,N/A,0.5875
ShahAgent,0.2677,0.1994,0.6464,0.6659,0.6136,0.5177,0.5908
Sontag,0.2589,0.2320,0.7765,0.7077,0.7178,0.6251,0.5956
\end{filecontents*}
\csvautotabular{\results/only_social/all.csv}
    \end{adjustbox}
        
    \caption{Social welfare in the only social population\footref{empty_values}}
    \label{only_social}
\end{table}

According to Table~\ref{only_social}, when the entire population tries to maximize social welfare, our agent managed to achieve the highest social welfare than all the other agents in all the domains (including domain D, in which our agent did not achieve the highest social welfare with the original population,  as shown in Table~\ref{social_welfare_results}). In addition, every agent achieved higher social welfare in this population than it achieved in the original population in most domains. This makes sense since when all the agents in the negotiation want to maximize the social welfare, the social welfare will naturally be higher than the opposite case.

\paragraph{Self Negotiation}
Next, we evaluated how well our agent does in a negotiation against itself, relative to other agents, i.e., negotiation when all the negotiators are instances of the same agent (each such instance still has different preferences, but the strategy is the same). For this evaluation, we used only the negotiations who contain 3 instances of the same agents. The beta score of each agent is provided in Table~\ref{same_agent}.

\begin{table}[H]
    \centering
    
    \begin{subtable}[b]{\columnwidth}
        \centering
        \begin{adjustbox}{max width=\columnwidth}
        \csvautotabular{\results/same_agent/domain_a.csv}
        \end{adjustbox}
        \caption{Domain A}
    \end{subtable}
\end{table}
\begin{table}[H]\ContinuedFloat
    \begin{subtable}[b]{\columnwidth}
        \centering
        \begin{adjustbox}{max width=\columnwidth}
        \csvautotabular{\results/same_agent/domain_b.csv}
        \end{adjustbox}
        \caption{Domain B}
    \end{subtable}
    \caption{Beta score of each agent against only instances of itself}
\end{table}
\begin{table}[H]\ContinuedFloat
    
    \begin{subtable}[b]{\columnwidth}
        \centering
        \begin{adjustbox}{max width=\columnwidth}
        \csvautotabular{\results/same_agent/domain_c.csv}
        \end{adjustbox}
        \caption{Domain C}
    \end{subtable}
    
\end{table}
\begin{table}[H]\ContinuedFloat
    \begin{subtable}[b]{\columnwidth}
        \centering
        \begin{adjustbox}{max width=\columnwidth}
        \csvautotabular{\results/same_agent/domain_d.csv}
        \end{adjustbox}
        \caption{Domain D}
    \end{subtable}
    \caption{Beta score of each agent against only instances of itself}

\end{table}
\begin{table}[H]\ContinuedFloat
    \begin{subtable}[b]{\columnwidth}
        \centering
        \begin{adjustbox}{max width=\columnwidth}
        \begin{filecontents*}{\results/same_agent/domain_e.csv}
Agent,0,0.1,0.2,0.3,0.4,0.5,0.6,0.7,0.8,0.9,1
HerbT+,0.7615,0.7694,0.7838,0.7891,0.7999,0.8007,0.8094,0.8187,\textbf{0.8255},\textbf{0.8305},\textbf{0.8320}
Agent33,0.5635,0.4979,0.4264,0.1659,0.3638,0.3645,0.3546,0.5033,0.4897,0.4241,0.2995
Agent36,0.4395,0.4403,0.4432,0.4457,0.4421,0.4430,0.4409,0.4429,0.4428,0.4421,0.4418
AgentH,0.5598,0.5497,0.5486,0.5599,0.5586,0.5503,0.5485,0.5366,0.5641,0.5513,0.5478
AgentKN,0.4688,0.4569,0.4625,0.4720,0.4694,0.4621,0.4647,0.4678,0.4643,0.4552,0.4701
AgentX,0.8013,0.8204,0.8059,0.8079,0.7838,0.7951,0.8191,0.8106,0.8089,0.8088,0.8040
AgreeableAgent2018,0.4427,0.4391,0.4428,0.4299,0.4297,0.4311,0.4368,0.4358,0.4295,0.4418,0.4286
Atlas3,\textbf{0.8287},\textbf{0.8258},\textbf{0.8284},\textbf{0.8246},\textbf{0.8352},\textbf{0.8279},\textbf{0.8365},\textbf{0.8334},0.8227,0.8247,0.8304
Caduceus,0.5190,0.5114,0.5096,0.5143,0.5135,0.5101,0.5202,0.5096,0.5198,0.5166,0.5091
JonnyBlack,0.8230,0.8230,0.8230,0.8230,0.8230,0.8230,0.8230,0.8230,0.8230,0.8230,0.8230
ParsAgent,0.5029,0.5079,0.5076,0.5010,0.5031,0.4981,0.5021,0.5011,0.5031,0.4999,0.5048
PonPokoAgent,0.3621,0.3905,0.4030,0.4521,0.4895,0.4870,0.3616,0.5000,0.4063,0.4539,0.4214
Rubick,0.1004,0.1004,0.1004,0.1004,0.1004,0.1004,0.1004,0.1004,0.1004,0.1004,0.1004
ShahAgent,0.5684,0.5650,0.5611,0.5636,0.5647,0.5570,0.5694,0.5610,0.5578,0.5647,0.5675
Sontag,0.7440,0.7526,0.7394,0.7474,0.7636,0.7567,0.7476,0.7421,0.7483,0.7606,0.7533
YXAgent,0.7815,0.7810,0.8109,0.8021,0.7755,0.8041,0.8112,0.8200,0.7842,0.8159,0.8104
\end{filecontents*}
\csvautotabular{\results/same_agent/domain_e.csv}
        \end{adjustbox}
        \caption{Domain E}
    \end{subtable}
    
\end{table}
\begin{table}[H]\ContinuedFloat
    \begin{subtable}[b]{\columnwidth}
        \centering
        \begin{adjustbox}{max width=\columnwidth}
        \begin{filecontents*}{\results/same_agent/domain_f.csv}
Agent,0,0.1,0.2,0.3,0.4,0.5,0.6,0.7,0.8,0.9,1
HerbT+,0.6481,0.6656,0.6656,\textbf{0.7181},\textbf{0.7370},\textbf{0.7581},\textbf{0.7871},\textbf{0.8031},\textbf{0.8277},\textbf{0.8471},\textbf{0.8543}
Agent36,0.1083,0.1802,0.1805,0.0723,0.1085,0.1085,0.1443,0.0722,0.1083,0.1807,0.0722
AgreeableAgent2018,0.3609,0.3609,0.3609,0.3609,0.3609,0.3609,0.3609,0.3609,0.3609,0.3609,0.3609
Atlas3,\textbf{0.7098},\textbf{0.7094},\textbf{0.7046},0.7089,0.7101,0.7108,0.7086,0.7071,0.7091,0.7072,0.7071
ParsAgent,0.3624,0.3235,0.3983,0.2819,0.3246,0.3992,0.2812,0.3646,0.2761,0.4369,0.3945
PonPokoAgent,0.0000,0.0000,0.0000,0.0000,0.0000,0.0000,0.0000,0.0000,0.0000,0.0000,0.0000
ShahAgent,0.0465,0.0414,0.0000,0.0436,0.0463,0.0000,0.0465,0.0427,0.0402,0.0000,0.0405
Sontag,0.6103,0.6183,0.6176,0.6135,0.6039,0.6023,0.6151,0.6029,0.6164,0.6204,0.6187
\end{filecontents*}
\csvautotabular{\results/same_agent/domain_f.csv}
        \end{adjustbox}
        \caption{Domain F}
    \end{subtable}
\end{table}
\begin{table}[H]\ContinuedFloat
    
    \begin{subtable}[b]{\columnwidth}
        \centering
        \begin{adjustbox}{max width=\columnwidth}
        \begin{filecontents*}{\results/same_agent/domain_g.csv}
Agent,0,0.1,0.2,0.3,0.4,0.5,0.6,0.7,0.8,0.9,1
HerbT+,0.6484,0.6638,0.6717,\textbf{0.6719},\textbf{0.6790},\textbf{0.6849},\textbf{0.6987},\textbf{0.6969},\textbf{0.7107},\textbf{0.7117},\textbf{0.7142}
Agent33,0.5662,0.5569,0.5598,0.5587,0.5587,0.5621,0.5587,0.5588,0.5642,0.5607,0.5753
Agent36,0.5607,0.5607,0.5607,0.5607,0.5607,0.5607,0.5607,0.5607,0.5607,0.5607,0.5607
AgentH,0.6088,0.6033,0.6022,0.5819,0.5950,0.5928,0.5870,0.5962,0.6014,0.5765,0.5809
AgentKN,0.5759,0.5748,0.5828,0.5704,0.5910,0.5675,0.5804,0.5805,0.5661,0.5703,0.5545
AgentX,\textbf{0.6760},\textbf{0.6834},\textbf{0.6899},0.6709,0.6780,0.6596,0.6652,0.6772,0.6766,0.6658,0.6646
AgreeableAgent2018,0.5607,0.5607,0.5607,0.5607,0.5607,0.5607,0.5607,0.5607,0.5607,0.5607,0.5607
Atlas3,0.6582,0.6582,0.6582,0.6582,0.6582,0.6582,0.6582,0.6582,0.6582,0.6582,0.6582
Caduceus,0.5607,0.5607,0.5607,0.5607,0.5607,0.5607,0.5607,0.5607,0.5607,0.5607,0.5607
JonnyBlack,0.5607,0.5607,0.5607,0.5607,0.5607,0.5607,0.5607,0.5607,0.5607,0.5607,0.5607
ParsAgent,0.5607,0.5607,0.5607,0.5607,0.5607,0.5607,0.5607,0.5607,0.5607,0.5607,0.5607
PonPokoAgent,0.5607,0.5607,0.5607,0.5607,0.5607,0.5607,0.5607,0.5607,0.5607,0.5607,0.5607
Rubick,0.5607,0.5607,0.5607,0.5607,0.5607,0.5607,0.5607,0.5607,0.5607,0.5607,0.5607
ShahAgent,0.5607,0.5607,0.5607,0.5607,0.5607,0.5607,0.5607,0.5607,0.5607,0.5607,0.5607
Sontag,0.5607,0.5607,0.5607,0.5607,0.5607,0.5607,0.5607,0.5607,0.5607,0.5607,0.5607
YXAgent,0.5607,0.5607,0.5607,0.5607,0.5607,0.5607,0.5607,0.5607,0.5607,0.5607,0.5607
\end{filecontents*}
\csvautotabular{\results/same_agent/domain_g.csv}
        \end{adjustbox}
        \caption{Domain G}
    \end{subtable}
    
    \caption{Beta score of each agent against only instances of itself}
    \label{same_agent}
\end{table}

According to Table~\ref{same_agent}, when competing against itself and when $\beta=1$, our agent achieved the highest beta score in all domains (including domain D). Furthermore, when only negotiations of three instances of the same agent are considered, an agent's average beta score is equal to the average individual utility and the average social welfare of the agent. This means that when $\beta=1$, in this kind of population, our agent outperforms all the other agents in all of the domains, both in its individual utility and social welfare. In addition, as can be seen in the table, unlike our agent, a lot of agents do not reach an agreement with themselves in some of the domains (when the beta score is the discounted reservation value of the domain).

}

\notaamas{
\subtopic{Correlation Assumption}
\label{correlation_assumption}

In our strategy, we assume that there is a strong correlation between an agent's utility function and the actions it takes\notaamas{, i.e., choosing a bid to offer and deciding whether to accept or reject another agent's offer}. We test this assumption by comparing the output of the model we trained for each agent on each possible bid with the output of the real utility function of the agent. We compare the outputs using two measures: mean absolute error \cite{MeanAbsoluteError} and Pearson's correlation coefficient \cite{PearsonsCorrelationCoefficient}.\notaamas{ From the mean absolute error of the output of our model and the real utility function, we will be able to tell how close our model is to the real utility function and from the correlation between the output of our model and the output of the real utility function we will be able to verify the correlation assumption we use in our strategy.} In addition, using these measures we compare our logistic regression model to other machine learning models: SVM \cite{SVM}, Decision tree \cite{DecisionTree}, and Random forests \cite{RandomForests}. We made the comparison using all of the negotiations in domain B. We chose domain B since it has the most samples of negotiations which proceeded to late rounds.
The results are provided in Figure~\ref{pearson_correlation} and Figure~\ref{mean_absolute_error}:

\notaamas{

\begin{figure}[H]
    \centering
    \includegraphics[width=0.9\columnwidth]{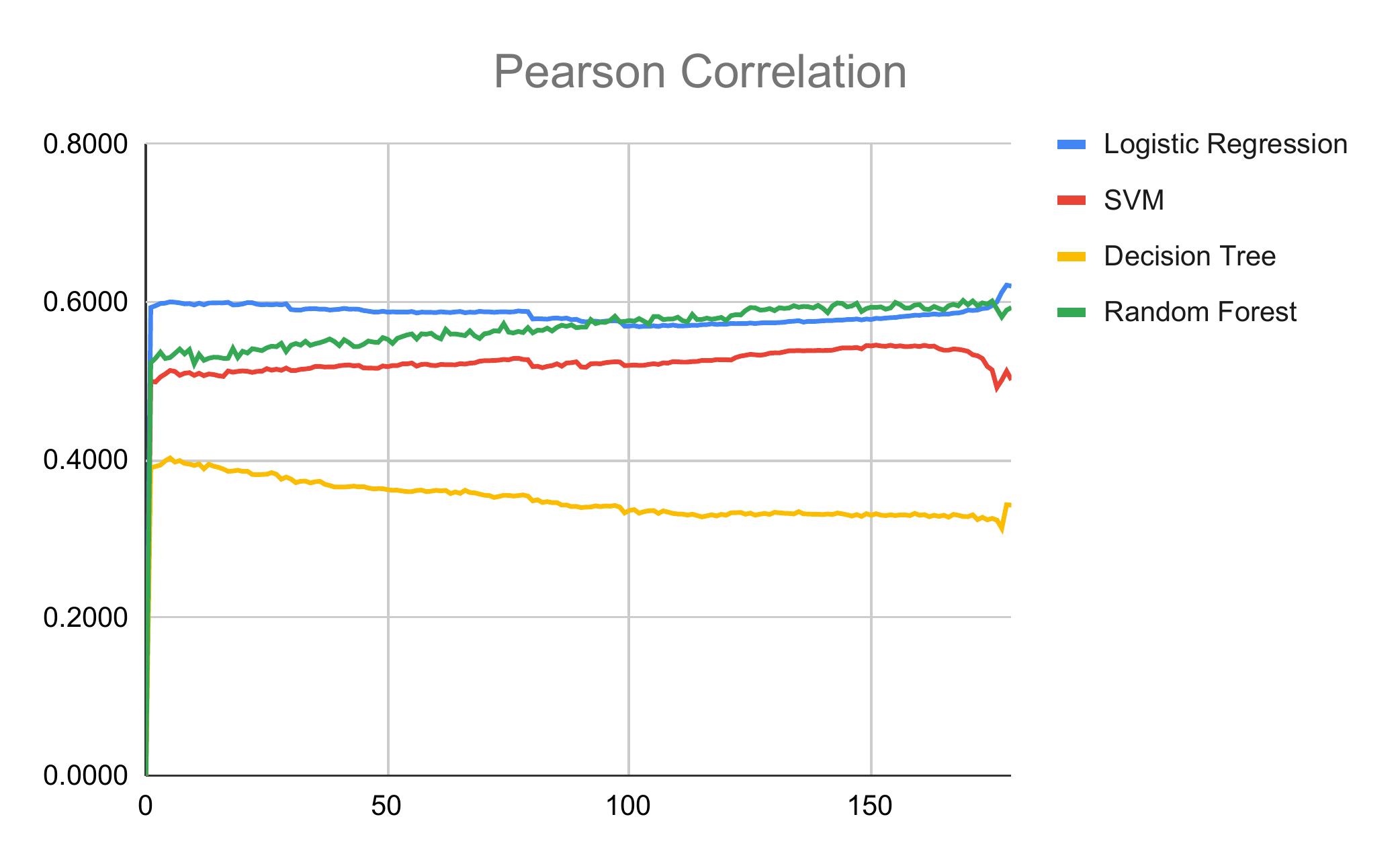}
    \caption{Pearson's correlation coefficient on each round in the negotiation}
    \label{pearson_correlation}
\end{figure}

\begin{figure}[H]
    \centering
    \includegraphics[width=0.9\columnwidth]{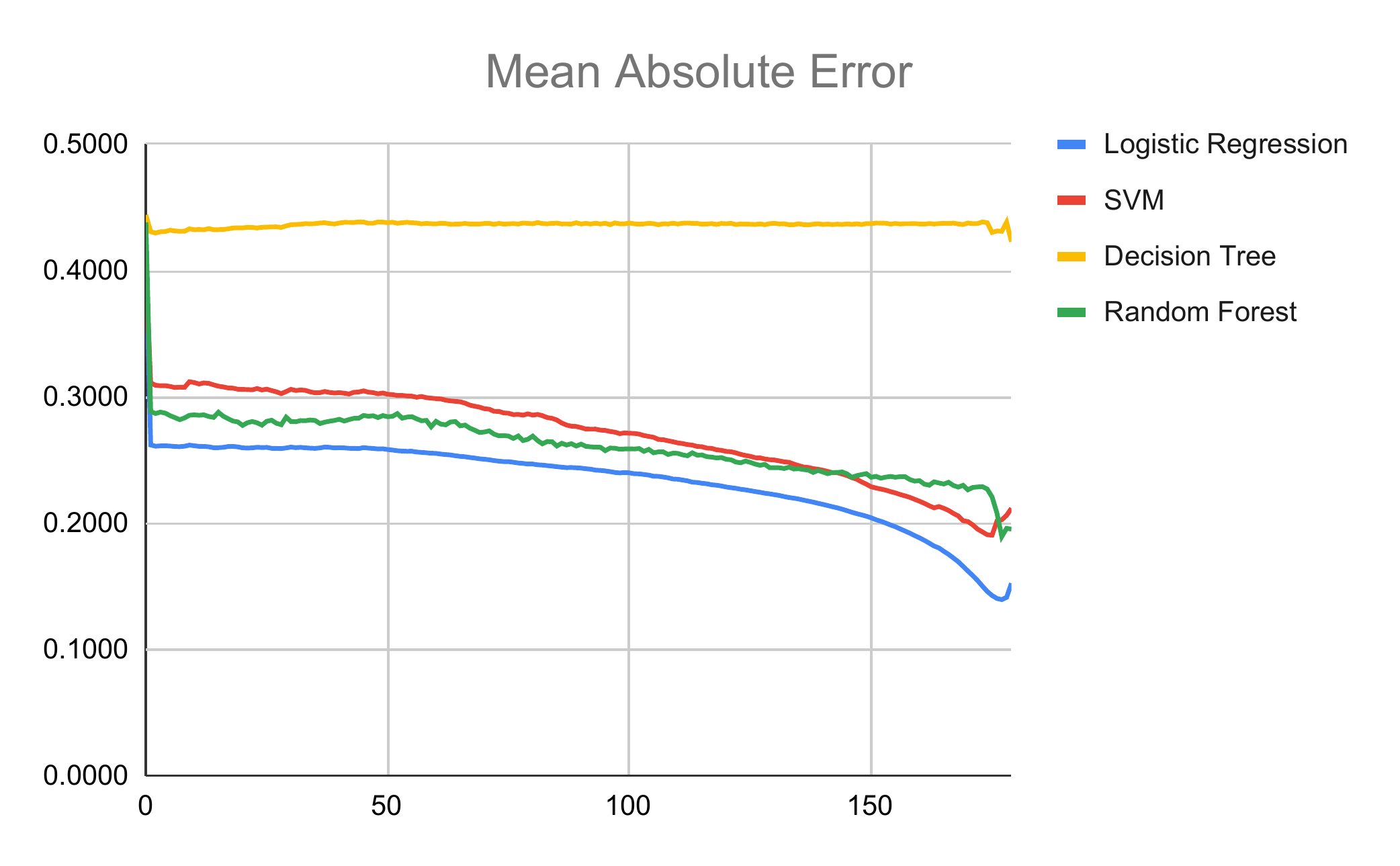}
    \caption{Mean absolute error on each round in the negotiation}
    \label{mean_absolute_error}
\end{figure}

}

\onlyaamas{
\begin{figure}[ht]
    \centering
    \begin{subfigure}[b]{0.49\columnwidth}
       \centering
        \includegraphics[width=\columnwidth]{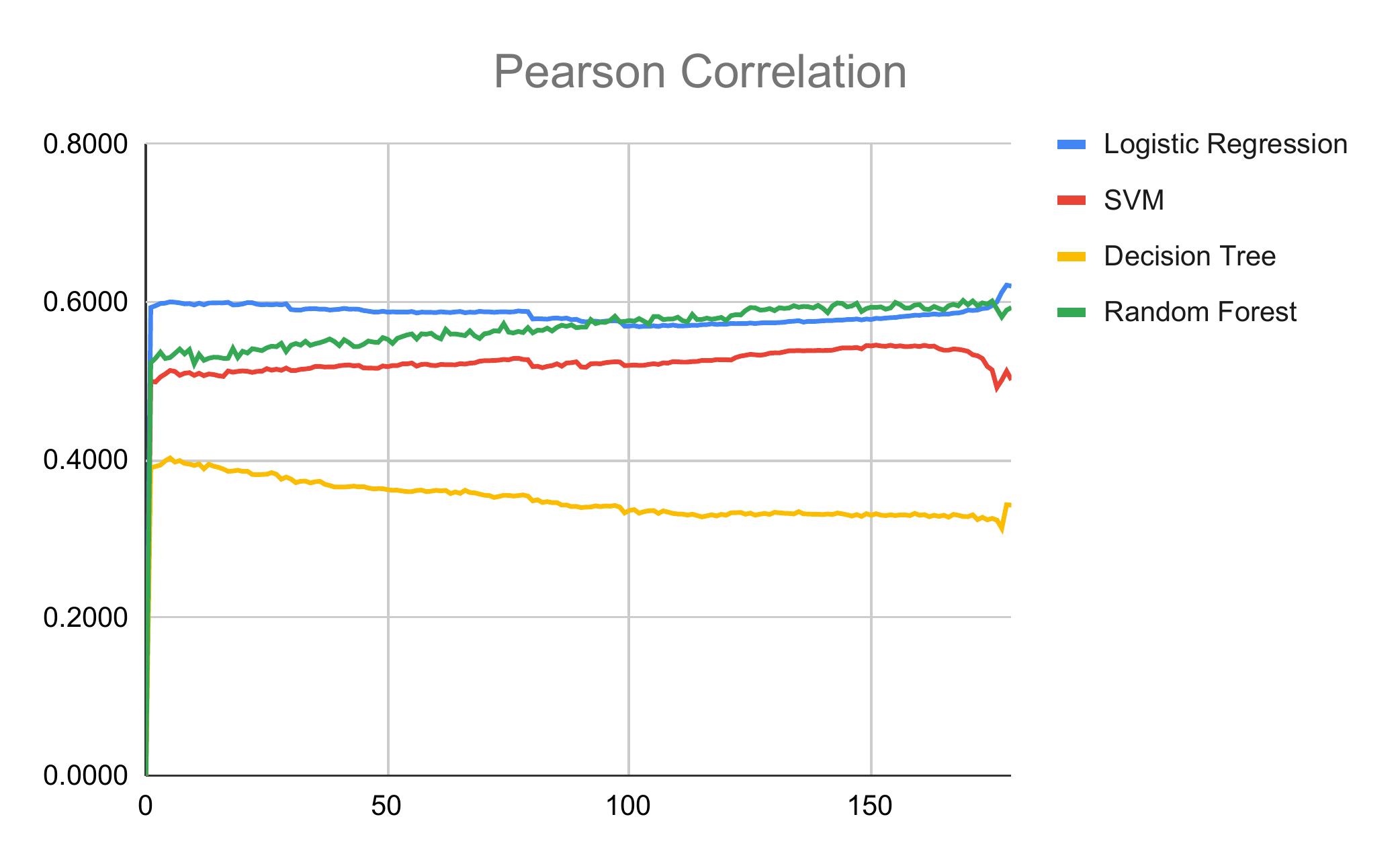}
        \caption{Pearson's correlation on each round in the negotiation}
        \label{pearson_correlation}
    \end{subfigure}
    \hfill
    \begin{subfigure}[b]{0.49\columnwidth}
        \centering
        \includegraphics[width=\columnwidth]{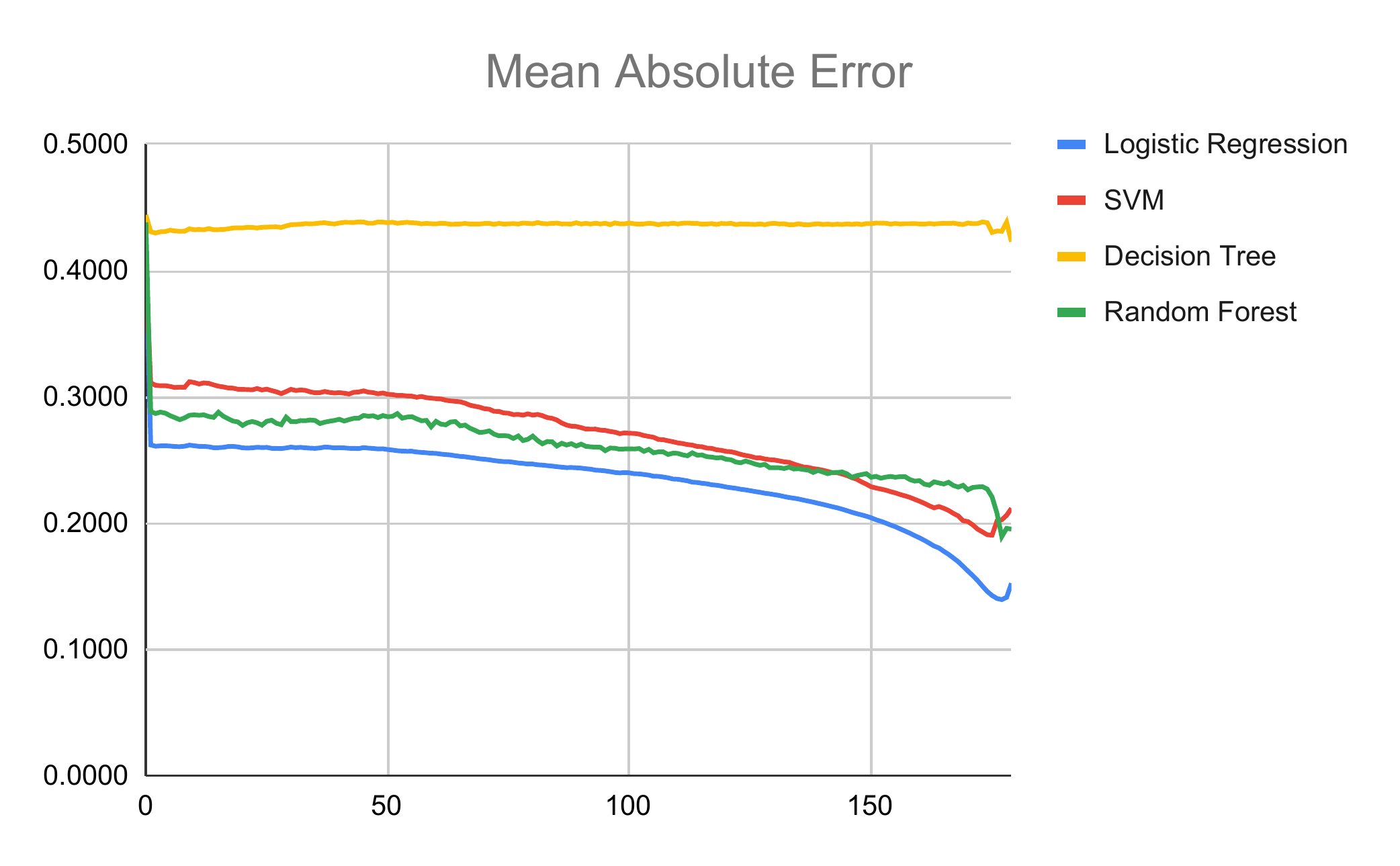}
        \caption{Mean absolute error on each round in the negotiation}
        \label{mean_absolute_error}
    \end{subfigure}
    \caption{}
\end{figure}
}

According to Figure \ref{pearson_correlation}, by training a logistic regression model with an agent's actions, we managed to build a model where the Pearson correlation coefficient of its output with the output of the real utility function is above 0.5. This indeed verifies the correlation assumption we use in our strategy. As expected, since the correlation is high, our model manages to achieve a relatively low mean absolute error with the real utility function. In addition, as can be seen in Figure \ref{mean_absolute_error}, a logistic regression model does manage to achieve the lowest mean absolute error between the compared methods. 

}

\notaamas{
\subtopic{Random Initialization}
\label{random_initialization}

In order to measure the effectiveness of re-initializing the logistic regression model with random weights in each turn, we compared our model with random initialization against our model which uses the same logistic regression model in all the negotiation's rounds and updates the model with each new observation. Table~\ref{random_initialization_table} provides the average agreement round and the achieved social welfare of each one of the models.
\begin{table}[H]
    \vspace{-6mm}

    \centering
    \begin{filecontents*}{\results/random_initialization/random_initialization.csv}
Agent,Rounds,Social Welfare
Continuous Training,81.670,0.712
Random Initialization,72.923,0.751
\end{filecontents*}
\csvautotabular{\results/random_initialization/random_initialization.csv}
    \caption{Comparison of random initialization and continuous training approaches on domain C}
    \label{random_initialization_table}
\end{table}
From Table~\ref{random_initialization_table}, we can see that when using the random initialization for the logistic regression model, our agent does manage to reach agreements faster, and therefore its achieved social welfare is higher.
}

\section{Conclusion and Future Work}


In this paper, we introduced an automated negotiation strategy for an environment with preference uncertainty that can be tuned by the user to balance between maximizing individual utility to maximizing social welfare.
The proposed design includes several innovative aspects such as combining two score functions in order to build a strategy for any linear tradeoff between individual utility and social welfare, using an opponent's acceptance probability as an estimate for its utility, a random initialization mechanism specifically suited for automated negotiation opponent's modeling, and an effective non-threshold approach for maximizing social welfare. 

Through an extensive evaluation, that includes 63 agents from ANAC 2015 - 2018, we show that in most scenarios, our agent achieves the highest scores, for various tradeoffs between social welfare and individual utility. When the goal is to fully maximize social welfare our agent outperforms all the top agents in almost every domain.

\notaamas{

Our motivation for developing such a strategy is to create a strategy that can simulate better the will of the user which in most cases is neither fully self-interested nor fully social. The majority of automated negotiation strategies today mainly focus on maximizing the individual utility of the user. In order to develop a strategy that can achieve a good agreement for the user both in the individual aspect of the agreement and the social aspect of the agreement, a study for social welfare strategies must be made as well.

In order to evaluate our strategy, we used 63 agents from ANAC 2015 - 2018 and compared our performance against the performance of the top 15 agents from these years. In order to cover as many scenarios as possible, including but not limited to the scenarios that were used in ANAC, we evaluated our agent using 7 domains as described in table \ref{domains}. From the results we made the following insights:

\begin{itemize}
    \item Our agent's achievements are stable and persistent on many domains and only influenced by the discount factor of the domain.

    \item Our agent, when fully maximizing social welfare, is able to achieve higher social welfare than all the top agents in all of the domains except for domains with a discount factor of 1.
    
    \item Our agent manages to achieve a higher beta score (i.e., the combination of individual utility and social welfare according to the user will) than most of the top agents in all of the domains except for domains with a discount factor of 1.
    
    \item Our agent's success is highly influenced by its low average agreement round which also explains why it does better in domains with discount factors different than 1.
    
    \item Our agent's performance is more affected by the offers it accepts rather than the offers it makes.
    
    \item Accepting most bids during the negotiation has a significant positive impact on the achieved social welfare.
\end{itemize}

\notaamas{
In addition, in our strategy, we make an assumption which the likelihood of a bid to be accepted is strongly correlated to the agent's utility function, and therefore we can use our Logistic Regression model as an approximation of the utility function of an agent. In our evaluation, we compared the correlation and the accuracy of the model we trained against the real utility function. Indeed with the results, we were able to verify our correlation assumption.
}
}

We see several interesting extension of this work to be carried out in future work. Among these we emphasize the need to evaluate our strategy with non-linear utility functions and non-linear tradeoff functions (which should not change much in terms of agent design, but requires further evaluation), extending the design to support settings where the agent has partial information about its own utility function, and domains with an extremely large solution space. In addition, future research can focus on extending our evaluation for more than 3-agent scenarios.

\bibliography{references}

\begin{thebibliography}{50}
\providecommand{\natexlab}[1]{#1}
\providecommand{\url}[1]{\texttt{#1}}
\expandafter\ifx\csname urlstyle\endcsname\relax
  \providecommand{\doi}[1]{doi: #1}\else
  \providecommand{\doi}{doi: \begingroup \urlstyle{rm}\Url}\fi

\bibitem[An et~al.(2010)An, Lesser, Irwin, and
  Zink]{automatednegotiationwithdecommitmentfordynamicresourceallocationincloudcomputing}
Bo~An, Victor Lesser, David Irwin, and Michael Zink.
\newblock Automated negotiation with decommitment for dynamic resource
  allocation in cloud computing.
\newblock volume~2, pages 981--988, 01 2010.
\newblock \doi{10.1145/1838206.1838338}.

\bibitem[Aydo{\u{g}}an et~al.(2017)Aydo{\u{g}}an, Festen, Hindriks, and
  Jonker]{AlternatingOffersProtocolsforMultilateralNegotiation}
Reyhan Aydo{\u{g}}an, David Festen, Koen~V. Hindriks, and Catholijn~M. Jonker.
\newblock \emph{Alternating Offers Protocols for Multilateral Negotiation},
  pages 153--167.
\newblock Springer International Publishing, Cham, 2017.
\newblock ISBN 978-3-319-51563-2.
\newblock \doi{10.1007/978-3-319-51563-2\_10}.
\newblock URL \url{https://doi.org/10.1007/978-3-319-51563-2\_10}.

\bibitem[Aydogan et~al.(2020)Aydogan, Fujita, Baarslag, Jonker, and
  Ito]{ANAC2018}
Reyhan Aydogan, Katsuhide Fujita, Tim Baarslag, Catholijn Jonker, and Takayuki
  Ito.
\newblock \emph{ANAC 2018: Repeated Multilateral Negotiation League}, pages
  77--89.
\newblock 02 2020.
\newblock ISBN 978-3-030-39877-4.
\newblock \doi{10.1007/978-3-030-39878-1\_8}.

\bibitem[Baarslag et~al.(2012)Baarslag, Hindriks, Jonker, Kraus, and
  Lin]{baarslag2012first}
Tim Baarslag, Koen Hindriks, Catholijn Jonker, Sarit Kraus, and Raz Lin.
\newblock \emph{The First Automated Negotiating Agents Competition (ANAC
  2010)}, pages 113--135.
\newblock Springer Berlin Heidelberg, Berlin, Heidelberg, 2012.
\newblock ISBN 978-3-642-24696-8.
\newblock \doi{10.1007/978-3-642-24696-8\_7}.
\newblock URL \url{https://doi.org/10.1007/978-3-642-24696-8\_7}.

\bibitem[Baarslag et~al.(2013)Baarslag, Fujita, Gerding, Hindriks, Ito,
  Jennings, Jonker, Kraus, Lin, Robu, et~al.]{baarslag2013evaluating}
Tim Baarslag, Katsuhide Fujita, Enrico~H Gerding, Koen Hindriks, Takayuki Ito,
  Nicholas~R Jennings, Catholijn Jonker, Sarit Kraus, Raz Lin, Valentin Robu,
  et~al.
\newblock Evaluating practical negotiating agents: Results and analysis of the
  2011 international competition.
\newblock \emph{Artificial Intelligence}, 198:\penalty0 73--103, 2013.

\bibitem[Baarslag et~al.(2014)Baarslag, Hindriks, Hendrikx, Dirkzwager, and
  Jonker]{DecouplingNegotiatingAgentstoExploretheSpaceofNegotiationStrategies}
Tim Baarslag, Koen Hindriks, Mark Hendrikx, Alexander Dirkzwager, and Catholijn
  Jonker.
\newblock \emph{Decoupling Negotiating Agents to Explore the Space of
  Negotiation Strategies}, pages 61--83.
\newblock Springer Japan, Tokyo, 2014.
\newblock ISBN 978-4-431-54758-7.
\newblock \doi{10.1007/978-4-431-54758-7\_4}.
\newblock URL \url{https://doi.org/10.1007/978-4-431-54758-7\_4}.

\bibitem[Bickerton(2002)]{GettingtoYes}
Jeff Bickerton.
\newblock Getting to yes: Negotiating agreement without giving in.
\newblock \emph{Quality Management Journal}, 9:\penalty0 73--74, 01 2002.
\newblock \doi{10.1080/10686967.2002.11919015}.

\bibitem[Breiman(2001)]{RandomForests}
Leo Breiman.
\newblock Random forests.
\newblock \emph{Mach. Learn.}, 45\penalty0 (1):\penalty0 5–32, October 2001.
\newblock ISSN 0885-6125.
\newblock \doi{10.1023/A:1010933404324}.
\newblock URL \url{https://doi.org/10.1023/A:1010933404324}.

\bibitem[Chen et~al.(2013)Chen, Ammar, Tuyls, and
  Weiss]{OptimizingcomplexautomatednegotiationusingsparsepseudoinputGaussianprocesses}
Siqi Chen, Haitham Ammar, Karl Tuyls, and Gerhard Weiss.
\newblock Optimizing complex automated negotiation using sparse pseudo-input
  gaussian processes.
\newblock volume~1, 05 2013.

\bibitem[Dirkzwager and Hendrikx(2014)]{TNR}
Alexander Dirkzwager and Mark Hendrikx.
\newblock \emph{An Adaptive Negotiation Strategy for Real-Time Bilateral
  Negotiations}, pages 163--170.
\newblock Springer Japan, Tokyo, 2014.
\newblock ISBN 978-4-431-54758-7.
\newblock \doi{10.1007/978-4-431-54758-7\_10}.

\bibitem[Endriss(2006)]{Monotonicconcessionprotocolformultilateralnegotiation}
Ulle Endriss.
\newblock Monotonic concession protocol for multilateral negotiation.
\newblock volume 2006, pages 392--399, 01 2006.
\newblock \doi{10.1145/1160633.1160702}.

\bibitem[Faratin and
  Sierra(1998)]{NegotiationDecisionFunctionsforAutonomousAgents}
Peyman Faratin and Carles Sierra.
\newblock Jennings: Negotiation decision functions for autonomous agents.
\newblock \emph{Robotics and Autonomous Systems - RaS}, 24, 01 1998.

\bibitem[Fujita et~al.(2017)Fujita, Aydo{\u{g}}an, Baarslag, Hindriks, Ito, and
  Jonker]{ANAC2015}
Katsuhide Fujita, Reyhan Aydo{\u{g}}an, Tim Baarslag, Koen Hindriks, Takayuki
  Ito, and Catholijn Jonker.
\newblock \emph{The Sixth Automated Negotiating Agents Competition (ANAC
  2015)}, pages 139--151.
\newblock Springer International Publishing, Cham, 2017.
\newblock ISBN 978-3-319-51563-2.
\newblock \doi{10.1007/978-3-319-51563-2\_9}.
\newblock URL \url{https://doi.org/10.1007/978-3-319-51563-2\_9}.

\bibitem[F{\"u}rnkranz(2010)]{DecisionTree}
Johannes F{\"u}rnkranz.
\newblock \emph{Decision Tree}, pages 263--267.
\newblock Springer US, Boston, MA, 2010.
\newblock ISBN 978-0-387-30164-8.
\newblock \doi{10.1007/978-0-387-30164-8\_204}.
\newblock URL \url{https://doi.org/10.1007/978-0-387-30164-8\_204}.

\bibitem[Ghosh et~al.(2014)Ghosh, Kyaw, and
  Verbrugge]{ConditionalPreferenceNetworksSupportMultiissueNegotiationswithMediator}
Sujata Ghosh, Thiri~Haymar Kyaw, and Rineke Verbrugge.
\newblock \emph{Conditional Preference Networks Support Multi-issue
  Negotiations with Mediator}, pages 171--195.
\newblock Springer Berlin Heidelberg, Berlin, Heidelberg, 2014.
\newblock ISBN 978-3-662-44994-3.
\newblock \doi{10.1007/978-3-662-44994-3\_9}.
\newblock URL \url{https://doi.org/10.1007/978-3-662-44994-3\_9}.

\bibitem[Gu and Ito(2017)]{AgentX}
Wen Gu and Takayuki Ito.
\newblock \emph{Agent X}, pages 239--242.
\newblock Springer International Publishing, Cham, 2017.
\newblock ISBN 978-3-319-51563-2.
\newblock \doi{10.1007/978-3-319-51563-2\_19}.
\newblock URL \url{https://doi.org/10.1007/978-3-319-51563-2\_19}.

\bibitem[Hayashi and Ito(2017)]{AgentH}
Masayuki Hayashi and Takayuki Ito.
\newblock \emph{AgentH}, pages 251--255.
\newblock Springer International Publishing, Cham, 2017.
\newblock ISBN 978-3-319-51563-2.
\newblock \doi{10.1007/978-3-319-51563-2\_21}.
\newblock URL \url{https://doi.org/10.1007/978-3-319-51563-2\_21}.

\bibitem[Hearst(1998)]{SVM}
Marti~A. Hearst.
\newblock Support vector machines.
\newblock \emph{IEEE Intelligent Systems}, 13\penalty0 (4):\penalty0 18–28,
  July 1998.
\newblock ISSN 1541-1672.
\newblock \doi{10.1109/5254.708428}.
\newblock URL \url{https://doi.org/10.1109/5254.708428}.

\bibitem[Hindriks and Tykhonov(2008)]{OpponentModellingBayesianLearning}
Koen Hindriks and Dmytro Tykhonov.
\newblock Opponent modelling in automated multi-issue negotiation using
  bayesian learning.
\newblock volume~1, pages 331--338, 05 2008.

\bibitem[Jonker et~al.(2017)Jonker, Aydogan, Baarslag, Fujita, Ito, and
  Hindriks]{ANAC}
Catholijn~M Jonker, Reyhan Aydogan, Tim Baarslag, Katsuhide Fujita, Takayuki
  Ito, and Koen Hindriks.
\newblock Automated negotiating agents competition (anac).
\newblock In \emph{Thirty-first AAAI conference on artificial intelligence},
  2017.

\bibitem[Kalpi{\'{c}} et~al.(2011)Kalpi{\'{c}}, Hlupi{\'{c}}, and
  Lovri{\'{c}}]{ttest}
Damir Kalpi{\'{c}}, Nikica Hlupi{\'{c}}, and Miodrag Lovri{\'{c}}.
\newblock \emph{Student's t-Tests}, pages 1559--1563.
\newblock Springer Berlin Heidelberg, Berlin, Heidelberg, 2011.
\newblock ISBN 978-3-642-04898-2.
\newblock \doi{10.1007/978-3-642-04898-2\_641}.
\newblock URL \url{https://doi.org/10.1007/978-3-642-04898-2\_641}.

\bibitem[Kavzoglu(1999)]{DeterminingOptimumStructureforArtificialNeuralNetworks}
Taskin Kavzoglu.
\newblock Determining optimum structure for artificial neural networks.
\newblock In \emph{Proceedings of the 25th Annual Technical Conference and
  Exhibition of the Remote Sensing Society}, pages 675--682. Remote Sensing
  Society Nottingham, UK Cardiff, UK, 1999.

\bibitem[Khosravimehr and Nassiri-Mofakham(2017)]{ParsAgent}
Zahra Khosravimehr and Faria Nassiri-Mofakham.
\newblock \emph{Pars Agent: Hybrid Time-Dependent, Random and Frequency-Based
  Bidding and Acceptance Strategies in Multilateral Negotiations}, pages
  175--183.
\newblock Springer International Publishing, Cham, 2017.
\newblock ISBN 978-3-319-51563-2.
\newblock \doi{10.1007/978-3-319-51563-2\_12}.

\bibitem[Kirch(2008)]{PearsonsCorrelationCoefficient}
Wilhelm Kirch, editor.
\newblock \emph{Pearson's Correlation Coefficient}, pages 1090--1091.
\newblock Springer Netherlands, Dordrecht, 2008.
\newblock ISBN 978-1-4020-5614-7.
\newblock \doi{10.1007/978-1-4020-5614-7\_2569}.
\newblock URL \url{https://doi.org/10.1007/978-1-4020-5614-7\_2569}.

\bibitem[Klein et~al.(2003)Klein, Faratin, Sayama, and
  Bar-Yam]{ProtocolsforNegotiatingComplexContracts}
Mark Klein, Peyman Faratin, Hiroki Sayama, and Yaneer Bar-Yam.
\newblock Protocols for negotiating complex contracts.
\newblock \emph{IEEE Expert / IEEE Intelligent Systems}, 18:\penalty0 32--38,
  12 2003.
\newblock \doi{10.1109/MIS.2003.1249167}.

\bibitem[{Krainin} et~al.(2007){Krainin}, {An}, and
  {Lesser}]{AnApplicationofAutomatedNegotiationtoDistributedTaskAllocation}
M.~{Krainin}, B.~{An}, and V.~{Lesser}.
\newblock An application of automated negotiation to distributed task
  allocation.
\newblock In \emph{2007 IEEE/WIC/ACM International Conference on Intelligent
  Agent Technology (IAT'07)}, pages 138--145, 2007.
\newblock \doi{10.1109/IAT.2007.28}.

\bibitem[Krainin et~al.(2007)Krainin, An, and
  Lesser]{ApplicationofAutomatedNegotiationtoDistributedTaskAllocation}
Michael Krainin, Bo~An, and Victor Lesser.
\newblock An application of automated negotiation to distributed task
  allocation.
\newblock In \emph{2007 IEEE/WIC/ACM International Conference on Intelligent
  Agent Technology (IAT'07)}, pages 138--145. IEEE, 2007.

\bibitem[Lam and Leung(2017)]{Phoenix}
Max Lam and Ho-fung Leung.
\newblock \emph{Phoenix: A Threshold Function Based Negotiation Strategy Using
  Gaussian Process Regression and Distance-Based Pareto Frontier
  Approximation}, volume 674, pages 201--212.
\newblock 04 2017.
\newblock ISBN 978-3-319-51561-8.
\newblock \doi{10.1007/978-3-319-51563-2\_15}.

\bibitem[Lin et~al.(2006)Lin, Kraus, Wilkenfeld, and
  Barry]{AnAutomatedAgentforBilateralNegotiationwithBoundedRationalAgentswithIncompleteInformation}
Raz Lin, Sarit Kraus, Jonathan Wilkenfeld, and James Barry.
\newblock An automated agent for bilateral negotiation with bounded rational
  agents with incomplete information.
\newblock volume 141, pages 270--274, 01 2006.

\bibitem[Lin et~al.(2014)Lin, Kraus, Baarslag, Tykhonov, Hindriks, and
  Jonker]{Genius}
Raz Lin, Sarit Kraus, Tim Baarslag, Dmytro Tykhonov, Koen Hindriks, and
  Catholijn~M. Jonker.
\newblock Genius: An integrated environment for supporting the design of
  generic automated negotiators.
\newblock \emph{Computational Intelligence}, 30\penalty0 (1):\penalty0 48--70,
  2014.
\newblock ISSN 1467-8640.
\newblock \doi{10.1111/j.1467-8640.2012.00463.x}.
\newblock URL \url{http://dx.doi.org/10.1111/j.1467-8640.2012.00463.x}.

\bibitem[Liu et~al.(2018)Liu, Moustafa, and Ito]{Agent33}
Shan Liu, Ahmed Moustafa, and Takayuki Ito.
\newblock Agent33: An automated negotiator with heuristic method for searching
  bids around nash bargaining solution.
\newblock In Tim Miller, Nir Oren, Yuko Sakurai, Itsuki Noda, Bastin Tony~Roy
  Savarimuthu, and Tran Cao~Son, editors, \emph{PRIMA 2018: Principles and
  Practice of Multi-Agent Systems}, pages 519--526, Cham, 2018. Springer
  International Publishing.
\newblock ISBN 978-3-030-03098-8.

\bibitem[Matsune and
  Fujita(2018)]{WeightingEstimationMethodsforOpponentsUtilityFunctionsUsingBoostinginMultiTimeNegotiations}
Takaki Matsune and Katsuhide Fujita.
\newblock Weighting estimation methods for opponents' utility functions using
  boosting in multi-time negotiations.
\newblock \emph{IEICE Transactions on Information and Systems},
  E101.D:\penalty0 2474--2484, 10 2018.
\newblock \doi{10.1587/transinf.2018EDP7056}.

\bibitem[McCullagh and Nelder(1989)]{GeneralizedLinearModels}
P.~McCullagh and J.A. Nelder.
\newblock \emph{Generalized Linear Models, Second Edition}.
\newblock Chapman and Hall/CRC Monographs on Statistics and Applied Probability
  Series. Chapman \& Hall, 1989.
\newblock ISBN 9780412317606.
\newblock URL \url{http://books.google.com/books?id=h9kFH2\_FfBkC}.

\bibitem[Meier(2006)]{SurveyofEconomic}
Stephan Meier.
\newblock A survey of economic theories and field evidence on pro-social
  behavior.
\newblock \emph{SSRN Electronic Journal}, 02 2006.
\newblock \doi{10.2139/ssrn.917187}.

\bibitem[Mirzayi et~al.(2018)Mirzayi, Taghiyareh, and Kazemi]{IQSon}
Sahar Mirzayi, Fattaneh Taghiyareh, and Seyed Mohammad~Hussein Kazemi.
\newblock Iqson: A context-aware negotiator agent with enhanced utility and
  decision making speed.
\newblock In \emph{2018 9th International Symposium on Telecommunications
  (IST)}, pages 603--608. IEEE, 2018.

\bibitem[Mori and Ito(2017)]{Atlas3}
Akiyuki Mori and Takayuki Ito.
\newblock \emph{Atlas3: A Negotiating Agent Based on Expecting Lower Limit of
  Concession Function}, pages 169--173.
\newblock Springer International Publishing, Cham, 2017.
\newblock ISBN 978-3-319-51563-2.
\newblock \doi{10.1007/978-3-319-51563-2\_11}.
\newblock URL \url{https://doi.org/10.1007/978-3-319-51563-2\_11}.

\bibitem[Nongaillard and
  Mathieu(2009)]{MultiagentResourceNegotiationforSocialWelfare}
Antoine Nongaillard and Philippe Mathieu.
\newblock A multi-agent resource negotiation for social welfare.
\newblock volume~2, pages 58--61, 01 2009.
\newblock \doi{10.1109/WI-IAT.2009.126}.

\bibitem[Nongaillard et~al.(2008)Nongaillard, Mathieu, and
  Jaumard]{MultiAgentResourceNegotiationfortheUtilitarianWelfare}
Antoine Nongaillard, Philippe Mathieu, and Brigitte Jaumard.
\newblock A multi-agent resource negotiation for the utilitarian welfare.
\newblock In \emph{International Workshop on Engineering Societies in the
  Agents World}, pages 208--226. Springer, 2008.

\bibitem[Ounpraseuth(2008)]{GaussianProcessesforMachineLearning}
Songthip Ounpraseuth.
\newblock Gaussian processes for machine learning.
\newblock \emph{Journal of the American Statistical Association}, 103:\penalty0
  429--429, 03 2008.
\newblock \doi{10.1198/jasa.2008.s219}.

\bibitem[Potdar et~al.(2017)Potdar, Pardawala, and
  Pai]{ComparativeStudyofCategoricalVariableEncodingTechniquesforNeuralNetworkClassifiers}
Kedar Potdar, Taher Pardawala, and Chinmay Pai.
\newblock A comparative study of categorical variable encoding techniques for
  neural network classifiers.
\newblock \emph{International Journal of Computer Applications}, 175:\penalty0
  7--9, 10 2017.
\newblock \doi{10.5120/ijca2017915495}.

\bibitem[Purrington and
  Durfee(2009)]{AgreeingOnSocialOutcomesUsingIndividualCPnets}
Keith Purrington and Edmund Durfee.
\newblock Agreeing on social outcomes using individual cp-nets.
\newblock \emph{Multiagent and Grid Systems}, 5:\penalty0 409--425, 12 2009.
\newblock \doi{10.3233/MGS-2009-0136}.

\bibitem[Ruder(2016)]{SGD}
Sebastian Ruder.
\newblock An overview of gradient descent optimization algorithms.
\newblock \emph{arXiv preprint arXiv:1609.04747}, 2016.

\bibitem[Sammut and Webb(2010)]{MeanAbsoluteError}
Claude Sammut and Geoffrey~I. Webb, editors.
\newblock \emph{Mean Absolute Error}, pages 652--652.
\newblock Springer US, Boston, MA, 2010.
\newblock ISBN 978-0-387-30164-8.
\newblock \doi{10.1007/978-0-387-30164-8\_525}.
\newblock URL \url{https://doi.org/10.1007/978-0-387-30164-8\_525}.

\bibitem[Sosale et~al.(2017)Sosale, Satish, and An]{AgentBuyog}
Bhargav Sosale, Swarup Satish, and Bo~An.
\newblock \emph{Agent Buyog: A Negotiation Strategy for Tri-Party Multi Issue
  Negotiation}, volume 674, pages 191--199.
\newblock 04 2017.
\newblock \doi{10.1007/978-3-319-51563-2\_14}.

\bibitem[Williams et~al.(2010)Williams, Robu, Gerding, and
  Jennings]{IAMhaggler}
Colin Williams, Valentin Robu, Enrico Gerding, and Nicholas Jennings.
\newblock Iamhaggler: A negotiation agent for complex environments.
\newblock \emph{Studies in Computational Intelligence}, 383, 10 2010.
\newblock \doi{10.1007/978-3-642-24696-8\_10}.

\bibitem[Williams et~al.(2011)Williams, Robu, Gerding, and
  Jennings]{UsingGaussianProcessestoOptimiseConcessioninComplexNegotiationsagainstUnknownOpponents}
Colin Williams, Valentin Robu, Enrico Gerding, and Nicholas Jennings.
\newblock Using gaussian processes to optimise concession in complex
  negotiations against unknown opponents.
\newblock \emph{IJCAI International Joint Conference on Artificial
  Intelligence}, 01 2011.
\newblock \doi{10.5591/978-1-57735-516-8/IJCAI11-080}.

\bibitem[Williams et~al.(2013)Williams, Robu, Gerding, and
  Jennings]{IAMhaggler2011}
Colin Williams, Valentin Robu, Enrico Gerding, and Nicholas Jennings.
\newblock Iamhaggler2011: A gaussian process regression based negotiation
  agent.
\newblock 435:\penalty0 209--212, 01 2013.
\newblock \doi{10.1007/978-3-642-30737-9-14}.

\bibitem[Yucel et~al.(2017)Yucel, Hoffman, and Sen]{JonnyBlack}
Osman Yucel, Jon Hoffman, and Sandip Sen.
\newblock \emph{Jonny Black: A Mediating Approach to Multilateral
  Negotiations}, pages 231--238.
\newblock Springer International Publishing, Cham, 2017.
\newblock ISBN 978-3-319-51563-2.
\newblock \doi{10.1007/978-3-319-51563-2\_18}.
\newblock URL \url{https://doi.org/10.1007/978-3-319-51563-2\_18}.

\bibitem[Zafari and Nassiri-Mofakham(2016)]{BraveCat}
Farhad Zafari and Faria Nassiri-Mofakham.
\newblock \emph{BraveCat: Iterative Deepening Distance-Based Opponent Modeling
  and Hybrid Bidding in Nonlinear Ultra Large Bilateral Multi Issue Negotiation
  Domains}, pages 285--293.
\newblock Springer International Publishing, Cham, 2016.
\newblock ISBN 978-3-319-30307-9.
\newblock \doi{10.1007/978-3-319-30307-9\_21}.
\newblock URL \url{https://doi.org/10.1007/978-3-319-30307-9\_21}.

\bibitem[Zheng et~al.(2013)Zheng, Dai, Chakraborty, and
  Sycara]{Multiagentnegotiationonmultipleissueswithincompleteinformation}
Ronghuo Zheng, T.~Dai, Nilanjan Chakraborty, and Katia Sycara.
\newblock Multiagent negotiation on multiple issues with incomplete
  information.
\newblock \emph{12th International Conference on Autonomous Agents and
  Multiagent Systems 2013, AAMAS 2013}, 2:\penalty0 1279--1280, 01 2013.

\end{thebibliography}

\begin{appendices}
\section*{Excepted Utility Approach}

A different approach from the one we describe in the Section~\ref{bid_valuation}, for assigning a social welfare-based score, is to calculate the score based on the expected social welfare of proposing the bid. However, calculating the expected social welfare requires a lot of knowledge, which is unavailable to us. In order to calculate the expected social welfare from proposing a bid, we need to be able the calculate the probability of the bid to be accepted by all opponents. While we model the probability of each one of the opponents to accept a bid, we do not know the correlation between the profiles and the acceptance strategies of the agents. For example, if we have two opponents with the sample profile and same strategy and we assume their acceptances are independent, given a bid with a probability of 0.5 to be accepted by each one of the opponents, we will assume the likelihood of both the opponents to accept the bid is 0.25 while it is actually 0.5.

If we still assume the events of opponents accepting bids are independent for each opponent, we can calculate the expected social welfare of proposing a bid (i.e., the expected social welfare of the next agent receiving a bid) by observing at the two outcomes of proposing the bid: an acceptance of the bid by the next agent or rejection (and a counteroffer) of the bid. The expected social welfare of the next opponent $a$ from receiving a bid $b$, at a certain round $r$, after the bid was accepted by $c$ opponents, in negotiation with $n$ agents, knowing the opponent will accept it is:

\[
expected\_social\_welfare_A(a, b, c, r) = expected\_social\_welfare((a + 1) \% n, b, c + 1, r + 1)
\]

The expected social welfare knowing the next opponent is going to reject the bid and propose a counteroffer is:

\[expected\_social\_welfare_R(a, b, c, r) = expected\_social\_welfare((a + 1) \% n, next\_bid(a), 0, r + 1)\]

Where $next\_bid(a)$ is the next bid agent $a$ is going to propose.
Therefore, the expected social welfare of the next agent receiving a bid, knowing only the probability of his acceptance of the bid, is:

\[
expected\_social\_welfare(a, b, c, r) = 
\]

\[
acceptance\_likelihood_{a, r}(b) * expected\_social\_welfare_A(a, b, c, r) + 
\]

\[
(1 - acceptance\_likelihood_{a, r}(b)) *
expected\_social\_welfare_R(a, b, c, r) 
\]

Where $acceptance\_likelihood_{a, r}(bid)$ is the likelihood of agent $a$ to accept a bid $b$ at round $r$. The base case of this recursion is either all the agents accepted a bid (i.e., $c = n - 1$) or no agreement was found within the round limit.

\[base\_case(a, b, c, r)=
\begin{cases}
    d_r * \sum_{i}^{n}utility_i(b) & \text{if }c = n - 1\\
    d_r * \sum_{i}^{n}reservation_i  & \text{if }round = limit
\end{cases} 
\]

Where $utility_i(bid)$ the utility of the i-th agent from bid and $reservation_i$ is the reservation of the i-th agent. 

Multiple issues arise when trying to calculate the score of a bid using this equation:
\begin{itemize}
    \item The opponent's counteroffer is unknown -- The function $next\_bid(a)$ is unknown to us, and therefore, we do not know what the agent will offer in case he will decide to reject our offer.
    
    \item Our next actions are unknown -- Since our actions are influenced by the offers the opponents make, since we do not know what offers the opponent are going to offer, we also do not know what actions we are going to take in the next turns.
    
    \item Exponential time complexity -- Even given all the missing information mentioned above, the complexity of calculating this equation is at least $O(2^{limit})$. In our settings, the round limit of the negotiations is 180. Therefore, the computation becomes unfeasible.
    
    \item Compatibility with real-time limit -- In our settings, the limit of each negotiation is 180 rounds. However, the limit can also be a real-time value. In ANAC the limit of each negotiation is 180 seconds. In such settings we would also not be able to know the number of rounds and therefore would not be able to calculate the base case (although we can try to estimate it by looking at the time of the previous rounds).
\end{itemize}

Nevertheless, we made an attempt to create a scoring function based on expected social welfare. In order the handle the mentioned issues, we made the following assumptions:

\begin{itemize}
    \item The events of each agent accepting a certain bid are independent.
    
    \item The actions of the opponents in the next turns are not going to affect our model in a way that will change our behavior.
    
    \item The negotiation is going to end with either an agreement on a bid that was proposed by our agent or a disagreement.
\end{itemize}

Of course, those assumptions will not always be correct, but this is a compromise we take in order to handle the issues in calculating the expected social welfare. 

Since we assume the next offers of the agents will not affect our behavior and the negotiation will not end with an agreement on an offer of an opponent, the offers of the opponent will not influence the rest of the negotiation and we might as well ignore them. In addition, since the opponents' actions will not affect our behavior, the next offer we are going to make is the same offer we are going to make until the end of the negotiation. Finally, in order to calculate the probability of a bid to be accepted by all opponents, we can use the independence assumptions and multiply the likelihood of each agent separately. Therefore, our scoring function will be calculated as follows: 

\[
social\_score(b, r)=
\begin{cases}
    (\prod_{i}^{n}acceptance\_likelihood_i(b)) * d_r * \sum_{i}^{n}utility_i(b) + \\
    (1 - \prod_{i}^{n}acceptance\_likelihood_i(b)) * social\_score(b, r + 1) & \text{if } r \neq limit\\
    d_r * \sum_{i}^{n}reservation_i  & \text{else}
\end{cases} 
\]

In order to calculate the acceptance likelihood of each agent, we can use our opponent model. To calculate the utility each agent gains from a certain bid, we can use the assumptions of the strong correlation between an agent's acceptance likelihood and an agent's utility function and calculate the utility using the output of the opponent model.

When evaluating our agent, we found this scoring function's performance to be less well than our original scoring function. A comparison of the two scoring functions is shown in table~\ref{expected_social_welfare}. 

\begin{table}[H]
    \centering
    \csvautotabular{\results/expected_social_welfare/social_welfare.csv}
    \caption{Social welfare for both scoring functions on domain C}
    \label{expected_social_welfare}
\end{table}

\end{appendices}

\end{document}